\documentclass[prd,aps,showpacs,nofootinbib,preprintnumbers,onecolumn]{revtex4-1}
\usepackage{amsfonts,amsmath,amssymb,bm,dsfont,eucal}
\usepackage{multirow}
\usepackage[english]{babel}
\usepackage[latin1]{inputenc}
\usepackage[T1]{fontenc}
\usepackage{ae,aecompl}
\usepackage{graphicx,graphics,epsfig,epstopdf,subfigure,color,psfrag}
\usepackage{hyperref}

\newcommand{\spur}[1]{\not\! #1 \,}
\newcommand{\be}{\begin{equation}}
\newcommand{\ee}{\end{equation}}
\newcommand{\bea}{\begin{eqnarray}}
\newcommand{\eea}{\end{eqnarray}}
\newcommand{\nn}{\nonumber}

\newcommand{\pVvec}{|{\vec p}_V|}

\begin{document}

\preprint{BARI-TH/718-2018}
\title{Excited heavy meson  decays   to light vector mesons: \\ \vspace*{0.2cm} implications for spectroscopy}
\author{S.~Campanella, P.~Colangelo and  F.~De~Fazio}
\affiliation{
Istituto Nazionale di Fisica Nucleare, Sezione di Bari, Via Orabona 4, I-70126 Bari, Italy}

\begin{abstract}
We analyze strong decays of excited charmed and beauty mesons into a light vector meson,  exploiting the  effective field theory  based on heavy quark (HQ) symmetries for heavy mesons,  and on the hidden gauge symmetry   to incorporate  light vector mesons. HQ symmetries allow to classify the heavy mesons in spin doublets, and to  relate decays of  excited  states. 
We build  effective Lagrangian terms governing the  ${\cal H}_i \to P^{(*)} V$ modes, with ${\cal H}_i$  an excited  $s$, $p$, $d$ and $f$-wave heavy-light quark meson,  $P,\,P^*$  the lowest-lying $J^P=(0^-,\,1^-)$ heavy-light mesons, and
V a light vector meson. Predictions are provided for ratios of decay widths  that are independent of the strong couplings  in the effective Lagrangian terms.
A classification of  the newly observed heavy-light mesons is proposed.
  \end{abstract}
\maketitle

\section{Introduction}
Great progress has been  achieved in heavy hadron spectroscopy,  thanks to the efforts of several experimental groups at different facilities which have provided new  pieces of information  \cite{Chen:2016spr}.
In the open-charm meson  spectrum the two lowest-lying (1S)  and the  four 1P orbital excitations  are  identified,  both for non-strange and strange mesons \cite{Patrignani:2016xqp}.
Information is available for larger mass states which could be identified either with higher orbital  or radial excitations.  Experimental observations are less abundant  in the case of beauty mesons: the established states are the two lowest-lying  (1S) states and   two among the four 1P orbital excitations, both with and without strangeness  \cite{Patrignani:2016xqp}.
There is progress also in baryon spectroscopy, with the observation of  five new narrow  $\Omega_c$ resonances \cite{Aaij:2017nav} and of the  doubly-charmed $\Xi_{cc}$ \cite{Aaij:2017ueg}:
however, in this paper we are only  concerned with  mesons.

Prompt production and production in $B$ decays, the  main production mechanisms of excited charmed mesons,  provide us with different and complementary information. Prompt production allows to establish  if a state has natural ($J^P=0^+,\,1^-,\,2^+,\, \dots$) or unnatural ($J^P=0^-,\,1^+,\,2^-,\, \dots$) parity, while  spin-parity can be determined  by Dalitz plot analyses in $B$ decay production.
On the other hand, it is possible to measure ratios of branching fractions of strong decay modes, an information that can be used to classify the decaying meson,  as we are going to discuss.
\begin{table}[b]
 \centering \caption{Mass, width and spin-parity of  charmed resonances observed by BaBar Collaboration  \cite{delAmoSanchez:2010vq}.}
 \label{tabBaBar}
 \begin{tabular}{c c c c}
   \hline \hline
   Resonance & mass (MeV) & $\Gamma$ (MeV) & $J^P$ \\ 
   \hline
   $D^0(2550) $ & $2539.4 \pm 4.5 \pm 6.8$ & $130 \pm 12 \pm 13$ & $0^-$ \\
  \hline 
   $D^{*0}(2600)$ & $2608.7 \pm 2.4 \pm 2.5$ & $93 \pm 6 \pm 13$ & natural \\
   $D^{*+}(2600)$ & $2621.3 \pm 3.7 \pm 4.2$ & $93 \ \text{(fixed)}$ & natural \\
\hline
   $D^0(2750)$ & $2752.4 \pm 1.7 \pm 2.7$ & $71 \pm 6 \pm 11$ & \\
  \hline
   $D^{*0}(2760)$ & $2763.3 \pm 2.3 \pm 2.3$ & $60.9  \pm 5.1 \pm 3.6$ & natural \\
   $D^{*+}(2760)$ & $2769.7 \pm 3.8 \pm 1.5$ & $60.9 \ \text{(fixed)}$ & natural \\
   \hline \hline
 \end{tabular}
\end{table}
\begin{table}[b]
 \centering \caption{Mass, width and spin-parity of   charmed mesons from the  LHCb  analysis of inclusive $D^{(*)} \pi$ production \cite{Aaij:2013sza}.}
 \label{tabLHCb13}
 \begin{tabular}{c c c c}
   \hline \hline
   Resonance & mass (MeV) & $\Gamma$ (MeV) & $J^P$ \\ 
   \hline
   $D_J^0(2580)$     & $2579.5 \pm 3.4 \pm 5.5$  & $177.5 \pm 17.7 \pm 46.0$ & unnatural \\
\hline
   $D^{*0}_J(2650)$   & $2649.2 \pm 3.5 \pm 3.5$  & $140.2 \pm 17.1 \pm 18.6$ & natural \\
\hline
   $D_J^0(2740)$     & $2737.0 \pm 3.5 \pm 11.2$ & $73.2 \pm 13.4 \pm 25.0$  & unnatural \\
  \hline
   $D^{*0}_J(2760)$   & $2761.1 \pm 5.1 \pm 6.5$  & $74.4 \pm 4.3 \pm 37.0$   & natural \\
   $D^{*0}_J(2760)$   & $2760.1 \pm 1.1 \pm 3.7$  & $74.4 \pm 3.4 \pm 19.1$   & natural \\
   $D^{*+}_J(2760)$ & $2771.7 \pm 1.7 \pm 3.8$  & $66.7 \pm 6.6 \pm 10.5$   & natural \\
   $D_J^0(3000)$     & $2971.8 \pm 8.7$          & $188.1 \pm 44.8$          & unnatural \\
\hline
   $D^{*0}_J(3000)$   & $3008.1 \pm 4.0$          & $110.5 \pm 11.5$          & natural \\
   $D^{*+}_J(3000)$ & $3008.1 \ \text{(fixed)}$ & $110.5 \ \text{(fixed)}$  & natural \\
   \hline \hline
 \end{tabular}
\end{table}
\begin{table}
 \centering \caption{Mass, width and spin-parity of  charmed mesons observed by LHCb in  Dalitz plot analysis of $B^- \to D^+ \pi^- \pi^-$ \cite{Aaij:2016fma}.}
 \label{tabLHCb16}
 \begin{tabular}{c c c c}
   \hline \hline
   Resonance & mass (MeV) & $\Gamma$ (MeV) & $J^P$ \\ 
   \hline
   $D^{*0}_1(2680)$   & $2681.1 \pm 5.6 \pm 4.9 \pm 13.1$ & $186.7 \pm 8.5 \pm 8.6 \pm 8.2$  & $1^-$ \\
   \hline
   $D^{*0}_3(2760)$   & $2775.5 \pm 4.5 \pm 4.5 \pm 4.7$  & $95.3  \pm 9.6 \pm 7.9 \pm 33.1$ & $3^-$ \\
   \hline
   $D^{*0}_2(3000)$   & $3214   \pm 29  \pm 33  \pm 36$   & $186   \pm 39  \pm 34  \pm 63$   & $2^+$ \\
   \hline \hline
 \end{tabular}
\end{table}
\begin{table}
 \centering \caption{Mass, width and spin-parity of the latest observed  strange-charmed mesons. }
 \label{tabStrani}
 \begin{tabular}{c c c c c}
   \hline \hline
   Resonance & mass (MeV) & $\Gamma$ (MeV) & $J^P$ & ref.\\ 
   \hline
 $D^*_{s1}(2700)$& $2709.2 \pm 1.9 \pm4.5$ & $115.8 \pm 7.3 \pm 12.1$  & $1^-$ & \cite{Aaij:2012pc} \\
  $D^*_{s1}(2700)$& $2699 \pm^{14}_7$ & $127 \pm^{24}_{19}$  & $1^-$ & \cite{Lees:2014abp} \\
      \hline 
      $D^*_{s1}(2860)$   & $2859 \pm 12 \pm 6 \pm 23$ & $159 \pm 23 \pm 27 \pm72$  & $1^-$ & \cite{Aaij:2014xza} \\
   \hline
   $D^*_{s3}(2860)$   & $2860.5 \pm 2.6\pm 2.5 \pm 6$  & $53  \pm 7 \pm 4 \pm 6$ & $3^-$ & \cite{Aaij:2014xza} \\
   \hline
   $D_{sJ}(3040)$   & $3044   \pm 8  \pm^{30}_5 $   & $239   \pm 35  \pm^{46}_{42}$   &  & \cite{Aubert:2009ah} \\
   \hline \hline
 \end{tabular}
\end{table}
\begin{table}[b]
 \centering \caption{PDG fit for the mass and  width of  non-strange  beauty mesons with uncertain classification \cite{Patrignani:2016xqp}. }
 \label{tabB}
 \begin{tabular}{c c c }
   \hline \hline
   Resonance & mass (MeV) & $\Gamma$ (MeV) \\ 
   \hline
 $B_J^+(5840)$& $5851 \pm 19$ & $571 \pm 19$   \\
 $B_J^0(5840)$& $5863 \pm 9$ & $584 \pm 9$   \\
  $B_J^+(5970)$& $5964 \pm 5$ & $685\pm 5$   \\
   $B_J^0(5970)$& $5971 \pm 5$ & $691\pm 5$   \\
   \hline \hline
 \end{tabular}
\end{table}

Several  observed open charm mesons  are awaiting for a proper identification.
In Table \ref{tabBaBar} we include the resonances observed by BaBar Collaboration (in 2010)  in  the inclusive production of $D^+ \pi^-$, $D^0 \pi^+$ and $D^{*+}\pi^-$ 
 \cite{delAmoSanchez:2010vq}.  LHCb Collaboration has  performed a similar analysis   (in 2013), with the findings  in Table \ref{tabLHCb13} \cite{Aaij:2013sza}. It has  also carried out  (in 2016) a  Dalitz plot analysis of $B^- \to D^+ \pi^- \pi^-$, reporting evidence of the resonances   in Table \ref{tabLHCb16}  \cite{Aaij:2016fma}.
 Many of the states found in the different analyses are likely to be the same,  namely the  BaBar  $D^0(2550)$ and $D^{*0}(2600)$  states  in Table \ref{tabBaBar} coincide with   the LHCb ones   $D_J^0(2580)$ and  $D_J^{*0}(2650)$  in Table \ref{tabLHCb13}.   $D_1^{*0}(2680)$ in Table \ref{tabLHCb16} is probably different from $D^{*0}(2600)$, although presumably they both  have $J^P=1^-$. The identification of $D^0(2750)$ in Table \ref{tabBaBar} with $D_J^0(2740)$ in Table \ref{tabLHCb13} is also plausible. Two different resonances are present in the mass range  around $2760$ MeV: one having $J^P=1^-$ and another one with  $J^P=3^-$. The state in Table \ref{tabLHCb16} is definitely the latter one,  reported by  LHCb   \cite{Aaij:2015sqa}. In the cases of mesons with mass close to  $3000$ MeV   LHCb  has not provided a systematic uncertainty for the parameters  in Table \ref{tabLHCb13}, since the states are observed at the limit of the considered mass range.
The  latest results  for strange-charmed mesons  are  in Table \ref{tabStrani}. 
 
 While spin-parity of charmed mesons can be established  by  the  amplitude analyses in  production  in $B$ decays, for beauty mesons the
quantum number assignment is more difficult.
In addition to the above mentioned established states, recent observations are due to CDF and  LHCb  Collaborations.
CDF found a state named $B(5970)$ \cite{Aaltonen:2013atp},  likely the same as $B_J^{0,+}(5960)$  observed by LHCb  together with  $B_J^{0,+}(5840)$   decaying to $B^+\pi^-$, $B^0\pi^+$ \cite{Aaij:2015qla}. 
Spin-parity is not established,   and   mass and width are affected by large uncertainties: the values  from PDG fits  \cite{Patrignani:2016xqp} are in Table \ref{tabB}.
The identification as  2S  excitations  has been proposed \cite{Aaij:2015qla}. New results on $B_{s1}(5830)$ and $B^*_{s2}(5840)$  have also been obtained by CMS \cite{Sirunyan:2018grk}.

In \cite{Colangelo:2012xi} a comprehensive analysis of  the  open charm and open beauty mesons was performed based on  the classification scheme  in the heavy quark limit,  attempting  to fit  the observed states in this  scheme. Information on  the strong decay modes to $D_{(s)} M$ or $D^*_{(s)}M$, with $M$  a light pseudoscalar meson, was exploited, 
and the states  in Table \ref{tabBaBar} and  most of those in Table \ref{tabStrani}  were considered. 
 In the same approach, studies for  the states in Tables \ref{tabLHCb13} and \ref{tabLHCb16}   observed after the analysis in \cite{Colangelo:2012xi}  have been carried out  in \cite{Wang:2016ewb,Gupta:2018zlg}.

More data are still needed for  classification, which is a non trivial task for  the newly observed mesons. If the resonance mass is large enough, several   decay channels are open, in particular those with a  light final vector meson which provide important new piece of information.
This paper is devoted to such a phenomenology.

In the next Section we restate the theoretical framework based on heavy quark (HQ) symmetries  to describe spectrum and decay processes. For transitions into  light pseudoscalars, effective Lagrangians are  written exploiting  the  HQ symmetries  and the  (spontaneoulsy broken) chiral symmetry holding in QCD for massless  $u,\,d,\,s$ quarks,  with the  light pseudoscalar mesons  being  the pseudo-Goldstone bosons.
The approach can be extended to  incorporate  the light vector mesons, treated  as gauge fields of a hidden local symmetry.
In Section \ref{sec:results} 
we construct  the effective Lagrangians describing strong  heavy-light meson decays with the emission of a light vector meson, generalizing the analysis in \cite{Casalbuoni:1992gi,Casalbuoni:1996pg}.  We give the   expressions  for   ${\cal H}_i \to P^{(*)} V$ decay rates, with  V  a light vector meson, $P^{(*)}$ the lowest-lying $J^P=(0^-,\,1^-)$ heavy-light mesons, and ${\cal H}_i$  either a orbital or a radial a heavy-light excitation. 
Sections \ref{sec:numerics}  and \ref{sec:numericsB}  contain  numerical analyses  for charmed and beauty mesons, considering states requiring identification and making predictions for heavier excitations.  Relations among decay rates, independent of the hadronic couplings,  are constructed:  they are suitable for experimental measurements and for classifications. The conclusions are presented in the last Section.

\section{Theoretical setup}\label{sec:th}
\subsection{Heavy-light meson decays to light pseudoscalar mesons}
The physics of hadrons containing a single heavy quark can be systematically analyzed considering the $m_Q \to \infty$ heavy quark (HQ) mass limit,  formalized in the Heavy Quark Effective Theory (HQET) \cite{Neubert:1993mb}. In such a limit, two symmetries emerging in QCD can be exploited:   the heavy quark spin symmetry, allowing to relate the properties of hadrons which only differ  for the heavy quark spin orientation, and the heavy quark flavour symmetry, relating the properties of hadrons which only differ  for the heavy quark flavour.
The classification of heavy-light  $Q{\bar q}$ mesons ($\bar q$  a light antiquark) in the HQ limit  is based on  the decoupling of the heavy quark spin $s_Q$ from the total angular momentum  $s_\ell$ of the light degrees of freedom (light quarks and gluons). Since such angular momenta are separately conserved in strong interaction processes,  the  heavy mesons  can be classified in doublets of different  $s_\ell$. 
Each doublet comprises two states,  spin partners,  with  total spin  $J=s_\ell \pm {1 \over 2}$ and parity $P=(-1)^{\ell +1}$,  $\ell$  being the orbital angular momentum of the light degrees of freedom and ${\vec s}_\ell={\vec \ell}+ {\vec s}_q$ ($s_q$  the light antiquark spin). 
In the  HQ limit the spin partners in each doublet are degenerate and,  due to flavour symmetry,  the properties of the states in a  doublet are related to those of the corresponding states differing for the heavy flavour.

Meson doublets corresponding to $\ell=0,1, 2$ and $3$ are those referred to as $s$, $p$, $d$ and $f$ wave states in the constituent quark model.
 The lowest lying $Q \bar q$ mesons correspond to $\ell=0$, hence
$ s_\ell^P={1 \over 2}^-$;
 the  doublet consists of   two states with  
$J^P=(0^-,1^-)$, denoted as $(P,P^*)$. For $\ell=1$ one has $ s_\ell^P={1 \over 2}^+$ and $ s_\ell^P={3 \over 2}^+$.
The  two  doublets have $J^P=(0^+,1^+)$ and
$J^P=(1^+,2^+)$, respectively;     the members of the
$J^P_{s_\ell}=(0^+,1^+)_{{1/2}}$ doublet are denoted  as
$(P^*_{0},P_{1}^\prime)$,  those of the
$J^P_{s_\ell}=(1^+,2^+)_{3/2}$  doublet as $(P_{1},P^*_{2})$.
For $\ell=2$ one has  $s_\ell^P={3 \over 2}^-$,  and  $s_\ell^P={5 \over 2}^-$;
the first doublet comprises   $(P_1^*,P_2)$ with $J^P=(1^-,\,2^-)$,
the second one  the states $(P_2^{\prime }, P_3^*)$ with $J^P=(2^-,\,3^-)$. 
One can continue with $\ell=3$, which gives $s_\ell^P=\frac{5}{2}^+$ and $s_\ell^P=\frac{7}{2}^+$: here we only consider the first one of such doublets, which comprises two states with $J^P=(2^+,\,3^+)$, denoted as $(P_2^{\prime *}, P_3)$. The same classification holds for  radial excitations: for them we 
use  the same  notation,   but for  a tilde  ($\tilde P$, $\tilde P^*$, ...).
 
Effective Lagrangians describing the strong interactions of such  mesons can be constructed introducing effective fields for each doublet, following  e.g. the procedure  based on the covariant representation of the states \cite{Falk:1991nq}.
We denote by  $H_a$ ($a=u,d,s$  a light flavour index) the $s_\ell^P={ 1 \over 2}^-$ doublet,  $S_a$ and $T_a$  the $ s_\ell^P={1 \over 2}^+$ and $s_\ell^P={3 \over 2}^+$ doublets,  $X_a$ the $ s_\ell^P={3 \over 2}^-$,   $X_a^\prime$  the  $s_\ell^P={5 \over 2}^-$ and $F_a$ the $s_\ell^P=\frac{5}{2}^+$ doublets.  The corresponding effective fields read: 
 \bea
H_a & =& \frac{1+{\rlap{v}/}}{2}[P_{a\mu}^*\gamma^\mu-P_a\gamma_5] \, , \label{neg} \nn  \\
S_a &=& \frac{1+{\rlap{v}/}}{2} \left[P_{1a}^{\prime \mu}\gamma_\mu\gamma_5-P_{0a}^*\right]   \, , \nn \\
T_a^\mu &=&\frac{1+{\rlap{v}/}}{2} \Bigg\{ P^{\mu\nu}_{2a}
\gamma_\nu - P_{1a\nu} \sqrt{3 \over 2} \gamma_5 \left[ g^{\mu
\nu}-{1 \over 3} \gamma^\nu (\gamma^\mu-v^\mu) \right]
\Bigg\}   \, , \label{pos2} \\
X_a^\mu &=&\frac{1+{\rlap{v}/}}{2} \Bigg\{ P^{*\mu\nu}_{2a}
\gamma_5 \gamma_\nu -P^{\prime *}_{1a\nu} \sqrt{3 \over 2}  \left[
g^{\mu \nu}-{1 \over 3} \gamma^\nu (\gamma^\mu+v^\mu) \right]
\Bigg\}  \, , \nn   \\
X_a^{\prime \mu\nu} &=&\frac{1+{\rlap{v}/}}{2} \Bigg\{
P^{*\mu\nu\sigma}_{3a} \gamma_\sigma -P^{\prime \alpha\beta}_{2a}
\sqrt{5 \over 3} \gamma_5 \Bigg[ g^\mu_\alpha g^\nu_\beta - {1
\over 5} \gamma_\alpha g^\nu_\beta (\gamma^\mu-v^\mu) -  {1 \over
5} \gamma_\beta g^\mu_\alpha (\gamma^\nu-v^\nu) \Bigg] \Bigg\}  \, ,\nn
\\
F_a^{ \mu\nu} &=&\frac{1+{\rlap{v}/}}{2} \Bigg\{
P^{\mu\nu\sigma}_{3a} \gamma_\sigma \gamma_5 -P^{\prime * \alpha\beta}_{2a}
\sqrt{5 \over 3} \Bigg[ g^\mu_\alpha g^\nu_\beta - {1
\over 5} \gamma_\alpha g^\nu_\beta (\gamma^\mu-v^\mu) -  {1 \over
5} \gamma_\beta g^\mu_\alpha (\gamma^\nu-v^\nu) \Bigg] \Bigg\} \nn
\,\,.\eea
$v$ is the meson four-velocity,  conserved in strong
interactions. The  operators in Eq.~(\ref{neg}) contain a factor $\sqrt{m_Q}$,  have dimension $3/2$ and 
annihilate mesons with four-velocity $v$.  

The octet of light pseudoscalar mesons is introduced defining $\displaystyle \xi=e^{i {\CMcal M}
\over f_\pi}$ and $\Sigma=\xi^2$,  with the matrix ${\CMcal M}$
comprising  $\pi, K$ and $\eta$ fields ($f_{\pi}=132 \; $ MeV):
\begin{equation}
{\CMcal M}= \left(\begin{array}{ccc}
\sqrt{\frac{1}{2}}\pi^0+\sqrt{\frac{1}{6}}\eta & \pi^+ & K^+\\
\pi^- & -\sqrt{\frac{1}{2}}\pi^0+\sqrt{\frac{1}{6}}\eta & K^0\\
K^- & {\bar K}^0 &-\sqrt{\frac{2}{3}}\eta
\end{array}\right) \,\,\,\, . \label{pseudo-octet}
\end{equation}
Vector and axial-vector currents can be defined:
 \begin{eqnarray}
{\CMcal V}_{\mu ba}&=&\frac{1}{2}\left(\xi^\dagger\partial_\mu
\xi
+\xi\partial_\mu \xi^\dagger\right)_{ba} \; , \label{Vmu} \\
{\CMcal A}_{\mu ba}&=&\frac{i}{2}\left(\xi^\dagger\partial_\mu
\xi-\xi
\partial_\mu \xi^\dagger\right)_{ba} \; , \label{Amu}
\end{eqnarray}
and  under the chiral group $SU(3)_L \times SU(3)_R$ the transformation properties are:
 \bea
 \xi &\to& L\xi U^\dagger=U \xi R^\dagger  \,\,\, , \label{transf-xi} \\
{\CMcal A}_{\mu} &\to & U {\CMcal A}_{\mu} U^\dagger  \,\,\, ,
\label{transf-A}
 \\
 {\CMcal V}_{\mu} &\to & U {\CMcal V}_{\mu} U^\dagger +U \partial_\mu U^\dagger \,\,\, . \label{transf-V} 
 \eea
With the definition 
\be
D_{\mu ba}=-\delta_{ba}\partial_\mu+{\CMcal V}_{\mu ba} \,\,\, , \label{partial}
\ee
at the leading order in the  heavy quark mass expansion and in the  light meson
momentum,   the effective Lagrangian terms   invariant under heavy quark spin-flavour  and light quark chiral transformations can be constructed 
 \cite{Wise:1992hn,Burdman:1992gh,Cho:1992gg,Yan:1992gz,Casalbuoni:1992dx}:
\bea
{\CMcal L}_H &=& \,  g \, Tr \Big[{\bar H}_a H_b \gamma_\mu\gamma_5 {\CMcal A}_{ba}^\mu \Big]  \,\,\, , \label{lagH} \\
{\CMcal L}_S &=& \,  h \, Tr \Big[{\bar H}_a S_b \gamma_\mu \gamma_5 {\CMcal A}_{ba}^\mu \Big]\, + \, h.c.  \,\,\, , \label{lagS} \\
{\CMcal L}_T &=&  {h^\prime \over \Lambda_\chi}Tr\Big[{\bar H}_a T^\mu_b (i D_\mu {\spur {\CMcal A}}+i{\spur D} { \CMcal A}_\mu)_{ba} \gamma_5\Big] + h.c.   \,\,\, ,  \label{lagT} \\
{\CMcal L}_X &=&  {k^\prime \over \Lambda_\chi}Tr\Big[{\bar H}_a X^\mu_b (i D_\mu {\spur {\CMcal A}}+i{\spur D} { \CMcal A}_\mu)_{ba} \gamma_5\Big] + h.c.   \,\,\, , \,\,\,\,\,\,\,\,\,\,\,\,   \label{lagX} \\
{\CMcal L}_{X^\prime} &=&  {1 \over {\Lambda_\chi}^2}Tr\Big[{\bar H}_a X^{\prime \mu \nu}_b \big[k_1 \{D_\mu, D_\nu\} {\CMcal A}_\lambda + k_2 (D_\mu D_\lambda { \CMcal A}_\nu + D_\nu
D_\lambda { \CMcal A}_\mu)\big]_{ba}  \gamma^\lambda \gamma_5\Big] + h.c.  \,\,\, , \label{lagXp}
\\
{\CMcal L}_{F} &=&  {1 \over {\Lambda_\chi}^2}Tr\Big[{\bar H}_a F^{ \mu \nu}_b \big[\hat p_1 \{D_\mu, D_\nu\} {\CMcal A}_\lambda + \hat p_2 (D_\mu D_\lambda { \CMcal A}_\nu + D_\nu
D_\lambda { \CMcal A}_\mu)\big]_{ba}  \gamma^\lambda \gamma_5\Big] + h.c.  \,\,\, , \label{lagF}
 \eea
with $\bar H = \gamma^0 H^\dagger \gamma^0$ and $\Lambda_\chi$ a scale parameter.
The coupling constants $g,\,h,\,h^\prime,\,k^\prime,k_{1,2} ,\,\hat p_{1,2}$  can be inferred from experiment, indeed bounds  have been found \cite{Colangelo:2012xi}.  Theoretical determinations 
using nonperturbative approaches are  available, namely  for $g$ and $h$ \cite{Colangelo:1994jc,Colangelo:1994es,Belyaev:1994zk,Colangelo:1995ph,Colangelo:1997rp,Becirevic:2009yb,Becirevic:2012zza}. 
 The expressions for the ${\cal H}_i \to P^{(*)}M$ decay widths, with $P^{(*)}$ in the $H$ doublet and $M$ a light pseudoscalar meson, can be found  in \cite{Colangelo:2012xi}, with the  exception  of  decaying mesons in the $F$ doublet,  obtained from  Eq.~(\ref{lagF}) \footnote{The expressions (\ref{g1F2}), (\ref{g2F2}) coincide with those given in \cite{Wang:2016ewb,Gupta:2018zlg}. Instead, our  Eq.~(\ref{gF3}) is different from the one quoted in \cite{Gupta:2018zlg}.}:
\bea
\Gamma (P_2^{\prime *}(v,\eta) \to P(v) M(p_M)) &=&C_M\,\frac{4 {\hat p}^2}{25 \pi f_\pi^2 \Lambda_\chi^4} \frac{m_P}{m_{P_2^{\prime *}}}\,
|{\vec p}_M|^5 \, \left(m_M^2+|{\vec p}_M|^2 \right) \,\,\, ,
\label{g1F2} \\
\Gamma (P_2^{\prime *}(v,\eta) \to P^*(v,\,\epsilon) M(p_M)) &=&C_M\,\frac{8 {\hat p}^2}{75 \pi f_\pi^2 \Lambda_\chi^4} \frac{m_{P^*}}{m_{P_2^{\prime *}}}\,
|{\vec p}_M|^5 \, \left(m_M^2+|{\vec p}_M|^2 \right) \,\,\, ,
\label{g2F2} \\
\Gamma (P_3(v,\eta) \to P^*(v,\,\epsilon) M(p_M)) &=&C_M \,\frac{4 {\hat p}^2}{15 \pi f_\pi^2 \Lambda_\chi^4} \frac{m_{P^*}}{m_{P_3}}\,
|{\vec p}_M|^5 \, \left(m_M^2+|{\vec p}_M|^2 \right) \,\,\, ,\label{gF3}
\eea
where $\eta$ ($\epsilon$)  is the polarization  tensor (vector), $\hat p$  the combination of the couplings $\hat p=\hat p_1+ \hat p_2$,    ${\vec p}_M$  the three-momentum of $M$, and  the factor $C_M$ is different for the various mesons,  $C_{\pi^+}=C_{K^+}=1$, $C_{\pi^0}=C_{K_S}={1 \over 2}$ and $C_{\eta}={2 \over 3}$.

 \subsection{Incorporating  light vector mesons}
 There are several  ways to incorporate the light vector mesons in the effective Lagrangian describing heavy meson decays. 
 Here we reconsider the hidden gauge symmetry approach \cite{Bando:1985rf,Bando:1987br,Georgi:1989gp,Georgi:1989xy}   applied in \cite{Ko:1993fn,Schechter:1992ue,Casalbuoni:1992gi,Kitazawa:1993bk}.

 The hidden local symmetry  method, which dates back to  applications to supergravity theories \cite{Cremmer:1978ds,Cremmer:1979up}, exploits the equivalence of  the non-linear sigma model based on a group $G$ spontaneously broken to a subgroup $H$, to another model having $G$ as global symmetry group and $H$ as a local symmetry. This allows to introduce the gauge bosons of the local  symmetry,
which   are identified with the light vector mesons in  applications to chiral theory. In this formulation the vector fields transform inhomogeneously under nonlinear realization of the chiral symmetry, 
while  in alternative approaches to incorporate the vector mesons (Weinber  \cite{Weinberg:1968de} and Callan, Coleman, Wess and Zumino  \cite{Callan:1969sn}), the vector fields transform homogeneously. The different methods are shown to be equivalent  \cite{Yamawaki:1986zz,Bando:1987br,Ecker:1989yg,Tanabashi:1995nz,Birse:1996hd,Harada:2003jx}.
 
 In the hidden gauge symmetry framework one writes 
 \be
 \Sigma=\xi_L \xi_R^\dagger \,\,\, .
 \ee
The fields $\xi_{L,R}$ transform under $SU(3)_L \times SU(3)_R \times SU(3)_H$ as
 \bea
 \xi_L \to U_L \xi_L U_H^\dagger(x) \,\,\, ,
 \nn \\
  \xi_R \to U_R \xi_R U_H^\dagger(x) \,\,\, ,
\nn 
\eea
 where $U_{L,R} \in SU(3)_{L,R}$, and $U_H(x) \in SU(3)_H$ is a local transformation.  The action of the group $SU(3)_H$ is {\it hidden} when one considers the field $\Sigma$.
One  now defines
  \begin{eqnarray}
{\CMcal A}_\mu&=&\frac{i}{2}\left(\xi^\dagger_L\partial_\mu \xi_L -\xi_R
\partial_\mu \xi^\dagger_R \right) \; ,\\
{\CMcal V}_\mu&=&\frac{1}{2}\left(\xi^\dagger_L\partial_\mu
\xi_L
+\xi_R\partial_\mu \xi^\dagger_R \right) \; .
\end{eqnarray}
Fixing the gauge in such a way that $\xi_L=\xi_R=\xi$,  these fields can be identified with the  ones in Eqs.~(\ref{Amu}) and (\ref{Vmu}).
 Their transformation properties under $SU(3)_H$  are  given by Eqs.~(\ref{transf-A})-(\ref{transf-V}), identifying $U$ with $U_H$.
 
 The  octet of light vector mesons plays the role of gauge fields of the a hidden symmetry, and is  introduced writing
 \be
 \rho_\mu=i \frac{g_V}{\sqrt{2}}{\hat \rho}_\mu ,
 \label{rho}
 \ee
 where  ${\hat \rho}_\mu$ is a Hermitian matrix defined in analogy to the matrix ${\CMcal M}$ of pseudoscalar fields (\ref{pseudo-octet}):
 \begin{equation}
{\hat \rho}_\mu= \left(\begin{array}{ccc}
\sqrt{\frac{1}{2}}\rho^0+\sqrt{\frac{1}{6}}\phi^{(8)} & \rho^+ & K^{*+}\\
\rho^- & -\sqrt{\frac{1}{2}}\rho^0+\sqrt{\frac{1}{6}}\phi^{(8)} & K^{*0}\\
K^{*-} & {\bar K}^{*0} &-\sqrt{\frac{2}{3}}\phi^{(8)}
\end{array}\right)_\mu \,\,\,\, . \label{matvecto1}
\end{equation}
The constant  $g_V$ is set to $g_V\simeq 5.8$  to satisfy the KSRF relations \cite{Kawarabayashi:1966kd,Riazuddin:1966sw}.
The observed vector mesons $\omega$ and $\phi$ correspond to a mixing between the octet component $\phi^{(8)}$  in  (\ref{matvecto1}) and the singlet component $\phi^{(0)}$:
\bea
\phi&=&\sin \theta_V \phi^{(0)}-\cos \theta_V \phi^{(8)} \nn \\
\omega&=& \cos \theta_V \phi^{(0)}+\sin \theta_V \phi^{(8)} \,\, .
\eea
 The  angle $\theta_V \simeq {\rm ArcTan} \displaystyle\frac{1}{\sqrt{2}}$  realizes the  ideal mixing allowing to identify $\omega$ and $\phi$ with the flavour eigenstates $\phi_q=\displaystyle\frac{{\bar u}u+{\bar d}d}{\sqrt{2}}$ and $\phi_s={\bar s}s$.
In terms of these,  in  (\ref{matvecto1}) one can replace $\displaystyle\frac{1}{\sqrt{3}}\phi^{(8)} =\sin \theta_V \phi^{(8)} \to \phi_q$, and $-\displaystyle\frac{2}{\sqrt{3}}\phi^{(8)} =-\cos \theta_V \phi^{(8)} \to \phi_s$: 
 \begin{equation}
{\hat \rho}_\mu= \left(\begin{array}{ccc}
\sqrt{\frac{1}{2}}\rho^0+\sqrt{\frac{1}{2}}\omega & \rho^+ & K^{*+}\\
\rho^- & -\sqrt{\frac{1}{2}}\rho^0+\sqrt{\frac{1}{2}}\omega & K^{*0}\\
K^{*-} & {\bar K}^{*0} &\phi
\end{array}\right)_\mu \,\,\,\, , \label{matvector}
\end{equation}
a replacement becoming  exact in the large $N_c$ limit  \cite{Jenkins:1995vb}.

The antisymmetric field tensor is defined:
\be
{\cal F}_{\mu \nu}=\partial_\mu \rho_\nu- \partial_\nu \rho_\mu+[\rho_\mu,\,\rho_\nu] \,\,.
\ee
$\rho_\mu$  transforms  as  ${\CMcal V}_\mu$:
 \be
\rho_{\mu} \to  U\rho_{\mu} U^\dagger +U \partial_\mu U^\dagger \label{transf-rho} \,,
 \ee
 while the difference ${\CMcal V}_\mu-\rho_\mu$, as well as  ${\cal F}_{\mu \nu}$  transform homogenously as ${\CMcal A}_\mu$:
 \bea
( {\CMcal V}_{\mu} -\rho_\mu) &\to&  U ( {\CMcal V}_{\mu} -\rho_\mu)  U^\dagger \,\, ,
\label{transf-diff}\\
{\cal F}_{\mu \nu} &\to  &U {\cal F}_{\mu \nu}  U^\dagger \,\,. \label{transf-Fmunu}
\eea
The covariant derivative $D_\alpha$ can be defined,  such that
 $D_\alpha {\cal F}_{\mu \nu} \to U (D_\alpha {\cal F}_{\mu \nu})  U^\dagger$.
 If $W_\alpha$ is a field transforming inhomogeneously, one  can  show that
 \be
 D_\alpha {\cal F}_{\mu \nu}=\partial_\alpha \,  {\cal F}_{\mu \nu}+ [{\cal F}_{\mu \nu},\,W_\alpha] \label{der-new}
 \ee
satisfies the previous relation.
$W_\alpha={\CMcal V}_\alpha$, or $W_\alpha=\rho_\alpha$, or a linear combination of them can be chosen, but 
for our purpose it is irrelevant to fix $W$, since at the  leading order in the effective theory and for processes describing  heavy-light meson decays  to another heavy one and a single light vector meson, only the partial derivative in (\ref{der-new}) contributes to the amplitude.

\section{Effective Lagrangian terms and strong decay widths}\label{sec:results}
We now construct the effective Lagrangian terms governing the decays ${\cal H}_i \to P^{(*)} V$, where ${\cal H}_i$ is a heavy-light meson,
V  a light vector meson and $P,\,P^*$  the lowest-lying heavy-light  $J^P=(0^-,\,1^-)$ mesons.  For the doublets corresponding to   $\ell=0$ and  $\ell=1$  such Lagrangians have been derived in \cite{Casalbuoni:1992gi,Casalbuoni:1996pg}.
We denote by ${\cal H}^{\mu_1 \mu_2\dots \mu_k}$ the spin doublet  which the decaying heavy meson  ${\cal H}_i $ belongs to.
The effective Lagrangian describing the transition ${\cal H}_i \to P^{(*)} V$ can have two structures:
\bea
{\cal L}_1&=-&\zeta \, Tr \left[ {\bar H} {\cal H}^{\mu_1 \dots \mu_k} \Gamma_{\mu_1 \dots \mu_k \alpha} ({\cal V}^\alpha-\rho^\alpha)\right]+h.c. \label{L1}
\\
{\cal L}_2&=&\mu \, Tr \left[ {\bar H} {\cal H}^{\mu_1 \dots \mu_k} \Gamma_{\mu_1 \dots \mu_k \alpha \beta}\, {\cal F}^{\alpha \beta})\right] +h.c. \,\, , \label{L2} 
\eea
with the minus sign in (\ref{L1}) included for later convenience.
The two structures $\Gamma_{\mu_1 \dots \mu_k \alpha}$ and $\Gamma_{\mu_1 \dots \mu_k \alpha \beta}$ are chosen in such a way that the  Lagrangians are invariant under heavy quark symmetry and  hidden gauge symmetry transformations, parity ($\cal P$), charge conjugation ($\cal C$) and  time reversal ($\cal T$).
Indeed, under such discrete transformations one has:
\bea
 {\cal V}^\alpha \stackrel{{\cal P}}{\rightarrow} {\cal V}_\alpha \,\,\,\,\,\,\,\,\,\,\,\,\,    {\cal V}^\alpha &\stackrel{\cal T}{\rightarrow} & {\cal V}_\alpha  \,\,\,\,\,\,\,\,\,\,\,\,\,\,\,\,   {\cal V}^\alpha \stackrel{\cal C}{\rightarrow}  -({\cal V}^\alpha)^T \nn \\
\rho^\alpha \stackrel{\cal P}{\rightarrow}  \rho_\alpha  \,\,\,\,\,\,\,\,\,\,\,\,\,\,\,    \rho^\alpha &\stackrel{\cal T}{\rightarrow}&  \rho_\alpha  \,\,\,\,\,\,\,\,\,\,\,\,\,\,\,\,\,\,   \rho^\alpha \stackrel{\cal C}{\rightarrow}  -(\rho^\alpha)^T  \nn \\
{\cal F}^{\alpha \beta} \stackrel{\cal P}{\rightarrow}  {\cal F}_{\alpha \beta}  \,\,\,\,\,\,\,    {\cal F}^{\alpha \beta} &\stackrel{\cal T}{\rightarrow} & -{\cal F}_{\alpha \beta} \,\,\,\,\,\,\,    {\cal F}^{\alpha \beta} \stackrel{\cal C}{\rightarrow}  -({\cal F}^{\alpha \beta})^T \,\,, \nn
\eea
where $T$ means transpose.
As for the heavy meson doublets,   they transform under $\cal P$ and $\cal T$ as \cite{Boyd:1994pa}:
\bea
{\cal H}^{\mu_1 \dots \mu_k}_v(x)  &&\stackrel{\cal P}{\rightarrow} \gamma^0 ({\cal H}_{\bar v})_{\mu_1 \dots \mu_k}({\bar x}) \gamma^0 \nn \\
{\cal H}^{\mu_1 \dots \mu_k}_v(x)  &&\stackrel{\cal T}{\rightarrow} {\hat T} ({\cal H}_{\bar v})_{\mu_1 \dots \mu_k}(-{\bar x}) {\hat T}^{-1} \,\,\, ,
\eea
where ${\bar v},\,{\bar x}$ denote the parity reflections of $v$ and $x$ (e.g. ${\bar v}^\mu=v_\mu$), and $\hat T=i \, \gamma^1 \gamma^3$.
Transforming all the fields according to these rules, it can be checked that all our effctive Lagrangian terms are invariant under parity and time reversal.
As for charge conjugation,  discussed e.g. in \cite{Kitazawa:1993bk}, the effective heavy meson fields transform into the corresponding fields that contain the negative energy component of the heavy quark.
For example, in the case of the lowest-lying doublet, denoting such field by $H_v^{(-)}(x)$ one has 
\be
H_v(x) \stackrel{\cal C}{\rightarrow} {\hat C} \left( {\overline H^{(-)}_v}(x) \right)^T {\hat C}^\dagger 
\ee 
where ${\hat C}=i \, \gamma^2 \gamma^0$. Invariance under charge conjugation is obtained adding to the effective Lagrangian an anti-particle part that has  the same form of  the particle part except for the substitutions: $H_v \to H_v^{(-)}$ and $v \to -v$. We  always imply that our effective Lagrangian terms include the corresponding antiparticle parts.

  Invariance under heavy quark velocity reparametrization must also be preserved \cite{Luke:1992cs}.
The heavy quark symmetry imposes further constraints, since  in  the decays of the two members in a  spin doublet to the  ones of the lowest-lying doublet, the  light meson must be emitted in the same orbital state.  This   reduces the number of terms in the effective Lagrangian. Beyond the leading order in the HQ expansion,  additional Lagrangian terms must be included  \cite{Falk:1995th,Colangelo:2005gb}. 

Considering the doublets  in (\ref{pos2}), in the effective Lagrangian terms  ~(\ref{L1}) and (\ref{L2})  we are  concerned  with indices having  $k=0,\,1,\,2$, that  we discuss  in turn.
 For the   decay mode $P_1 \to P_2 V$  we have $\pVvec=\displaystyle\frac{\lambda^{1/2} \left( m_{P_1}^2,\,m_{P_2}^2,\,m_V^2 \right)}{2 m_{P_1}}$ and $E_V=\displaystyle\frac{m_{P_1}^2-m_{P_2}^2+m_V^2}{2m_{P_1}}$, with $\lambda(x,\,y,\,z)=x^2+y^2+z^2-2xy-2xz-2yz$  the triangular function. 

 Before discussing in details the transitions of states in the various doublets, we remark that 
the effective Lagrangian  approach is in principle applicable when the emitted light particle is soft. This is guaranteed when the mass difference between the decaying  meson and the final heavy-light meson is not too large.
When decays of heavier  excitations are considered, it is possible that corrections  from higher order terms in the effective Lagrangians could become sizeable. Nevertheless, we push our  predictions also for large values of the mass of the decaying particle, considering  the  symmetries  as the main guidelines in the description of the heavy-light meson phenomenology.

\subsection{Transitions ${\tilde H} \to H  V$, with  ${\tilde H}=({\tilde P},\,{\tilde P}^*)$}\label{Hdecays}
When the decaying meson  belongs to the  $H$ doublet we have  $k=0$ in Eqs.~(\ref{L1}) and (\ref{L2}).
Decays to $P^{(*)} V$ are not kinematically allowed for the $n=1$ $H$ doublet, hence  we consider  the radially excited  $\tilde H$ doublet  ($n=1$ is relevant for processes with intermediate virtual mesons  \cite{Casalbuoni:1992gi}).
The decays occur in $p$-wave,  and the terms (\ref{L1}) and (\ref{L2})  fulfilling  the constraints are:
\bea
{\cal L}_1^H&=&-g_1^H \, Tr \left[ {\bar H} {\tilde H} \gamma^\alpha ({\cal V}_\alpha-\rho_\alpha)\right] +h.c.
\label{LHHV1} \\
{\cal L}_2^H&=& g_2^H \, \frac{1}{\Lambda}Tr \left[ {\bar H} {\tilde H}  \sigma^{\alpha \beta}{\cal F}_{\alpha \beta}\right] +h.c.  \,\, , \label{LHHV2}
\eea
with the parameter $\Lambda$  introduced  to render the  couplings  dimensionless. We set $\Lambda=1$ GeV.
In the previous expressions,  the replacement of  a single  $\gamma$ matrix with the four-velocity $v$   produces terms that  either give  the same result or  vanish,
a remark holding for all  cases considered below. The Lagrangians ~(\ref{LHHV1}) and (\ref{LHHV2}) coincide with those  obtained  in \cite{Casalbuoni:1992gi},
and from them  the decay widths are worked out:
\bea
\Gamma \left({\tilde P}(v) \to P(v) V(p_V,\,\epsilon_V)\right) &=&C_V\,\frac{g_V^2\left(g_1^H\right)^2}{4 \pi m_V^2}\frac{m_P}{m_{\tilde P}}|{\vec p}_V |^3 \,\,\,, \label{PtildePV} \\
\Gamma \left({\tilde P}(v) \to P^*(v,\epsilon) V(p_V,\,\epsilon_V)\right) &=&C_V\,
\frac{ 2 g_V^2 (g_2^{H})^2}{ \pi \Lambda^2 }\frac{m_{P^*}}{m_{\tilde P}}|{\vec p}_V |^3\,\,\,, \label{PtildePstarV} \\
\Gamma \left({\tilde P}^*(v,{\tilde \epsilon})\to P(v)  V(p_V,\,\epsilon_V)\right) &=&C_V\,
\frac{ 2 g_V^2 (g_2^{H})^2}{3 \pi \Lambda^2}\frac{m_{P}}{m_{{\tilde P}^*}}|{\vec p}_V |^3\,\,\,, \label{PtildestarPV} \\
\Gamma 
\left({\tilde P}^*(v,{\tilde \epsilon})\to P^*(v,\epsilon)  V(p_V,\,\epsilon_V)\right) &=&C_V\,
\frac{ g_V^2 }{12 \pi m_V^2}\frac{m_{P^*}}{m_{{\tilde P}^*}}\left(16 {\tilde m}_V^2(g_2^{H})^2+ 3 \left(g_1^H \right)^2 \right) |{\vec p}_V |^3 \,\,\,,  \label{PtildestarPstarV}
\eea
with $\epsilon_V$ and  $\epsilon, \tilde \epsilon$ light and heavy meson polarization vectors,  ${\tilde m}_V=\displaystyle{\frac{m_V}{\Lambda}}$, and  $C_V=1$ for $V=\rho^\pm,\,K^{*\pm},\,K^{*0},\,{\bar K}^{*0},\,\varphi$,  $C_V=\frac{1}{2}$ for $V=\rho^0, \omega $. 

Relations among the decay widths,  not involving the coupling constants, can be constructed:
\bea
R_H&=&\frac{\Gamma ({\tilde P} \to P^* V)}{\Gamma ({\tilde P}^*\to P  V)}=3 \frac{m_{P^*}}{m_P} \frac{m_{{\tilde P}^*}^4}{m_{{\tilde P}}^4}\, \frac{\lambda^{3/2}(m_{{\tilde P}}^2,\,m_{P^*}^2,\,m_V^2)}{\lambda^{3/2}(m_{{\tilde P}^*}^2,\,m_{P}^2,\,m_V^2)} \,  , \label{RH1}  \\
\Gamma ({\tilde P}^*\to P^*  V) &=&  \frac{m_{P^*}}{m_P} 
\left\{ 2\, \frac{\lambda^{3/2}(m_{{\tilde P}^*}^2,\,m_{P^*}^2,\,m_V^2)}{\lambda^{3/2}(m_{{\tilde P}^*}^2,\,m_{P}^2,\,m_V^2)} \,
\Gamma ({\tilde P}^*\to P V)+ \frac{m_{\tilde P}^4}{m_{{\tilde P}^*}^4}
\frac{\lambda^{3/2}(m_{{\tilde P}^*}^2,\,m_{P^*}^2,\,m_V^2)}{\lambda^{3/2}(m_{{\tilde P}}^2,\,m_{P}^2,\,m_V^2)} \,
\Gamma ({\tilde P} \to P V) \right\} . \,\,\, 
\label{RH2}
\eea
Other relations  independent of the couplings can be worked out considering  modes with different final light vector mesons, as  discussed  in  Sect. \ref{sec:numerics}.

\subsection{ $ {\tilde S} \to H  V$, with  ${\tilde S}=({\tilde P}_0^*,\,{\tilde P}_1^\prime)$ }\label{Sdecays}
When the decaying meson belongs to the $S$ doublet  one has  $k=0$ in Eqs.~(\ref{L1}) and (\ref{L2}). The
$P^{(*)} V$ phase space is closed for  $n=1$,   therefore we consider  radial excitations in  $\tilde S$ doublet. 
The transitions occur in $s$-wave, and the effective Lagrangian
 terms  (\ref{L1}) and (\ref{L2}) read:
\bea
{\cal L}_1^S&=&-g_1^S \, Tr \left[ {\bar H} {\tilde S} \gamma^\alpha ({\cal V}_\alpha-\rho_\alpha)\right] +h.c.
\label{LHSV1} \\
{\cal L}_2^S&=& g_2^S \, \frac{1}{\Lambda}Tr \left[ {\bar H} {\tilde S}  \sigma^{\alpha \beta}{\cal F}_{\alpha \beta}\right]  +h.c.  \,\, ,\label{LHSV2}
\eea
as also obtained in \cite{Casalbuoni:1996pg}.
The  decay widths read:
\bea
\Gamma \left({\tilde P}_0^*(v) \to P^*(v,\epsilon) V(p_V,\,\epsilon_V)\right) &=& \nn \\ 
&& \hskip -5cm C_V\,\frac{g_V^2}{4 \pi m_V^2 }\frac{ m_{P^*}}{m_{{\tilde P}^*_0}}  
\left\{(g_1^S)^2(3m_V^2+|{\vec p}_V|^2 ) +12 g_1^S \, g_2^S\, {\tilde m}_V \,m_V\sqrt{m_V^2+|{\vec p}_V|^2}+4(g_2^S)^2\,{\tilde m}_V^2(3m_V^2+2|{\vec p}_V|^2)\right\} |{\vec p}_V | \, , \hskip1cm  \label{SP0PstarV} \\
\Gamma\left({\tilde P}_1^\prime(v,\eta) \to P(v) V(p_V,\,\epsilon_V)\right) &=&  \nn \\ 
&& \hskip -5cm C_V\,\frac{g_V^2}{12 \pi m_V^2 }\frac{m_{P}}{ m_{{\tilde P}^\prime_1}} 
\left\{(g_1^S)^2(3m_V^2+|{\vec p}_V|^2 ) +12 g_1^S \, g_2^S\, {\tilde m}_V \,m_V\sqrt{m_V^2+|{\vec p}_V|^2}+4(g_2^S)^2\,{\tilde m}_V^2(3m_V^2+2|{\vec p}_V|^2)\right\}|{\vec p}_V | \, , \,\,\,\,\,\, \label{SP1PV} \\
\Gamma\left({\tilde P}_1^\prime(v,\eta) \to P^*(v,\,\epsilon) V(p_V,\,\epsilon_V)\right) &=&  \nn \\ 
&& \hskip -5cm C_V\,\frac{g_V^2}{6 \pi m_V^2 }\frac{ m_{P^*}}{m_{{\tilde P}^\prime_1}}
\left\{(g_1^S)^2(3m_V^2+|{\vec p}_V|^2 ) +12 g_1^S \, g_2^S\, {\tilde m}_V \,m_V\sqrt{m_V^2+|{\vec p}_V|^2}+4(g_2^S)^2\,{\tilde m}_V^2(3m_V^2+2|{\vec p}_V|^2)\right\} |{\vec p}_V | \, , \,\,\,\,\,\,\,\,   \label{SP1PstarV} 
\eea
with $\eta$ polarization vector.
The  transition $P_0^*(v) \to P(v) V(p_V,\,\epsilon_V)$  is forbidden.

\subsection{$ {\tilde T} \to H  V$, with  ${\tilde T}=({\tilde P}_1,\,{\tilde P}_2^*)$}\label{Tdecays}
For the decays of the states in the $T$ doublet one has  $k=1$ in (\ref{L1}) and (\ref{L2}).
The transitions to $P^{(*)} V$ is not kinematically allowed for  the $n=1$ $T$ doublet, hence consider $n=2$   $\tilde T$. 
The transitions proceed in $d$-wave, and  the effective Lagrangian reads:
\be
{\cal L}_2^T=i \,h^T \, \frac{1}{\Lambda^2} Tr \left[ {\bar H} T_\mu \sigma^{ \alpha \beta} {\cal D}^\mu {\cal F}_{\alpha \beta}\right]  +h.c. \,\, ,  \label{LHTV2}
\ee
with the covariant derivative acting on the light vector meson field tensor.
The  resulting decay widths are:
\bea
\Gamma\left({\tilde P}_1(v,\eta) \to P(v) V(p_V,\,\epsilon_V)\right) &=&C_V\,\frac{g_V^2\, \left(h^T \right)^2}{9 \pi \Lambda^4 }\frac{m_{P}}{ m_{{\tilde P}_1}} \pVvec^5  \,\, , \label{gammaP1TtildePV} \\
\Gamma\left({\tilde P}_1(v,\eta) \to P^*(v,\,\epsilon) V(p_V,\,\epsilon_V)\right) &=& C_V\,\frac{5 g_V^2 \, \left(h^T \right)^2}{9 \pi \Lambda^4 }\frac{m_{P^*}}{ m_{{\tilde P}_1}} \pVvec^5   \,\, ,
\label{gammaP1TtildePstarV} \\
\Gamma 
\left({\tilde P}_2^*(v,\eta) \to P(v) V(p_V,\,\epsilon_V)\right)&=& C_V\,\frac{g_V^2 \, (h^T)^2}{5 \pi \Lambda^4 }\frac{m_{P}}{ m_{{\tilde P}_2^*}} \pVvec^5  \,\, ,  \label{gammaP2TtildePV} \\
 \Gamma\left({\tilde P}_2^*(v,\eta) \to P^*(v,\,\epsilon) V(p_V,\,\epsilon_V)\right) &= &C_V\,
\frac{7g_V^2  \, (h^T)^2}{15 \pi  \Lambda^4}\frac{m_{P^*}}{ m_{{\tilde P}_2^*}} \pVvec^5 \,\,.
\label{gammaP2TtildePstarV}
\eea
The  relations are fulfilled:
\bea
\frac{\Gamma({\tilde P}_1 \to P^* V)}{\Gamma({\tilde P}_1 \to P V)} &=& 5 \frac{m_{P^*}}{m_P} \,
 \frac{\left(\pVvec^{({\tilde P}_1 \to P^* V)}\right)^5}{\left(\pVvec^{({\tilde P}_1 \to P V)}\right)^5}  \,\, ,
 \label{R1T}
 \\
\frac{\Gamma({\tilde P}_2^* \to P^* V)}{\Gamma({\tilde P}_2^* \to P V)} &=& \frac{7}{3} \frac{m_{P^*}}{m_P} \,
 \frac{\left(\pVvec^{({\tilde P}_2^* \to P^* V)}\right)^5}{\left(\pVvec^{({\tilde P}_2^* \to P V)}\right)^5}  \,\, .
 \label{R2T} 
\eea
Ratios of decay rates for modes with   different final  light vector mesons,  independent of the coupling constant,  will be constructed below.

\subsection{ $ X \to H  V$, with $X=(P_1^*,\,P_2)$ }\label{Xdecays}
For $X$ doublet one  has  $k=1$ in (\ref{L1}) and (\ref{L2}).
No candidates  belonging to such a  doublet have been observed,   and we do not know whether  the $P^{(*)} V$ channels are open for the 
 $n=1$ states. The transitions occur in $p$-wave and are governed by the Lagrangian
\be
{\cal L}_2^X=i \, h^X \, \frac{1}{\Lambda^2} Tr \left[ {\bar H} X_\mu \sigma^{ \alpha \beta} {\cal D}^\mu {\cal F}_{\alpha \beta}\right]  +h.c. \,\,\, , \label{LHXV2}
\ee
with the covariant derivative acting on the light vector meson field tensor. The  decay widths are given by:
\bea
\Gamma\left(P_1^*(v,\eta) \to P(v) V(p_V,\,\epsilon_V)\right) &= & C_V\,
\frac{g_V^2\, \left(h^X \right)^2 }{9 \pi \Lambda^4 }\frac{m_{P}}{ m_{P_1^*}} \pVvec^3 \left(m_V^2+\pVvec^2 \right) \,\,\, ,  \label{P1XPV}\\
 \Gamma \left(P_1^*(v,\eta) \to P^*(v,\,\epsilon) V(p_V,\,\epsilon_V)\right)&=& C_V\,\frac{g_V^2 \, \left(h^X \right)^2}{9 \pi  \Lambda^4}\frac{m_{P^*}}{ m_{P_1^*}} \pVvec^3 \,
 \left(8 m_V^2+5 \pVvec^2 \right) \,\,\, ,\label{P1XPstarV} \\
 \Gamma \left(P_2(v,\eta) \to P(v) V(p_V,\,\epsilon_V)\right)&=& C_V\, \frac{g_V^2  \, \left(h^X \right)^2}{15 \pi \Lambda^4 }\frac{m_{P}}{ m_{P_2}} \pVvec^3  \left(5m_V^2+3 \pVvec^2 \right) \,\,\, ,
 \label{P2XPV} \\
 \Gamma \left(P_2(v,\eta) \to P^*(v,\,\epsilon) V(p_V,\,\epsilon_V)\right) &=&C_V\,\frac{g_V^2 \, \left(h^X \right)^2 }{15 \pi  \Lambda^4}\frac{m_{P^*}}{ m_{P_2}} \pVvec^3
 \left(10 m_V^2+7 \pVvec^2 \right)\,\,\, .
 \label{P2XPstarV}
\eea
 Coupling-independent ratios of decay widths are:
\bea
\frac{\Gamma(P_1^* \to P^* V)}{\Gamma(P_1^* \to P V)} &=&  \frac{m_{P^*}}{m_P} \,
 \frac{\left(\pVvec^{(P_1^* \to P^* V)}\right)^3}{\left(\pVvec^{(P_1^*\to P V)}\right)^3}\, \frac{8 m_V^2+5\left(\pVvec^{(P_1^* \to P^* V)}\right)^2}{ m_V^2+\left(\pVvec^{(P_1^* \to P V)}\right)^2} \,\, , 
 \label{R1X}
 \\
\frac{\Gamma(P_2 \to P^* V)}{\Gamma(P_2  \to P V)} &=&  \frac{m_{P^*}}{m_P} \,
 \frac{\left(\pVvec^{(P_2  \to P^* V)}\right)^3}{\left(\pVvec^{(P_2  \to P V)}\right)^3}\, \frac{10 m_V^2+7\left(\pVvec^{(P_2  \to P^* V)}\right)^2}{ 5m_V^2+3\left(\pVvec^{(P_2 \to P V)}\right)^2} \,\,,
 \label{R2X} 
\eea
while ratios of decay widths for processes  with different  final  light vector meson are discussed in  Sect. \ref{sec:numerics}.
\subsection{$ X^\prime \to H  V$, with $X^\prime=(P_2^{\prime},\,P_3^*)$}\label{Xpdecays}
For the decays of the members of the $X^\prime$ doublet one has to consider  $k=2$   in (\ref{L1}) and (\ref{L2}). 
 The processes occur in $f$-wave, with Lagrangian
\be
{\cal L}_2^{X^\prime}=k^{X^\prime} \, \frac{1}{\Lambda^3}\, Tr \left[ {\bar H} X^\prime_{\mu \nu} \, {\cal D}^\mu {\cal D}^\nu \,\sigma^{ \alpha \beta}  {\cal F}_{\alpha \beta}\right]  +h.c.  \label{LHXpV2} \,\,.
\ee
The decay widths read:
\bea
\Gamma \left(P_2^{\prime }(v,\eta) \to P(v) V(p_V,\,\epsilon_V)\right) &= & C_V\,\frac{4g_V^2\, \left( k^{X^\prime} \right)^2}{75\pi \Lambda^6} \frac{m_P}{m_{P_2^{\prime }}} \pVvec^7  \,\,\, , \label{gammaP2XprimoPV} \\
\Gamma \left(P_2^{\prime }(v,\eta) \to P^*(v,\,\epsilon) V(p_V,\,\epsilon_V)\right)&=& C_V\,\frac{16g_V^2 \, \left( k^{X^\prime} \right)^2}{75\pi \Lambda^6} \frac{m_P^*}{m_{P_2^{\prime }}} \pVvec^7 \,\,\, ,
 \label{gammaP2XprimoPstarV} \\
\Gamma \left(P_3^*(v,\eta) \to P(v) V(p_V,\,\epsilon_V)\right) &=&C_V\,\frac{8g_V^2\, \left( k^{X^\prime} \right)^2}{105 \pi \Lambda^6} \frac{m_P}{m_{P_3^*}}\, \pVvec^7 \,\,\, ,
 \label{gammaP3XprimoPV} \\
\Gamma \left(P_3^*(v,\eta) \to P^*(v,\,\epsilon)V(p_V,\,\epsilon_V)\right) &=&C_V\,\frac{4g_V^2\, \left( k^{X^\prime} \right)^2}{21 \pi \Lambda^6} \frac{m_P^*}{m_{P_3^*}}\, \pVvec^7 \,\,\, .\label{gammaP3XprimoPstarV}
\eea
For this doublet the relations are fulfilled: 
\bea
\frac{\Gamma(P_2^\prime\to P^* V)}{\Gamma(P_2^\prime \to P V)} &=& 4 \frac{m_{P^*}}{m_P} \,
 \frac{\left(\pVvec^{(P_2^\prime \to P^* V)}\right)^7}{\left(\pVvec^{(P_2^\prime\to P V)}\right)^7} \,\,\, ,
 \label{R1Xp}
 \\
\frac{\Gamma(P_3^*\to P^* V)}{\Gamma(P_3^* \to P V)} &=& \frac{5}{2} \frac{m_{P^*}}{m_P} \,
 \frac{\left(\pVvec^{(P_3^* \to P^* V)}\right)^7}{\left(\pVvec^{(P_3^*\to P V)}\right)^7} \,. 
 \label{R2Xp}
 \eea

\subsection{ $ F \to H  V$, with $F=(P_2^{\prime*},\,P_3)$}\label{Fdecays}
The case of the $F$  doublet   requires  $k=2$  in (\ref{L1}) and (\ref{L2}). 
The  transitions occur in $d$-wave with effective Lagrangian
\bea
{\cal L}_2^{F}&=&k_1^{F} \, \frac{1}{\Lambda^2}\, Tr \left[ {\bar H} F_{\mu \nu} \, (g^{\mu \alpha}{\cal D}^\nu \gamma^\beta +g^{\nu \alpha}{\cal D}^\mu \gamma^\beta-g^{\mu \beta}{\cal D}^\nu \gamma^\alpha -g^{\nu \beta}{\cal D}^\mu \gamma^\alpha)  {\cal F}_{\alpha \beta}\right] \nn 
\\
 &+& k_2^F 
\, \frac{1}{\Lambda^3}\, Tr \left[ {\bar H} F_{\mu \nu}  {\cal D}^\mu {\cal D}^\nu \,  \sigma^{\alpha \beta}  {\cal F}_{\alpha \beta}\right]
\label{LHFV2} + h.c.  \,\,,
\eea
and decay widths
\bea
\Gamma \left(P_2^{\prime *}(v,\eta) \to P(v) V(p_V,\,\epsilon_V)\right) &=&C_V\,\frac{4g_V^2}{75 \pi \Lambda^4} \frac{m_P}{m_{P_2^{\prime *}}}\, \pVvec^5\, \Bigg( 3 k_1^F + \frac{k_2^F}{\Lambda}\sqrt{m_V^2+\pVvec^2} \,   \Bigg)^2 \,\,\, ,\label{gammaP2FPV} \\
\Gamma \left(P_2^{\prime *}(v,\eta) \to P^*(v,\,\epsilon) V(p_V,\,\epsilon_V)\right) &=&C_V\,\frac{2g_V^2}{75\pi \Lambda^4 } \frac{m_{P^*}}{m_{P_2^{\prime *}}} \pVvec^5 \nn \\ &&
 \left\{ 12\left(k_1^F \right)^2+\frac{8}{\Lambda} k_1^F\,k_2^F \sqrt{m_V^2+\pVvec^2}+\left(\frac{k_2^F}{\Lambda}\right)^2\,(13m_V^2+8 \pVvec^2)
 \right\}  \,\,\, ,\label{gammaP2FPstarV} \\
\Gamma \left(P_3(v,\eta) \to P(v)V(p_V,\,\epsilon_V)\right) &=&C_V\,\frac{2g_V^2}{105 \pi \Lambda^6} \frac{m_P}{m_{P_3}}\, \left(k_2^F \right)^2 \,\pVvec^5(7m_V^2+4 \pVvec^2) \,\,\, , \label{gammaP3FPV} \\
\Gamma \left(P_3(v,\eta) \to P^*(v,\,\epsilon)V(p_V,\,\epsilon_V)\right)&=&C_V\,\frac{4g_V^2}{105 \pi \Lambda^4} \frac{m_{P^*}}{m_{P_3}}\, \pVvec^5 
\nn \\&&
\left\{ 21\left(k_1^F \right)^2+\frac{14}{\Lambda} k_1^F\,k_2^F \sqrt{m_V^2+\pVvec^2}+\left(\frac{k_2^F}{\Lambda}\right)^2 \,(7m_V^2+5 \pVvec^2)
\right\}
\,\,\, .  \label{gammaP3FPstarV}
\eea

A relation independent of the couplings connects  various modes:
\bea
\Gamma\left(P_3\to P^*V\right) &=&\frac{\left(\pVvec^{(P_3 \to P^* V)}\right)^5}{\left(E_V^{(P_2^{\prime *} \to PV)}-E_V^{(P_2^{\prime *} \to P^*V)}\right)}
\nn \\ && \left\{C_1^{(P_2^{\prime *} \to P^* V)} \Gamma(P_2^{\prime *} \to P^* V)+C_2^{(P_2^{\prime *} \to P^* V)} \Gamma(P_2^{\prime *} \to P V)+C_3^{(P_2^{\prime *} \to P^* V)} \Gamma(P_3 \to P V) \right\} ,
\label{relF} \eea
where
\bea
C_1^{(P_2^{\prime *} \to P^* V)}&=&\frac{5}{2}\frac{ m_{P_2^{\prime *}}}{m_{P_3}}\frac{E_V^{(P_2^{\prime *} \to PV)}-E_V^{(P_3 \to P^*V)}}{ \left(\pVvec^{(P_2^{\prime *} \to P^* V)}\right)^5} \nn \\
C_2^{(P_2^{\prime *} \to P^* V)} &=&
\frac{ 5}{3}\frac{m_{P_2^{\prime *}}}{m_{P_3}}\frac{m_{P^*}}{m_P}\frac{E_V^{(P_3 \to P^*V)}-E_V^{(P_2^{\prime *} \to P^*V)}}{\left(\pVvec^{(P_2^{\prime *} \to P V)}\right)^5}  \\
C_3^{(P_2^{\prime *} \to P^* V)}&=& \frac{m_{P^*}}{m_{P_3}} \frac{1}{\left(\pVvec^{(P_3 \to P V)}\right)^5}\frac{1}{7m_V^2+4\left(\pVvec^{(P_3 \to P V)}\right)^2}  
 \Bigg\{28 \left(\pVvec^{(P_2^{\prime *} \to P^* V)}\right)^2\left(E_V^{(P_3 \to P^*V)}-E_V^{(P_2^{\prime *} \to PV)}\right)\nn \\
&+&\frac{14}{3}\left(\pVvec^{(P_2^{\prime *} \to P V)}\right)^2\left(E_V^{(P_2^{\prime *} \to P^*V)}-E_V^{(P_3 \to P^*V)}\right)+10\left(\pVvec^{(P_3 \to P^* V)}\right)^2  \left(E_V^{(P_2^{\prime *} \to PV)}-E_V^{(P_2^{\prime *} \to P^*V)}\right) \nn \\
&-&\frac{7}{6} m_V^2 \left(27 E_V^{(P_2^{\prime *} \to PV)}+8E_V^{(P_2^{\prime *} \to P^*V)}-35E_V^{(P_3 \to P^*V)} \right) 
\Bigg\}.\,\,\, \nn
\eea
Moreover, for $V_A$ and $V_B$ two light vector mesons one simply has: 
\be
R_{V_A V_B}^{P_3}= \frac{ \Gamma(P_3\to P_A  V_A) }{\Gamma(P_3 \to P_B  V_B) }=\frac{C_{V_A}}{C_{V_B}}\frac{|{\vec p}_{V_A} |^5\,(7m_{V_A}^2+4 |{\vec p}_{V_A} |^2)}{|{\vec p}_{V_B} |^5\,(7m_{V_B}^2+4 |{\vec p}_{V_B} |^2)} \,\,.
\label{RP3FV1V2} 
\ee

\section{Numerical Analysis: Charm }\label{sec:numerics}

The expressions  in the previous Section allow  to construct quantities  useful for the classification of  high mass charmed and beauty states.
In Table \ref{ObsDdoublets} we collect  the observed $h{\bar q}$, $h{\bar s}$  mesons, with  $q=u,d$  and $h=c,b$.  
For the various  ${\cal H}_i \to P^{(*)}V$  modes,
ratios of decay widths of ${\cal H}_i$ to different light  vector mesons,  and ratios of decay widths  involving the same light vector meson  and a different member of the final $H$ doublet can be constructed,
obtaining quantities independent of the coupling constants in the  effective Lagrangians.
For  non-strange decaying mesons we define
\bea
R^{{\cal H}_i^+ \to D}_{\omega \rho}&=&\frac{\Gamma ({\cal H}_i^+ \to D^+ \omega)}{\Gamma ({\cal H}_i^+ \to D^+ \rho^0)+\Gamma ({\cal H}_i^+ \to D^0 \rho^+)}  \,\,\, ,
\label{RDOmegaRhopiu}\\
R^{{\cal H}_i^+ \to D_{(s)}}_{K^* \rho}&=&\frac{\Gamma ({\cal H}_i^+ \to D_s {\bar K}^{*0})}{\Gamma ({\cal H}_i^+ \to D^+ \rho^0)+\Gamma ({\cal H}_i^+ \to D^0 \rho^+)}  \,\,\, ,
\label{RDKstarRhopiu}\\
R^{{\cal H}_i^+ \to D^*}_{\omega \rho}&=&\frac{\Gamma ({\cal H}_i^+ \to D^{*+} \omega)}{\Gamma ({\cal H}_i^+ \to D^{*+} \rho^0)+\Gamma ({\cal H}_i^+ \to D^{*0} \rho^+)}  \,\,\, ,
\label{RDstarOmegaRhopiu}\\
R^{{\cal H}_i^+ \to D_{(s)}^*}_{K^* \rho}&=&\frac{\Gamma ({\cal H}_i^+ \to D_s^* {\bar K}^{*0})}{\Gamma ({\cal H}_i^+ \to D^{*+} \rho^0)+\Gamma ({\cal H}_i^+ \to D^{*0} \rho^+)}  \,\,\, ,
\label{RDstarKstarRhopiu}
\eea
and
\bea
R^{{\cal H}_i^0 \to D}_{\omega \rho}&=&\frac{\Gamma ({\cal H}_i^0 \to D^0 \omega)}{\Gamma({\cal H}_i^0 \to D^0 \rho^0)+\Gamma ({\cal H}_i^0 \to D^+ \rho^-)} \,\,\, ,
\label{RDOmegaRho0}\\
R^{{\cal H}_i^0 \to D_{(s)}}_{K^* \rho}&=&\frac{\Gamma ({\cal H}_i^0 \to D_s K^{*-})}{\Gamma({\cal H}_i^0 \to D^0 \rho^0)+\Gamma ({\cal H}_i^0 \to D^+ \rho^-)} \,\,\, ,
\label{RDKstarRho0}\\
R^{{\cal H}_i^0 \to D^*}_{\omega \rho}&=&\frac{\Gamma ({\cal H}_i^0 \to D^{*0} \omega)}{\Gamma ({\cal H}_i^0 \to D^{*0} \rho^0)+\Gamma ({\cal H}_i^0 \to D^{*+} \rho^-)} \,\,\, ,
\label{RDstarOmegaRho0}\\
R^{{\cal H}_i^0 \to D_{(s)}^*}_{K^* \rho}&=&\frac{\Gamma ({\cal H}_i^0 \to D_s^* K^{*-})}{\Gamma ({\cal H}_i^0 \to D^{*0} \rho^0)+\Gamma ({\cal H}_i^0 \to D^{*+} \rho^-)} \,\,\, .
\label{RDstarKstarRho0}
\eea
For decaying strange mesons we define
\bea
R^{{\cal H}_{is}\to D_{(s)}}_{ \phi K^*}&=& \frac{\Gamma ({\cal H}_{is} \to D_s \phi)}{\Gamma ({\cal H}_{is} \to D^+ K^{*0})+\Gamma ({\cal H}_{is} \to D^0 K^{*+})} \,\,\, , \label{RsDOmegaKstar}\\
R^{{\cal H}_{is}\to D_{(s)}^*}_{ \phi K^*}&=& \frac{\Gamma ({\cal H}_{is} \to D_s^* \phi)}{\Gamma ({\cal H}_{is} \to D^{*+} K^{*0})+\Gamma ({\cal H}_{is} \to D^{*0} K^{*+})} \,\,\, .\label{RsDstarOmegaKstar}
\eea
For different final heavy mesons we consider
\bea
R^{{\cal H}_i^+}_{ \rho}&=&\frac{\Gamma ({\cal H}_i^+ \to D^{*+} \rho^0)+\Gamma ({\cal H}_i^+ \to D^{*0} \rho^+)} {\Gamma({\cal H}_i^+ \to D^+ \rho^0)+\Gamma ({\cal H}_i^+ \to D^0 \rho^+)}  \,\,\, ,
\label{RsameRhopiu}\\
R^{{\cal H}_i^0}_{ \rho}&=&\frac{\Gamma ({\cal H}_i^0 \to D^{*0} \rho^0)+\Gamma ({\cal H}_i^0 \to D^{*+} \rho^-)} {\Gamma({\cal H}_i^0 \to D^0 \rho^0)+\Gamma ({\cal H}_i^0 \to D^+ \rho^-)} \,\,\, ,
\label{RsameRho0}\\
R^{{\cal H}_{is}}_{K^*}&=&\frac{\Gamma ({\cal H}_{is} \to D^{*+} K^{*0})+\Gamma ({\cal H}_{is} \to D^{*0} K^{*+})}{\Gamma ({\cal H}_{is} \to D^+ K^{*0})+\Gamma ({\cal H}_{is} \to D^0 K^{*+})}   \,\,\, . \label{RsameKstar}
\eea
\begin{figure}[t!]
 \begin{center}
 \includegraphics[width=0.45\textwidth] {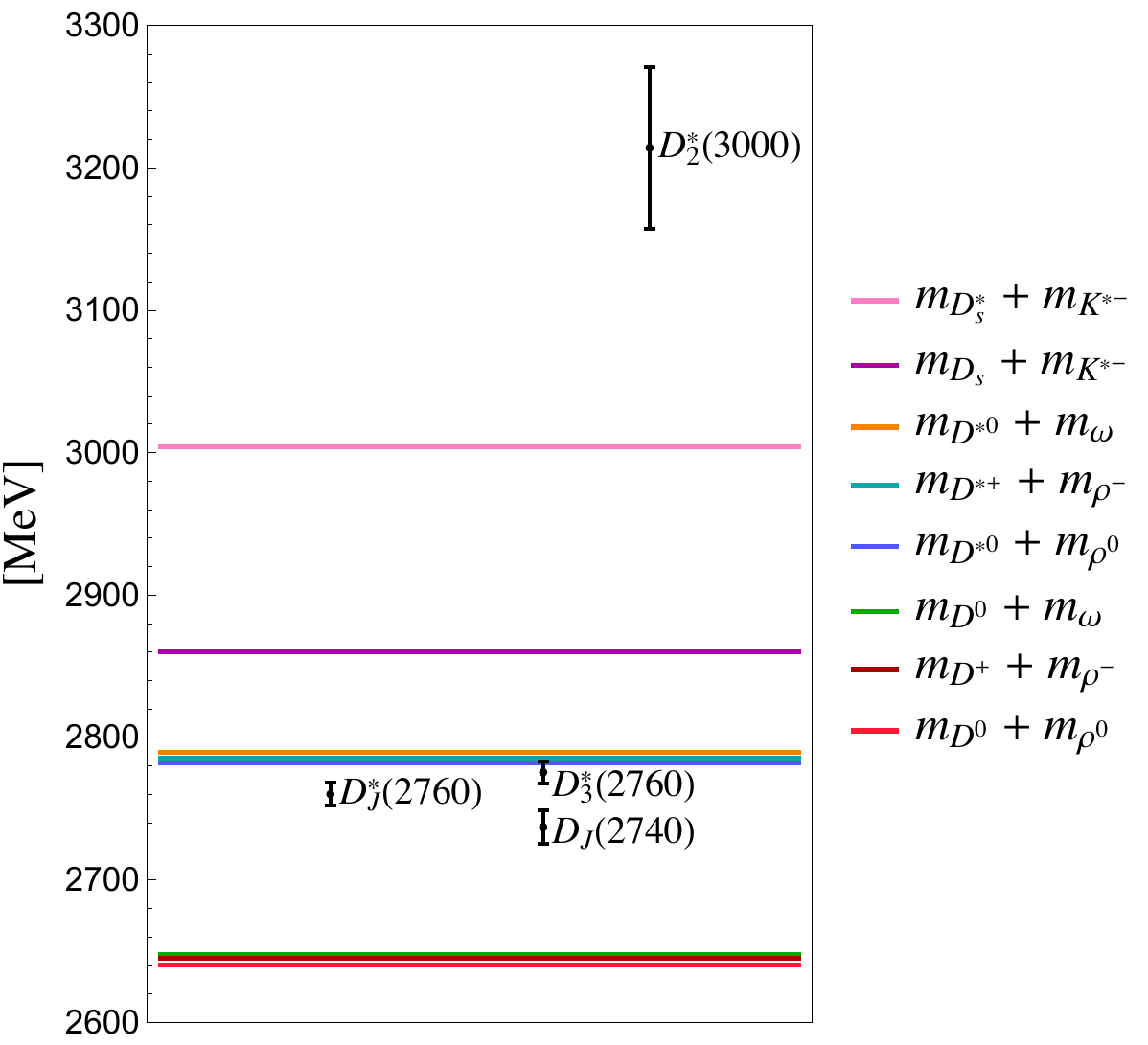}\hspace*{0.3cm}
      \includegraphics[width=0.45\textwidth] {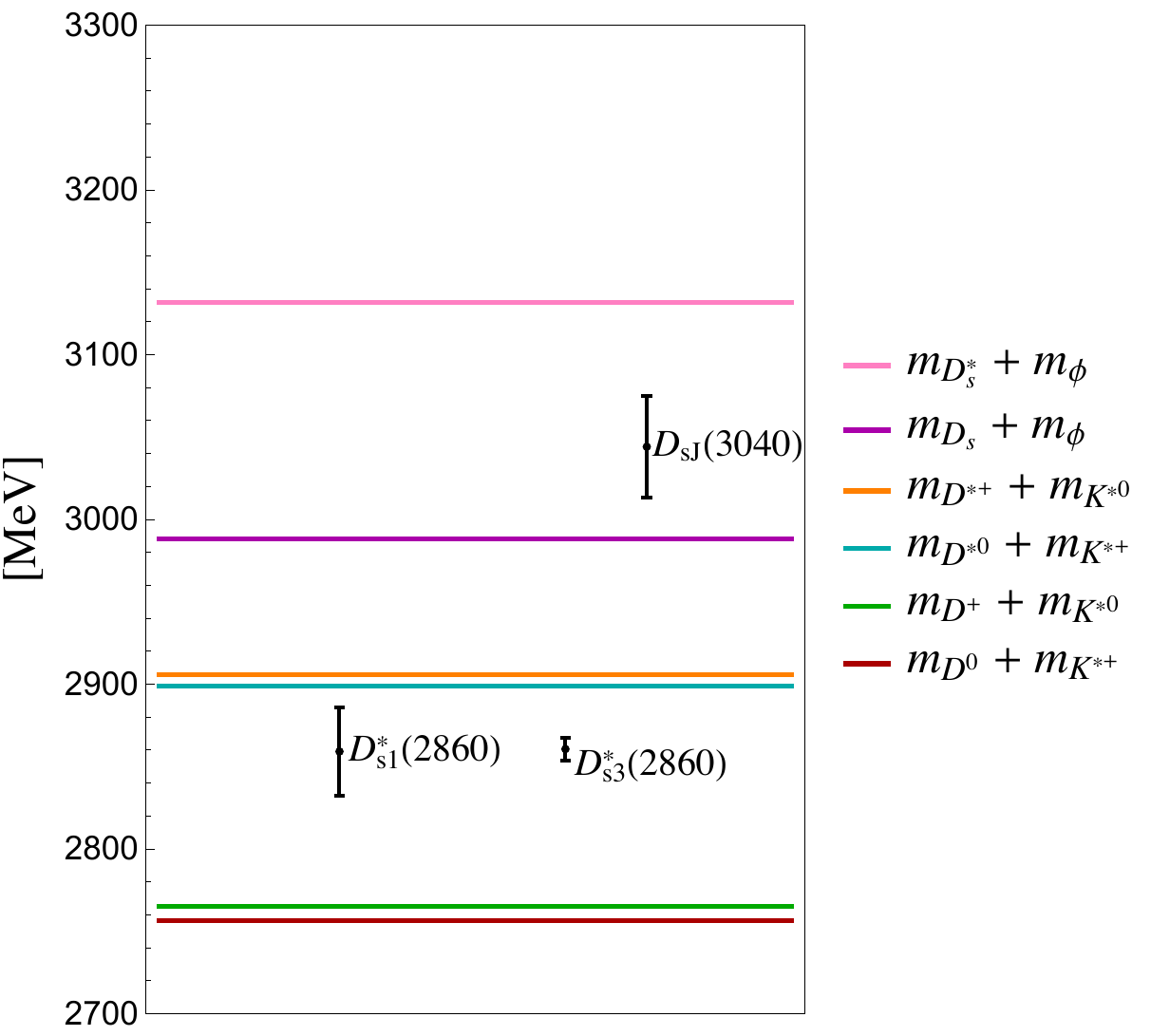}\hspace*{0.3cm}
 \caption{ $PV$ and $P^*V$ thresholds for decaying neutral non-strange  (left)  and for strange charmed  mesons (right).  The position of several resonances, with the  mass uncertainty, is indicated.}
  \label{thresholds}
 \end{center}
\end{figure}
The $PV$ and $P^*V$ thresholds of  neutral non-strange and of strange charmed mesons are  shown in Fig.~\ref{thresholds}. For charged non-strange charmed mesons the  thresholds are almost coincident with the neutral ones.

\begin{table*}[b]
\caption{Observed mesons with open charm and open beauty, classified in HQ doublets. The assignment for the states in boldface is uncertain. Two possible classifications are indicated for $D_2^*(3000)$.
}\label{ObsDdoublets}
\begin{tabular}{|c   |   c || c c |c c|| c c | c c | }\hline
doublet  \,\,$s_\ell^P$& \,\,\,$J^P$ \,\,\,& \,\,$c{\bar q}$ (n=1) \,\,\,& \,\,$c{\bar q}$ (n=2)\,\,\,& \,\, $c{\bar s}$  (n=1) \,\,\,& \,\, $c{\bar s}$ (n=2) \,\,\,
& \,\,$b{\bar q}$  (n=1) \,\,\,& \,\,$b{\bar q}$ (n=2) \,\,\,& \,\,\, $b{\bar s}$  (n=1) \,& \, $b{\bar s}$ (n=2)\,\,\\
\hline \hline
\multirow{2}{*}{\,\,\,\,\,\,\,\,$H$ \,\,\,\,\,\,\,\,\, $\frac{1}{2}^- $  }&  $0^-$ & $ D(1869)$ &
$\mathbf{ D(2550)}$   & $D_s(1968)$  & & $ B(5279)$ & $\mathbf{B_{J}(5840)}$& $ B_s(5366)$ &
\\  \cline{2-10}
   & $1^-$ & $ D^*(2010)$ & $\mathbf{D^* (2600)}$ & $D_s^*(2112)$  & $D_{s1}^{*}(2700)$ & $ B^*(5325)$ & $\mathbf{B_{J}(5960)}$& $ B_s^*(5415)$ &
\\ \hline \hline
\multirow{2}{*}{\,\,\,\,\,\,\,\,$S$ \,\,\,\,\,\,\,\,\, $\frac{1}{2}^+ $ }&   $0^+$ & $ D^*_0(2400)$ &     & $D_{s0}^*(2317)$  & & & & &
\\ \cline{2-10}
  & $1^+$ &   $D^{ \prime}_1(2430)$  & & $D_{s1}^\prime(2460)$  &   & & & &
\\ \hline \hline
\multirow{2}{*}{\,\,\,\,\,\,\,\,$T$ \,\,\,\,\,\,\,\,\, $\frac{3}{2}^+ $ }   & $1^+$ & $D_1(2420)$  & & $D_{s1}(2536)$  &  & $ B_1(5721)$ & & $ B_{s1}(5830)$ &
\\ \cline{2-10}
   & $2^+$ &   $D^*_2(2460)$ & $\mathbf{D_2^*(3000)}$& $D_{s2}^*(2573)$  & & $ B_2^*(5747)$ & & $ B_{s2}^*(5840)$ &
\\ \hline \hline
\multirow{2}{*}{\,\,\,\,\,\,\,\,$X$  \,\,\,\,\,\,\,\,\, $\frac{3}{2}^- $ }   & $1^-$ & $\mathbf{D_{J}^*(2760)}$    & &  $\mathbf{D_{s1}^*(2860)}$ & & & & &
\\ \cline{2-10}
   & $2^-$ &     & & & & & & &
\\ \hline \hline
  \multirow{2}{*}{\,\,\,\,\,\,\,\,$X^\prime$ \,\,\,\,\,\,\,\,\, $\frac{5}{2}^- $ }&   $2^-$ &   $\mathbf{D_2^{\prime}(2740)}$   & &  & & & & &
\\ \cline{2-10}
   & $3^-$ &  $\mathbf{D_3^*(2760)}$  & &  $D_{s3}^*(2860)$  & & & & &
\\ \hline \hline
  \multirow{2}{*}{\,\,\,\,\,\,\,\,$F$ \,\,\,\,\,\,\,\,\, $\frac{5}{2}^+ $ }&   $2^+$ &   $\mathbf{D_2^*(3000)}$   & &  & & & & &
\\ \cline{2-10}
   & $3^+$ &   & &  & & & & &
\\ \hline
  \end{tabular}
\end{table*}
\subsection{States in  ${\tilde H}$ doublets}
There are candidates of radial excitations of  $(D_{(s)},\,D^*_{(s)})$ in the  $H$ doublet. In particular, 
 $D_{s1}^*(2700)$  observed by Belle \cite{Brodzicka:2007aa} and BaBar \cite{Aubert:2006mh},  with  mass and width  in Table \ref{tabStrani}, can be identified with the $n=2$ excitation  of $D^*_s(2112)$.   Indeed, the measurement   \cite{Aubert:2009ah}
\be
{{\cal B}(D_{s1}^{*}(2700) \to D^*K) \over {\cal B }(D_{s1}^{*}(2700) \to
DK)} = 0.91 \pm 0.13_{stat} \pm 0.12_{syst} \,\, ,
\label{exp1}
\ee
with $D^{(*)}K= D^{(*)0}K^+ + D^{(*)^+}K_S^0$,  agrees  with the
prediction  for the first radial excitation of $D_s^*(2112)$ \cite{Colangelo:2007ds}.
The situation is unclear for the states without strangeness. Two  resonances  can be identified with the members of the $n=2$ ${\tilde H}$ doublet,
  ($D(2550)$,\, $D^*(2600)$) in Table \ref{tabBaBar},  most likely coinciding with  $(D_J^0(2580),\, D_J^{*0}(2650))$   in Table \ref{tabLHCb13}.
 However, this classification needs to be further corroborated  \cite{Colangelo:2012xi}. 

\begin{figure}[ht]
 \begin{center}
\includegraphics[width=0.45\textwidth] {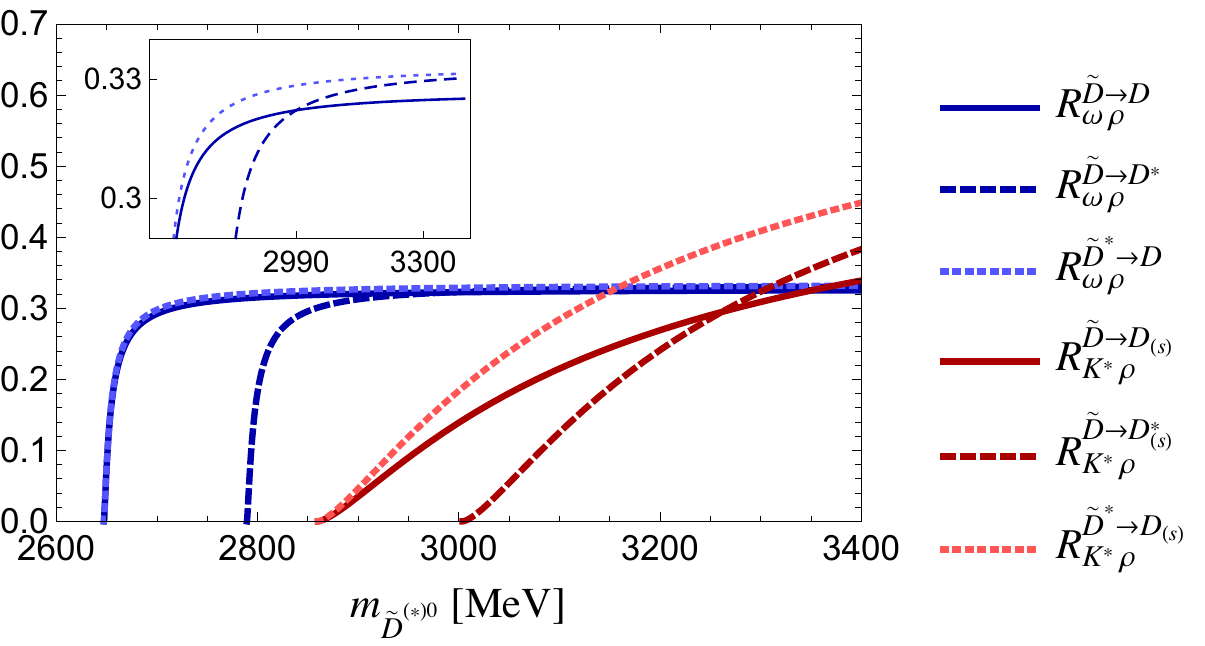}\hskip0.3cm
 \includegraphics[width=0.45\textwidth] {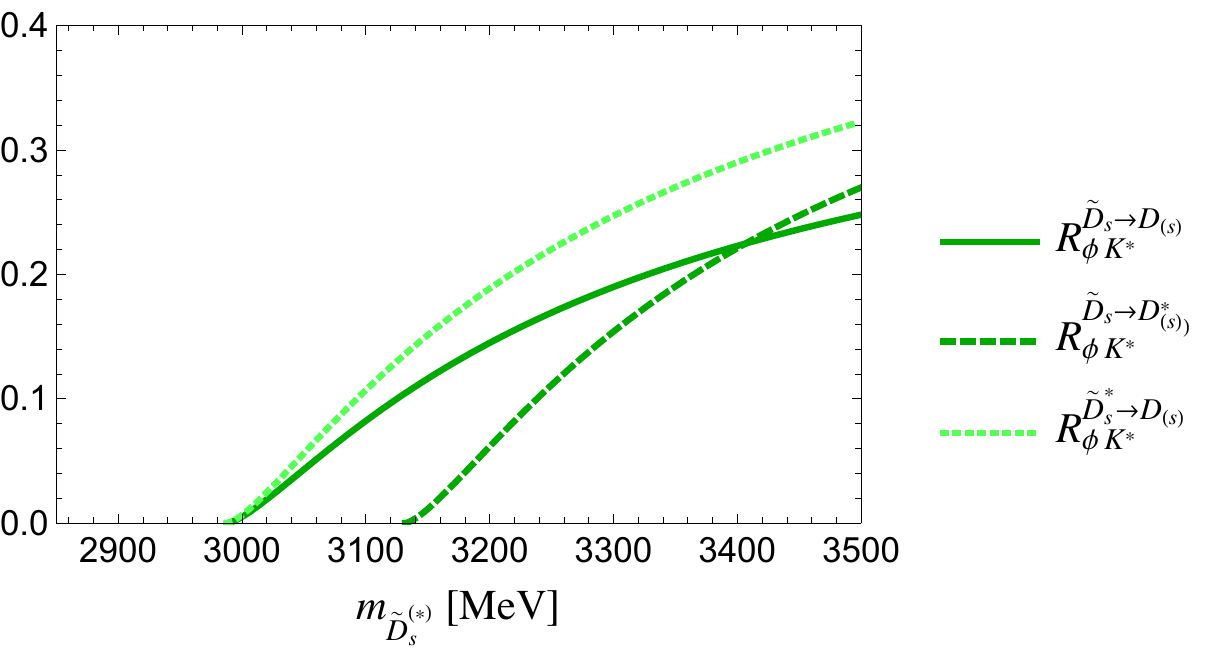}
 \caption{Ratios in Eqs.~(\ref{RDOmegaRho0})-(\ref{RDstarKstarRho0})   (left panel) 
and  ~(\ref{RsDOmegaKstar}),(\ref{RsDstarOmegaKstar}) (right panel),  evaluated  varying the mass of  the decaying meson belonging to  an excited ($n=3$) ${\tilde H}$ doublet.  
}  \label{ratioHtilde}
 \end{center}
\end{figure}

\begin{figure}[b!]
 \begin{center}
  \includegraphics[width=0.45\textwidth] {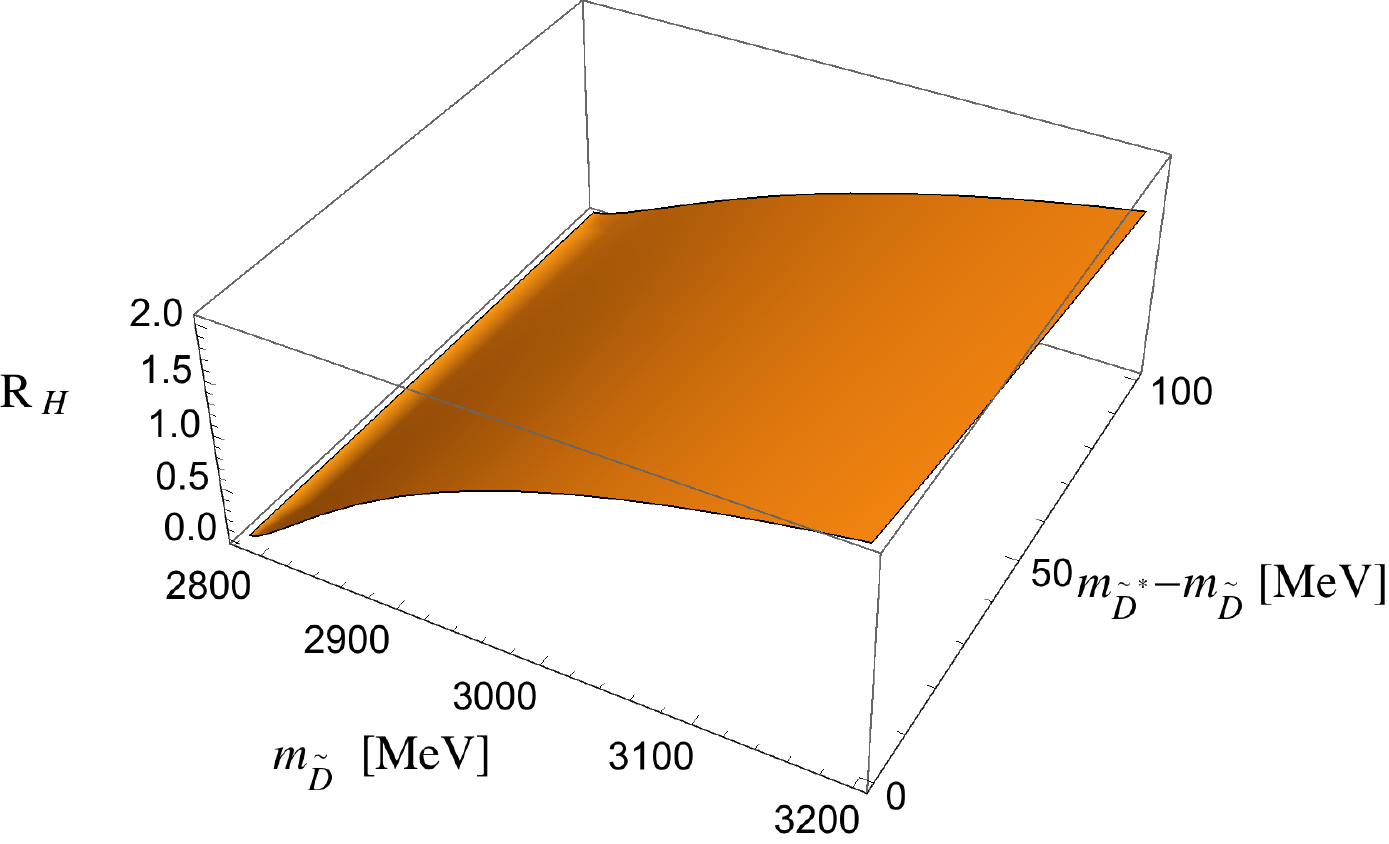}\hspace*{0.3cm}
\vspace*{0mm}
 \caption{Ratio ~(\ref{RH1}) for $V=\rho$,  varying $m_{\tilde D}$ and the mass splitting $m_{{\tilde D}^*}-m_{\tilde D}$. }
  \label{ratioHtilde3D}
 \end{center}
\end{figure}
\begin{figure}[h]
 \begin{center}
  \includegraphics[width=0.55\textwidth] {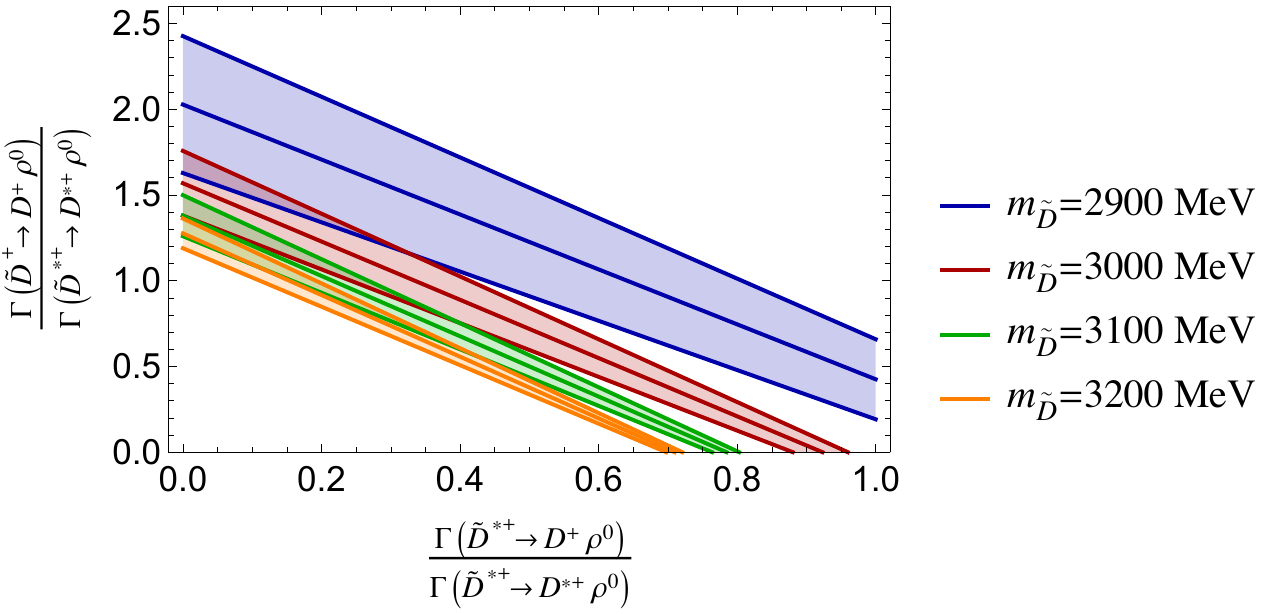}\hspace*{0.3cm}
\vspace*{0mm}
 \caption{Relation ~(\ref{relHtilde}) for several values of $m_{\tilde D}$.  The bands correspond to the chosen  $m_{\tilde D^*} - m_{\tilde D}$ spin splitting.  }
  \label{relazioneHtilde}
 \end{center}
\end{figure}
The masses of the three states are  below the $P^{(*)}V$ thresholds, hence higher radial excitations must be considered for decays into light vector mesons. 
Using  (\ref{PtildePV})-(\ref{PtildestarPstarV}), we compute   the ratios  $R^{{\tilde D} \to D}_{\omega \rho}$,  $R^{{\tilde D} \to D_{(s)}}_{K^* \rho}$, $R^{{\tilde D} \to D^*}_{\omega \rho}$, $R^{{\tilde D} \to D_{(s)}^*}_{K^* \rho}$, 
  $R^{{\tilde D}^*\to D}_{\omega \rho}$, and $R^{{\tilde D}^*\to D_{(s)}}_{K^* \rho}$ in Eqs.~(\ref{RDOmegaRho0})-(\ref{RDstarKstarRho0})
for the $({\tilde D}^0,\,{\tilde D}^{*0})$ excited doublet.
For the strange partners  we consider $R^{{\tilde D}_s \to D_{(s)}}_{ \phi K^*}$, 
  $R^{{\tilde D}_s \to D_{(s)}^*}_{ \phi K^*}$  and $R^{{\tilde D}_s^* \to D_{(s)}}_{ \phi K^*}$ in Eqs.~(\ref{RsDOmegaKstar}),(\ref{RsDstarOmegaKstar}).
In Fig.~\ref{ratioHtilde}   we depict such observables varying the mass of the decaying meson.  We find
 $R^{{\tilde D} \to D}_{\omega \rho}>R^{{\tilde D} \to D^*}_{\omega \rho}$   for $m_{{\tilde D}^0}>2990$ MeV, and 
 $R^{{\tilde D} \to D_{(s)}}_{K^* \rho}\simeq R^{{\tilde D} \to D_{(s)}^*}_{K^* \rho}$ for  $m_{{\tilde D}^0}\simeq 3260$ MeV.
The relation (\ref{RH1}),  varying $m_{\tilde D}$  and setting the mass difference between the two spin partners of radial excitations  in the range $0\leq m_{{\tilde D}^*}-m_{\tilde D} \leq 100$ MeV,  is shown in Fig.~\ref{ratioHtilde3D} for  $V=\rho$.

An interesting relation is obtained in terms  the  ratios
\be
R_1=\frac{\Gamma({\tilde D}^* \to DV)}{\Gamma({\tilde D}^* \to D^* V)} \,\, ,\hskip1cm 
R_2=\frac{\Gamma({\tilde D} \to DV)}{\Gamma({\tilde D}^* \to D^* V) } \,\, , 
\ee
 using    (\ref{RH2}): 
\be
R_2=\frac{m_{{\tilde D}^*}^4}{m_{{\tilde D}}^4}\Bigg\{\frac{m_D}{m_{D^*}}\frac{\lambda^{3/2}(m_{{\tilde D}}^2,\,m_D^2,m_V^2)}{\lambda^{3/2}(m_{{\tilde D}^*}^2,\,m_{D^*}^2,m_V^2)} -2 R_1\,
\frac{\lambda^{3/2}(m_{{\tilde D}}^2,\,m_D^2,m_V^2)}{\lambda^{3/2}(m_{{\tilde D}^*}^2,\,m_D^2,m_V^2)}
\Bigg\} \,\,. \label{relHtilde}
\ee
This relation is shown for $V=\rho$ in Fig.~\ref{relazioneHtilde},   varying the mass $m_{\tilde D}$ of $J^P=0^-$  radial excitation in the range $[2900,\,3200]$ MeV, and setting the spin splitting  $m_{{\tilde D}^*}-m_{{\tilde D}}=40 \pm 20$  MeV. 

\subsection{States in  $X$ doublets}
\begin{figure}[b]
 \begin{center}
 \includegraphics[width=0.45\textwidth] {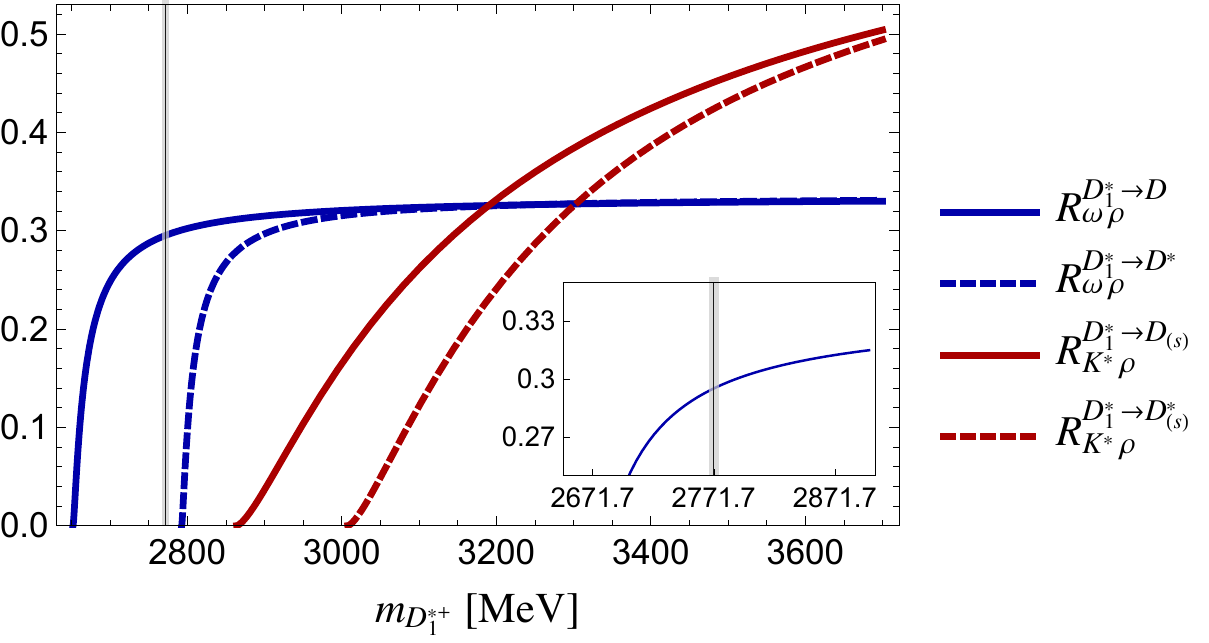}\hskip0.3cm
 \includegraphics[width=0.45\textwidth] {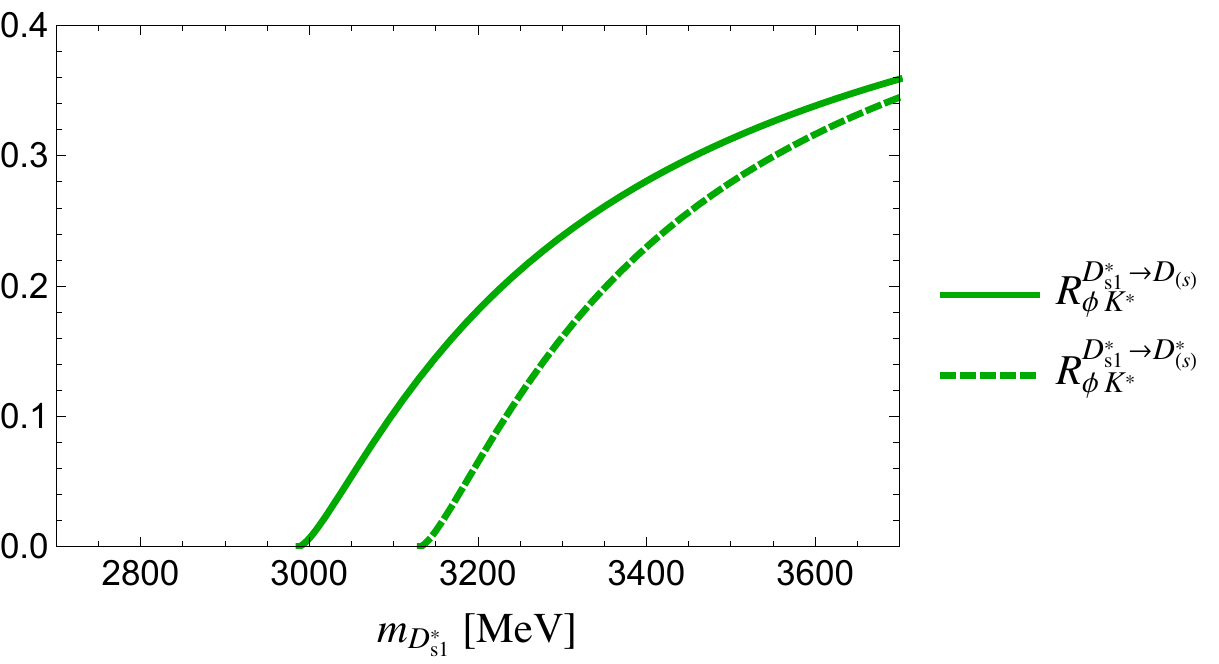}
\\
\includegraphics[width=0.45\textwidth] {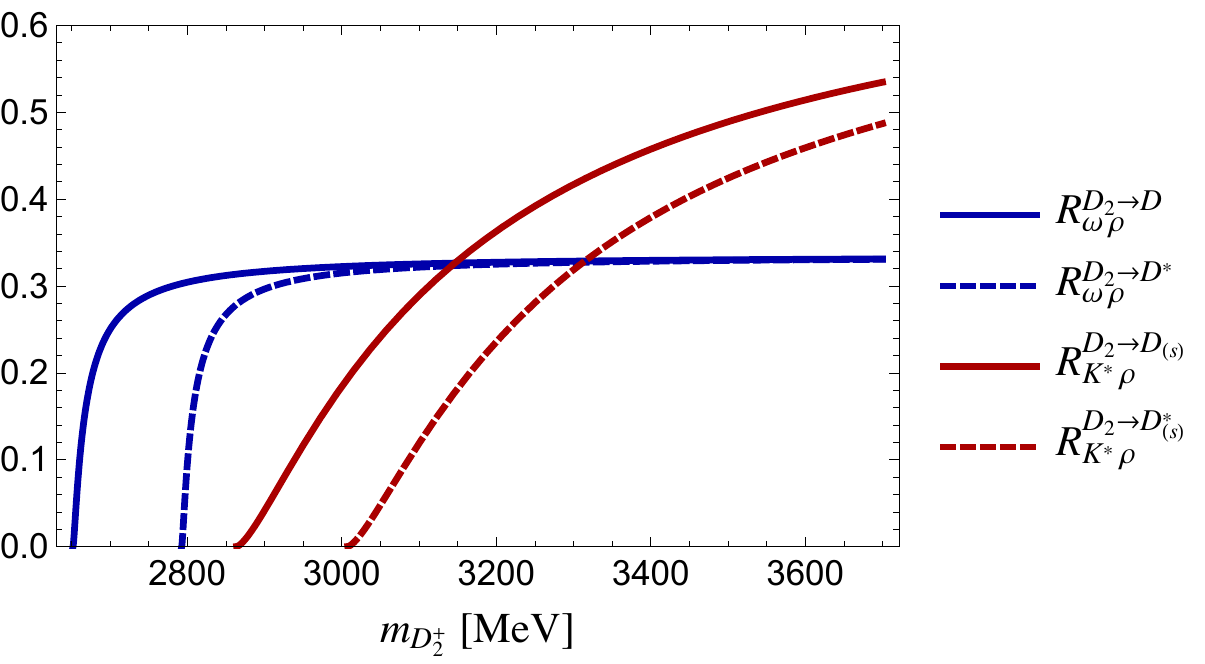}\hskip0.3cm
  \includegraphics[width=0.45\textwidth] {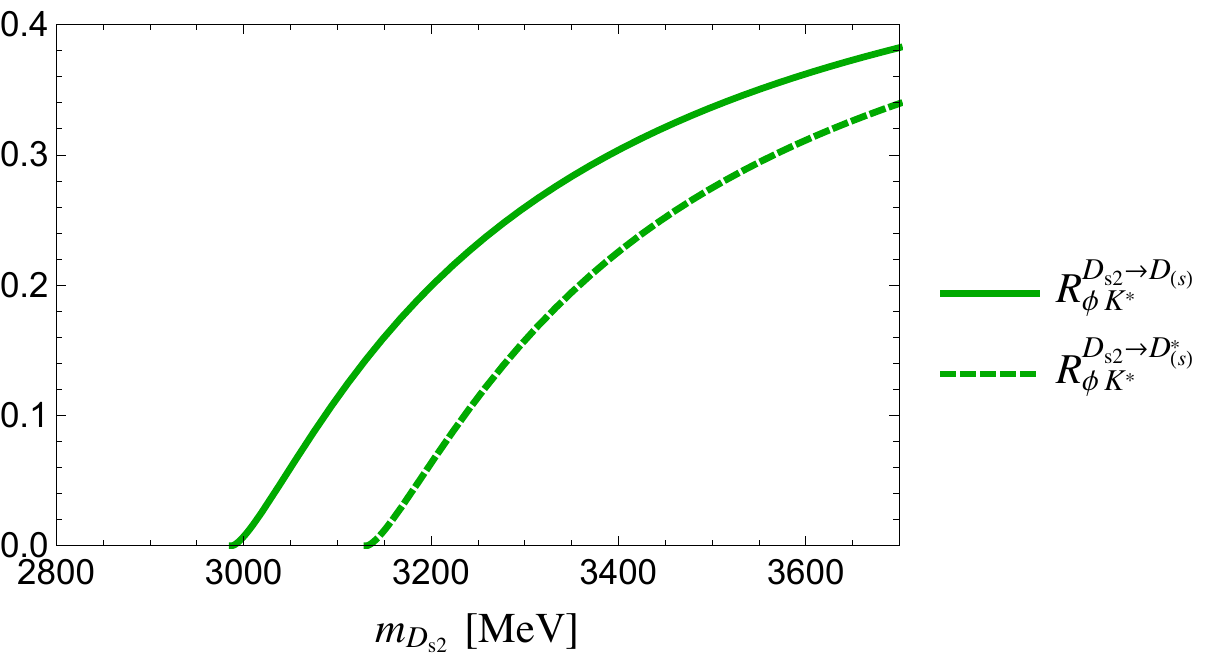}
 \caption{Ratios in  Eqs.~(\ref{RDOmegaRhopiu})-(\ref{RDstarKstarRhopiu})   (left panels) and  ~(\ref{RsDOmegaKstar})-(\ref{RsDstarOmegaKstar}) (right panels), evaluated when the decaying particle is  $D_{(s)1}^*$ (top row) and $D_{(s)2}$ (bottom row)  in the $X$ doublet. The gray region   corresponds to the measured mass of $D^{*+}_J(2760)$, candidate to be identified with  $D_1^*$. 
} \label{ratiosXD1VV}
 \end{center}
\end{figure}

Ratios of decay rates  independent of the coupling constant can be written for  $(D_1^*,\,D_2)$ belonging to the $X$ doublet. They are
 plotted  in Fig.~\ref{ratiosXD1VV} varying the  mass of the  decaying particle.
There are two  candidates for the  lowest-lying $X$ doublet: $D_J^{*+}(2760)$ observed  in the decay to $D^0 \pi^+$ \cite{Aaij:2013sza}, that is likely to have $J^P=1^-$,  and  $D_{s1}^*(2860)$   \cite{Aaij:2014xza}. Their  parameters are  in Tables \ref{tabLHCb16} and   \ref{tabStrani}.
Since the $PV$ and not $P^*V$ modes are kinematically allowed, we display in Fig.~\ref{ratiosXD1VV}
 only the ratio $R^{D_1^* \to D}_{\omega \rho}$    for $D_J^{*+}(2760)$, with  
the gray vertical line corresponding to the measured  $D_J^{*+}(2760)$  mass. 
Identifying $D_1^*$ with $D^{*+}_J(2760)$, we  predict 
\be
R^{D^{*+}(2760) \to D}_{\omega \, \rho}=(29.5\pm0.15) \times 10^{-2} \,\, .
  \ee 
The ratios $R^{D^*_1 \to D}_{\omega \, \rho}$ and $R^{D^*_1 \to D^*}_{\omega \, \rho}$  are nearly equal  for masses larger than $\simeq 3.25$ GeV, while $R^{D^*_1 \to D_{(s)}}_{K^* \, \rho}< R^{D^*_1 \to D_{(s)}^*}_{K^* \, \rho}$ .  The ratios  for $D_{s1}^*$ and the spin partner $D_{(s)2}$ are also in Fig.~\ref{ratiosXD1VV}.   

\begin{figure}[t]
 \begin{center}
  \includegraphics[width=0.4\textwidth] {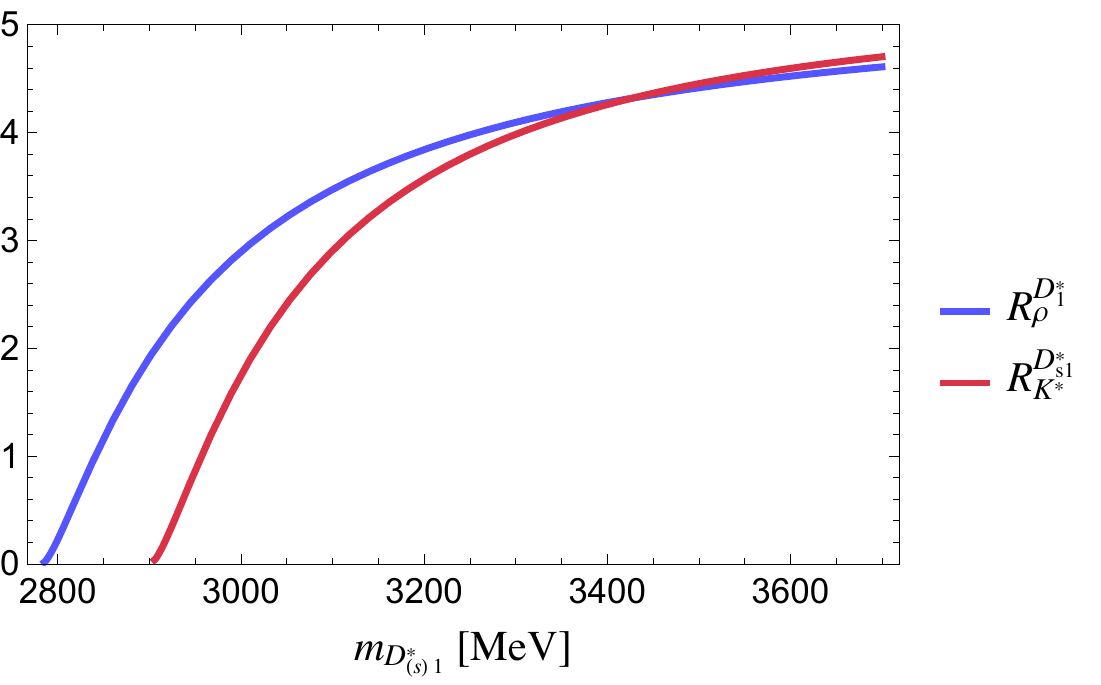}\hskip0.3cm
    \includegraphics[width=0.4\textwidth] {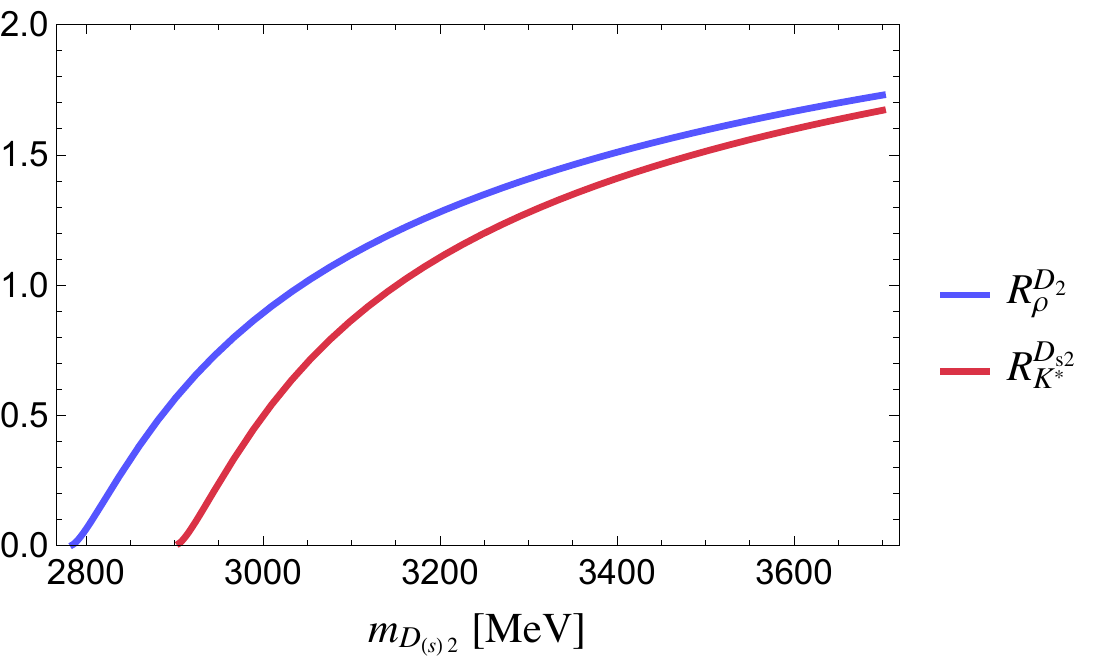}
\vspace*{0mm}
 \caption{Ratios  (\ref{RsameRhopiu})-(\ref{RsameKstar}) for decaying particle in  the $X$ doublet. }
  \label{ratiosXsameV}
 \end{center}
\end{figure}

The observables in  Eqs.~(\ref{RsameRhopiu})-(\ref{RsameKstar}), also independent of the coupling constant and involving the same final light vector meson, are
 displayed in Fig.~\ref{ratiosXsameV}.

\begin{figure}[b]
 \begin{center}
  \includegraphics[width=0.4\textwidth] {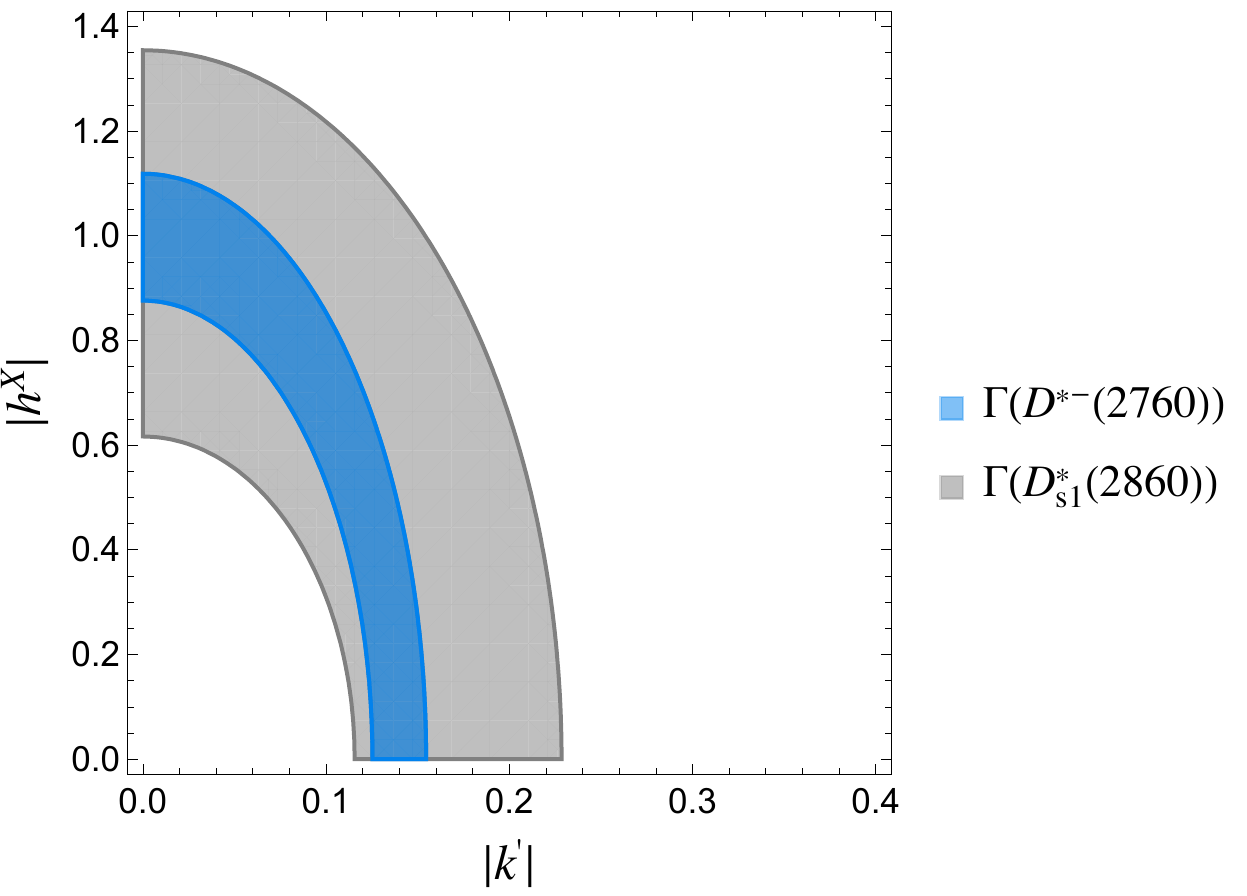}
\vspace*{0mm}
 \caption{Bounds for the couplings $k^\prime$ and $h^X$ from the  widths of  the $X$ doublet candidates  $D^{*+}(2760)$  and $D_{s1}^*(2860)$.  }
  \label{regplotX}
 \end{center}
\end{figure}

At the chosen order in the effective Lagrangian approach   the strong decay widths of the members of the $X$ doublet with a light final pseudoscalar meson depend on the  constant $k^\prime$  in Eq.~(\ref{lagX}).
Neglecting phase-space suppressed channels (e.g. decays to  excited doublets), the widths of the members of the $X$ doublet are determined by the couplings $k^\prime$ and $h^X$.
Saturating   the widths of $D^{*+}(2760)$ and $D_{s1}^{*+}(2860)$    by the  modes
\bea
D^{*+}(2760)&\to& D^{(*)+} \pi^0,\, D^{(*)0} \pi^+,\, D^{(*)+} \eta,\, D_s^{(*)} K_S,\, D^+ \rho^0,\, D^0\rho^+,\, D^+ \omega \nn \\
D_{s1}^{*+}(2860)&\to& D^{(*)+}K_S,\, D^{(*)0} K^+,\, D_s^{(*)} \eta,\, D^+  K^{*0},\, D^0 K^{*+} ,\nn 
\eea
the couplings $k^\prime$ and $h^X$ can be constrained in the region in Fig.~\ref{regplotX}, with the bound $|k^\prime|<0.16$.

\subsection{States in  $ X^\prime$ doublet}
In 2006 BaBar observed the $D_{sJ}(2860)$ meson decaying to $DK$ \cite{Aubert:2006mh}, which was proposed as the $J^P=3^-$ state in the $c \bar s$ $X^\prime $  doublet  \cite{Colangelo:2006rq}.  A subsequent LHCb analysis  supported this classification  and showed that another state, $D_{s1}^*(2860)$  with $J^P=1^-$, is present in same mass region,  likely the member of the $X$ doublet \cite{Aaij:2014xza}.
The parameters of the $J^P=3^-$ resonance are  in Table \ref{tabStrani}.
LHCb   observed another candidate for the $X^\prime$ doublet,  $D_3^{*0}(2760)$ with parameters  in Table \ref{tabLHCb16}, that can be identified with the non-strange partner of $D_{s3}^{*-}(2860)$  \cite{Aaij:2016fma}.
 Finally, BaBar and LHCb found  a resonance that might be the $J^P=2^-$  state  in the $X^\prime$ doublet: this is  $D^0(2750)$  decaying to $D^{*+}\pi^-$ \cite{delAmoSanchez:2010vq}, with parameters in Table \ref{tabBaBar}.   The LHCb $D^0(2740)$ state, observed in   $D^{*+} \pi^-$  \cite{Aaij:2013sza} (see  Table \ref{tabLHCb13}),   likely coincides with it. 

\begin{figure}[t]
 \begin{center}
\includegraphics[width=0.45\textwidth] {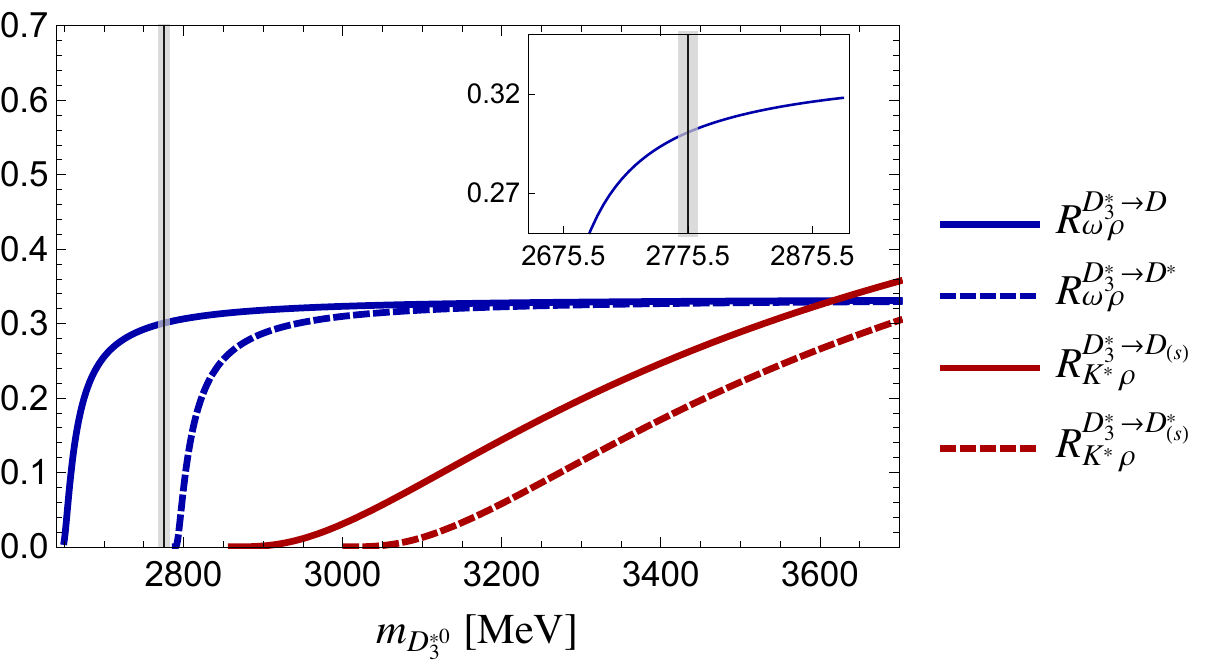} \hskip 0.3cm
 \includegraphics[width=0.45\textwidth] {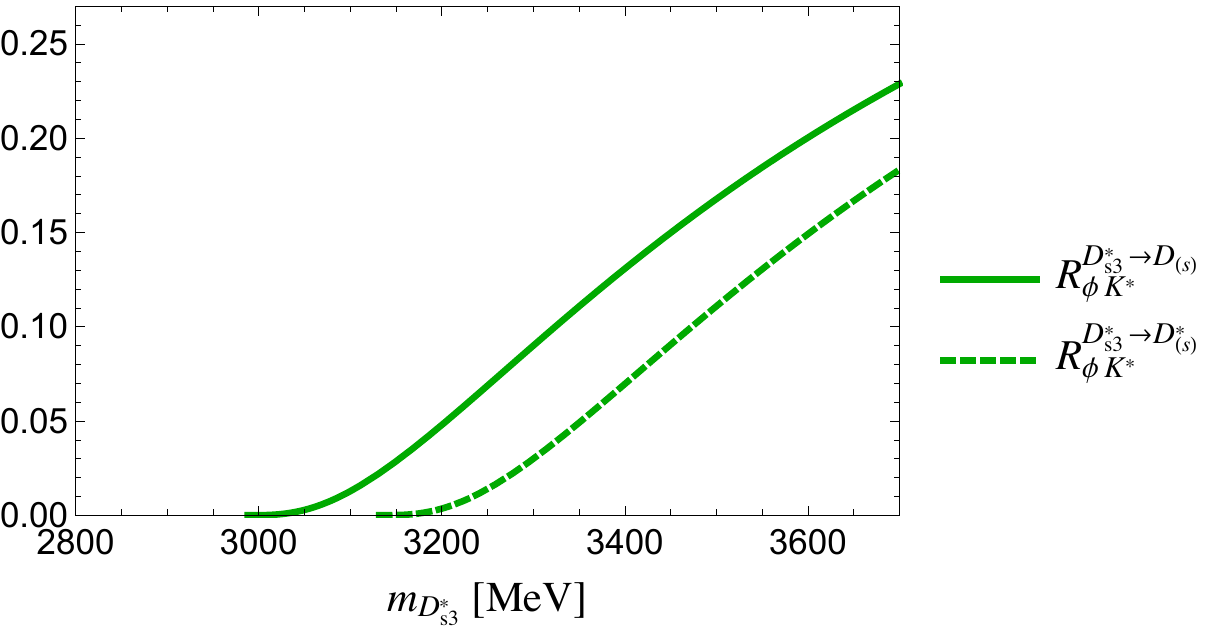} \\
  \includegraphics[width=0.45\textwidth] {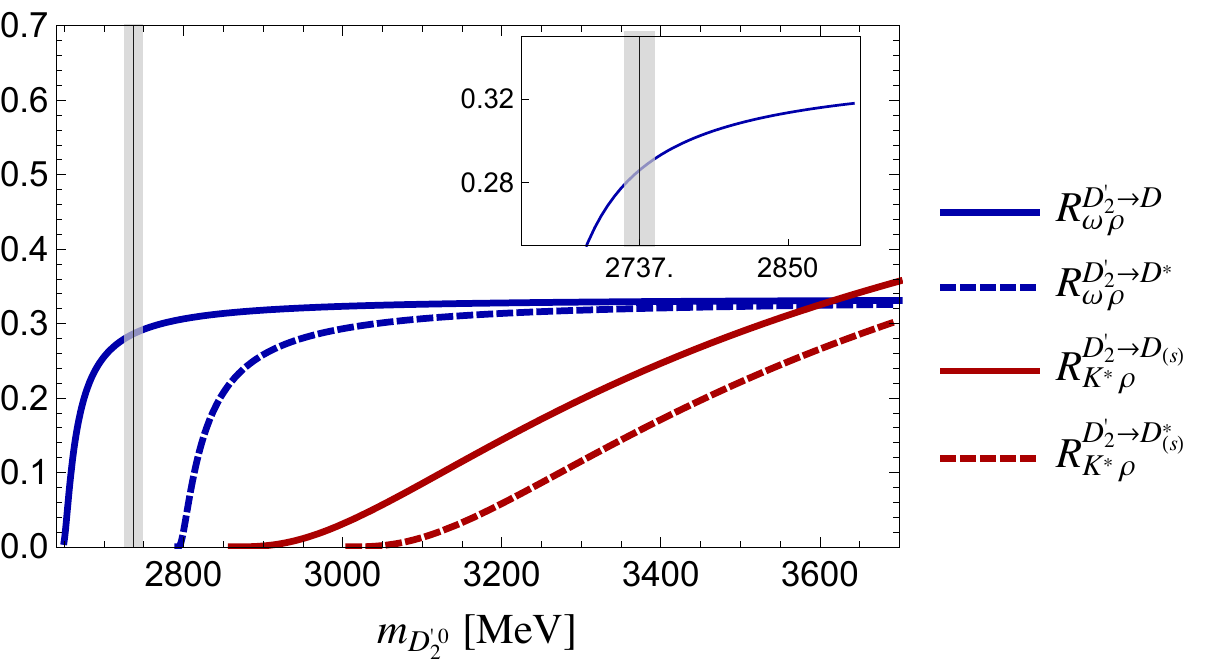} \hskip 0.3cm
    \includegraphics[width=0.45\textwidth] {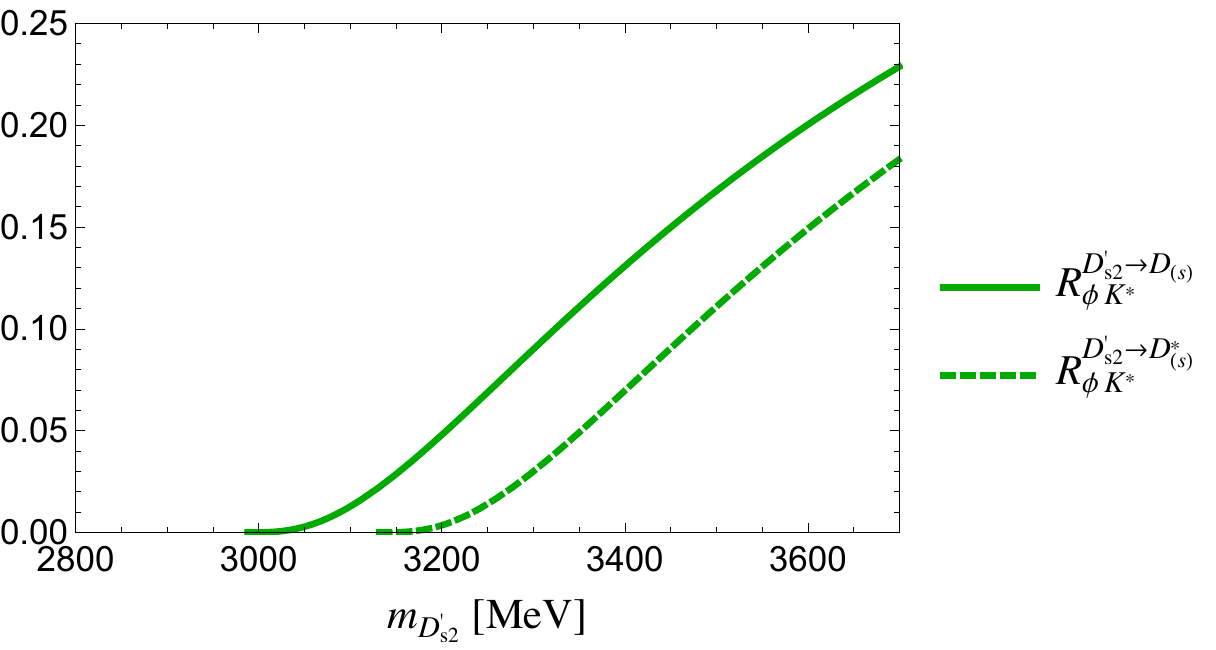}
 \caption{Ratios in  Eqs.~(\ref{RDOmegaRho0})-(\ref{RDstarKstarRho0})   (left panels) and  ~(\ref{RsDOmegaKstar}),(\ref{RsDstarOmegaKstar}) (right panels), evaluated when the decaying particle is the decaying particle $D_{(s)3}^*$ (top row) and $D_{(s)2}^\prime$ (bottom row) in the $X^\prime$ doublet. 
The gray regions (enlarged in the inset)  correspond to the measured mass of  $D^{*0}_3(2760)$ candidate as  $D_3^*$,  and of  $D^0(2740)$ candidate for the  $D_2^\prime$ in the $X^\prime$ doublet.}
  \label{ratiosXpD3VV}
 \end{center}
\end{figure}

\begin{figure}[b]
 \begin{center}
  \includegraphics[width=0.42\textwidth] {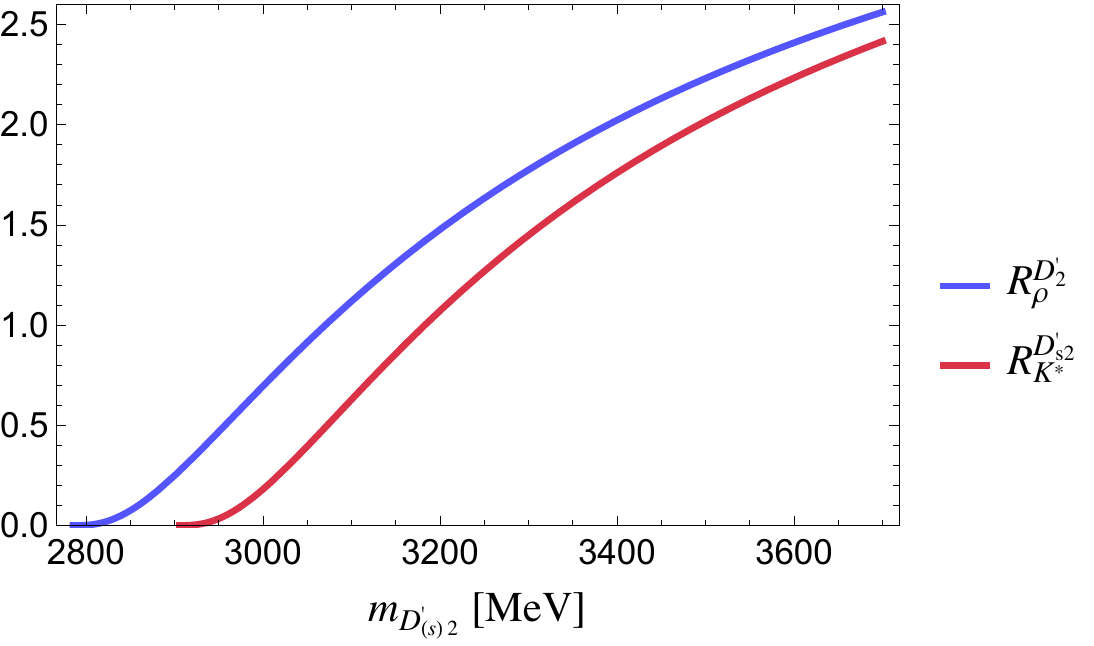}\hskip0.3cm
    \includegraphics[width=0.42\textwidth] {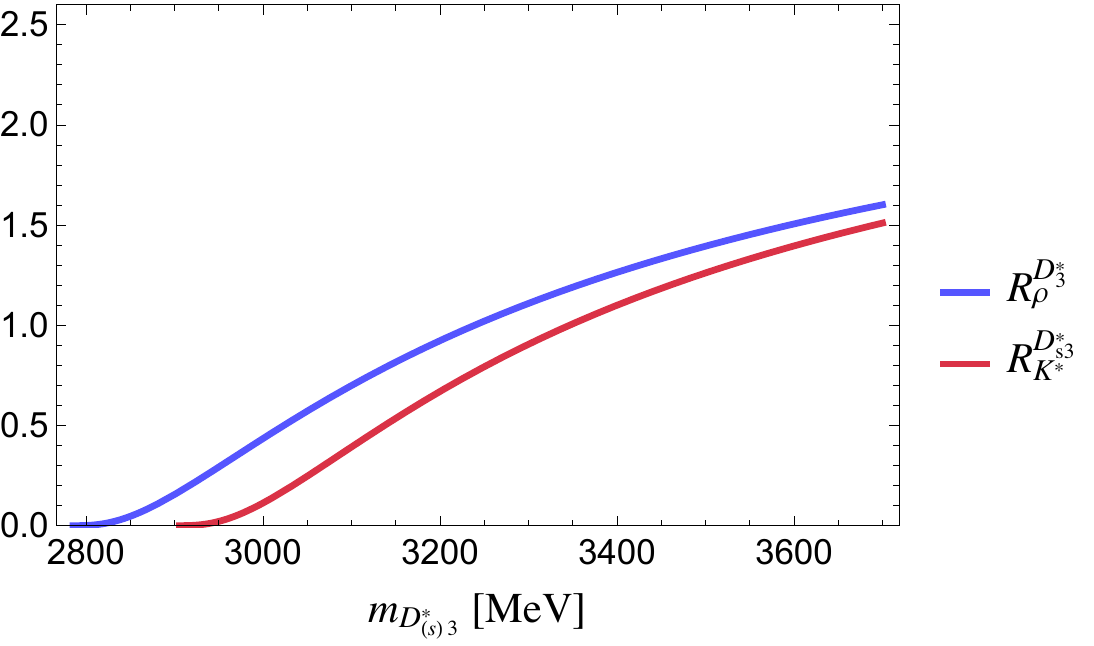}
\vspace*{0mm}
 \caption{Ratios (\ref{RsameRhopiu})-(\ref{RsameKstar}) for a  decaying particle belonging to  $X^\prime$.  }
  \label{ratiosXpsameV}
 \end{center}
\end{figure}

For the two $J^P=3^-$ states, allowed decays to light vector mesons are $D_3^{*0}(2760) \to D^+ \rho^-,\,D^0 \rho^0,\,D^0 \omega$ and $D_{s3}^{*+}\to {\bar D}^0K^{*+},\,D^- K^{*0}$. We plot in Fig.~\ref{ratiosXpD3VV}
ratios of widths independent of the  coupling constant,  varying  the mass of the decaying particle.
In correspondence to the measured  $D_3^{*0}(2760)$ mass we predict:
\be
R^{D_3^{*0 }(2760)\to D}_{\rho\,\omega}=(30.1 \pm 0.2) \times 10^{-2} \,\, 
\ee
and
\bea
R_{a}^{X^\prime}&=&\frac{\Gamma(D_3^{*0}(2760) \to D^+ \rho^-)+\Gamma(D_3^{*0}(2760) \to D^0 \rho^0)}{\Gamma(D_{s3}^{*-}(2860) \to {\bar D}^0 K^{*-})+\Gamma(D_{s3}^{*-}(2860) \to D^- K^{*0})}=1.6 \pm 0.5 \,\, , \label{R3rho} \\
R_b^{X^\prime}&=&\frac{\Gamma(D_3^{*0}(2760) \to D^0 \omega)}{\Gamma(D_{s3}^{*-}(2860) \to {\bar D}^0 K^{*-})+\Gamma(D_{s3}^{*-}(2860) \to D^- K^{*0})}=0.47\pm0.16 \,\,.\label{R3omega}
\eea
Analogous ratios  for the $J^P=2^-$ member of the $X$ doublet,  with and without strangeness,
are shown in Fig.~\ref{ratiosXpD3VV}.
In the non-strange case, the candidate is  $D^0(2740)$.     The $D^+ \rho^-,\,D^0 \rho^0,\,D^0 \omega$ channels are open,
and we predict:
\be
R^{D_2^{\prime 0}(2740) \to D}_{\rho\,\omega}=(28.6 \pm 0.6)\times 10^{-2} \,\, .
\label{RD2740}\ee
 In the same figure $R^{D_2^\prime \to D}_{\rho\,\omega}$ is plotted versus the mass of $D_2^\prime$,  with the gray region corresponding  to the  $D^0(2740)$  measured mass. 
Ratios  involving  the same final light vector meson   are displayed in Fig.~\ref{ratiosXpsameV}.

\begin{figure}[t]
 \begin{center}
  \includegraphics[width=0.4\textwidth] {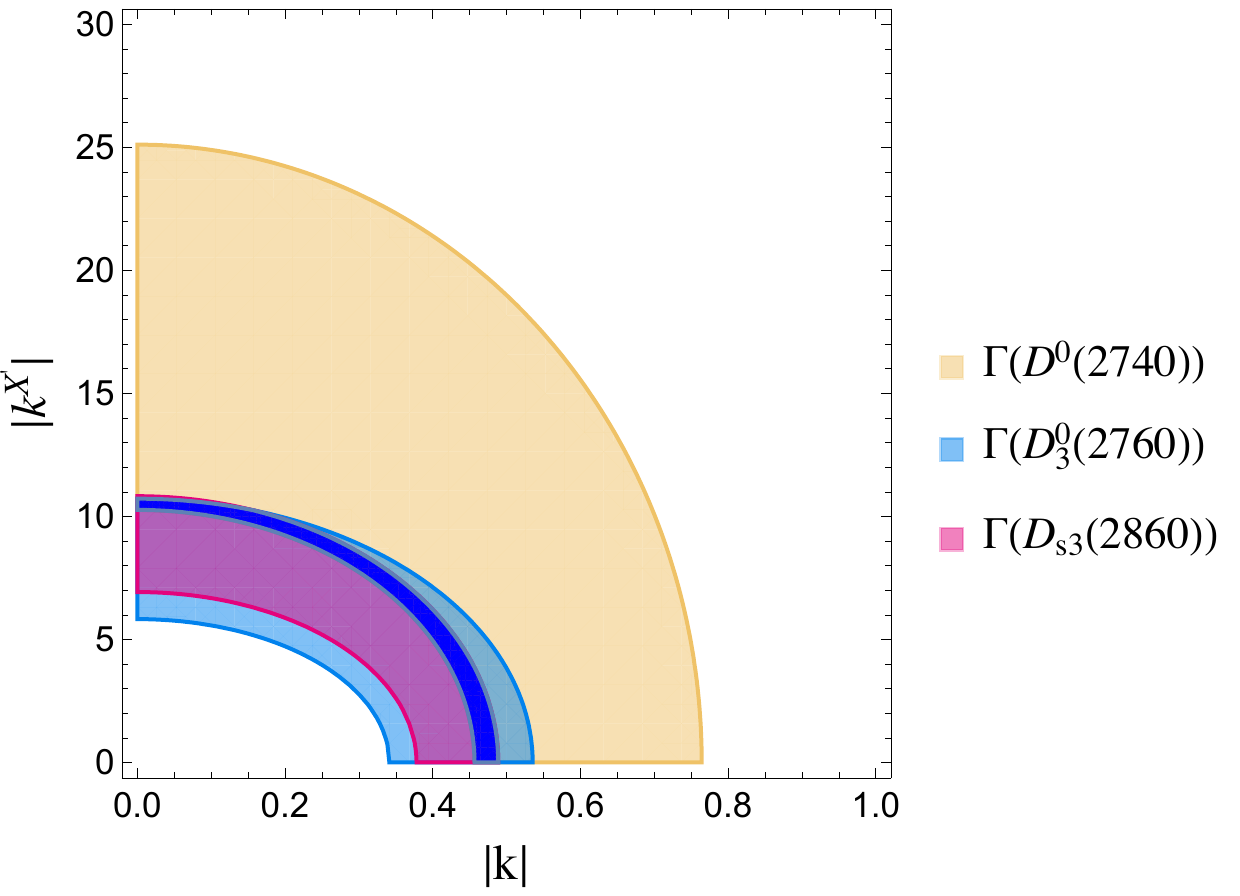}\hspace*{0.3cm}
   \includegraphics[width=0.3\textwidth] {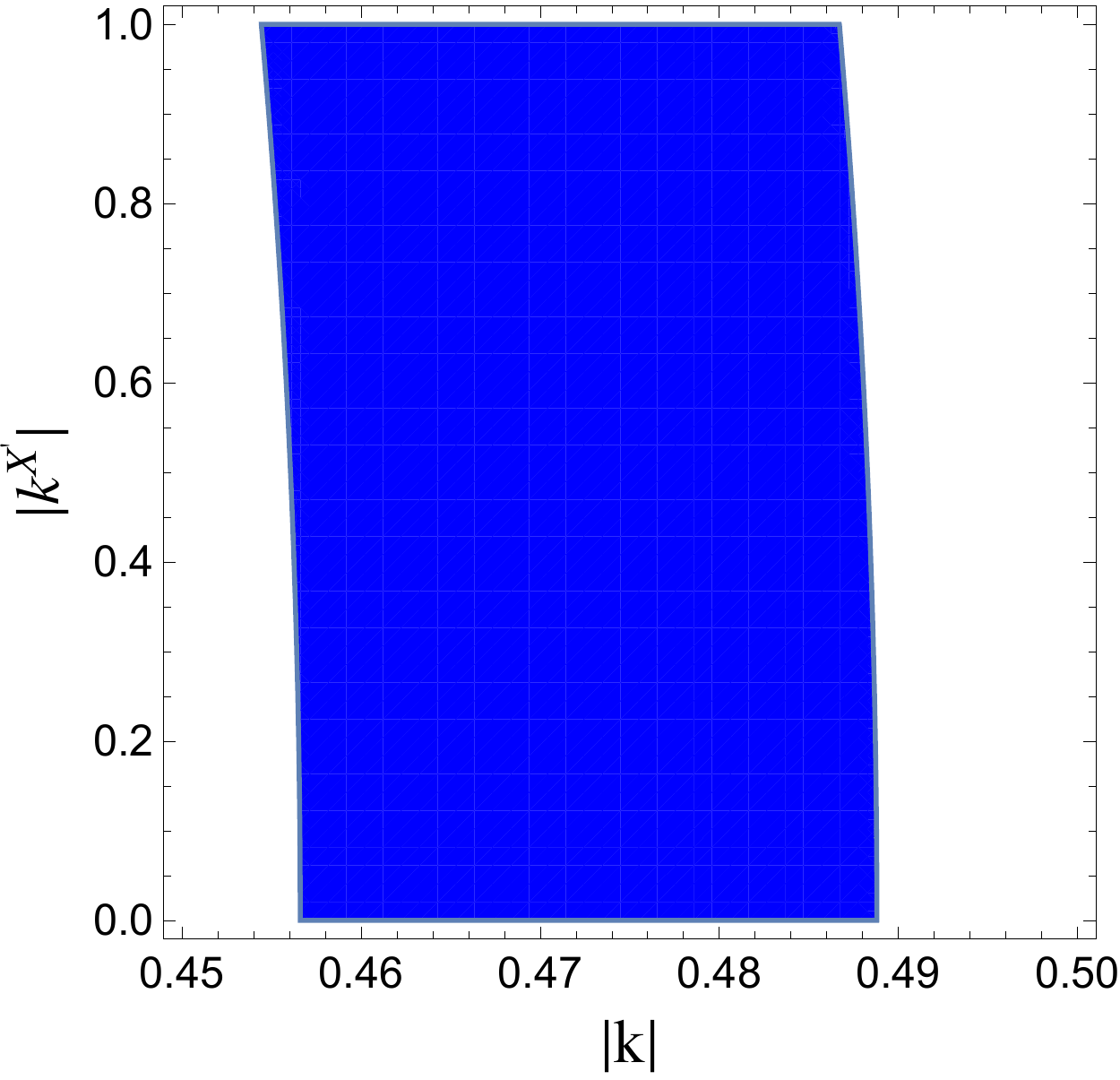}
\vspace*{0mm}
 \caption{Left: Constraints  on the couplings $k,\,k^{X^\prime}$ from the measured widths of  $D^0(2740)$ \cite{Aaij:2013sza}, $D_3^*(2760)$ \cite{Aaij:2016fma} and $D_{s3}^*(2860)$ \cite{Aaij:2014xza}, candidates for the $X^\prime$ doublet. In the dark blue region all  constraints are fulfilled. Right:  coupling region for $|k^{X^\prime}|<1$. }
  \label{regplotXp}
 \end{center}
\end{figure}

In the effective Lagrangian approach, the strong decay widths of the members of the $X^\prime$ doublet to a light pseudoscalar meson  are controlled by  $k=k_1+k_2$, with $k_1$ and $k_2$  in Eq.~(\ref{lagXp}).
Neglecting phase-space suppressed modes, the  widths of the members of the $X^\prime$ doublet are determined by the  couplings $k$ and $k^{X^\prime}$. 
If   $(D^0(2750),\,D_3^*(2760)) and \, D_{s3}^*(2860)$  belong to  $X^\prime$ doublet, their widths  impose constraints on the two constants,
as shown in the left panel of Fig.~\ref{regplotXp}  obtained assuming  the full widths  saturated by 
\bea
D^0(2740) &\to& D^{*0}\pi^0,\, D^{*+}\pi^-,\, D^{*0} \eta,\, D^*_s K^-,\, D^+ \rho^-,\, D^0\rho^0,\, D^0 \omega \nn \\
D_3^{*0}(2760) &\to& D^{(*)+} \pi^-,\,  D^{(*)0} \pi^0,\,D^{(*)0} \eta,\, D^{(*)}_s K^-,\, D^+ \rho^-,\, D^0\rho^0,\, D^0 \omega \nn \\
D_{s3}^*(2860) &\to& D^{(*)+}K_S,\, D^{(*)0} K^+,\,D_s^{(*)} \eta,\, D^+K^{*0},\, D^0 K^{*+} \,\,.
\nn 
\eea
For $|k^{X^\prime}|<1$ the coupling region is also shown in Fig.~\ref{regplotXp} (right panel):   $k^{X^\prime}$ is  unconstrained, while $|k|=0.47 \pm 0.02$, slightly above  the value obtained in \cite{Colangelo:2012xi} using the BaBar data \cite{Aubert:2006mh,delAmoSanchez:2010vq}.
\subsection{States in  $ {\tilde T}$ doublet}
\begin{figure}[t]
 \begin{center}
\includegraphics[width=0.45\textwidth] {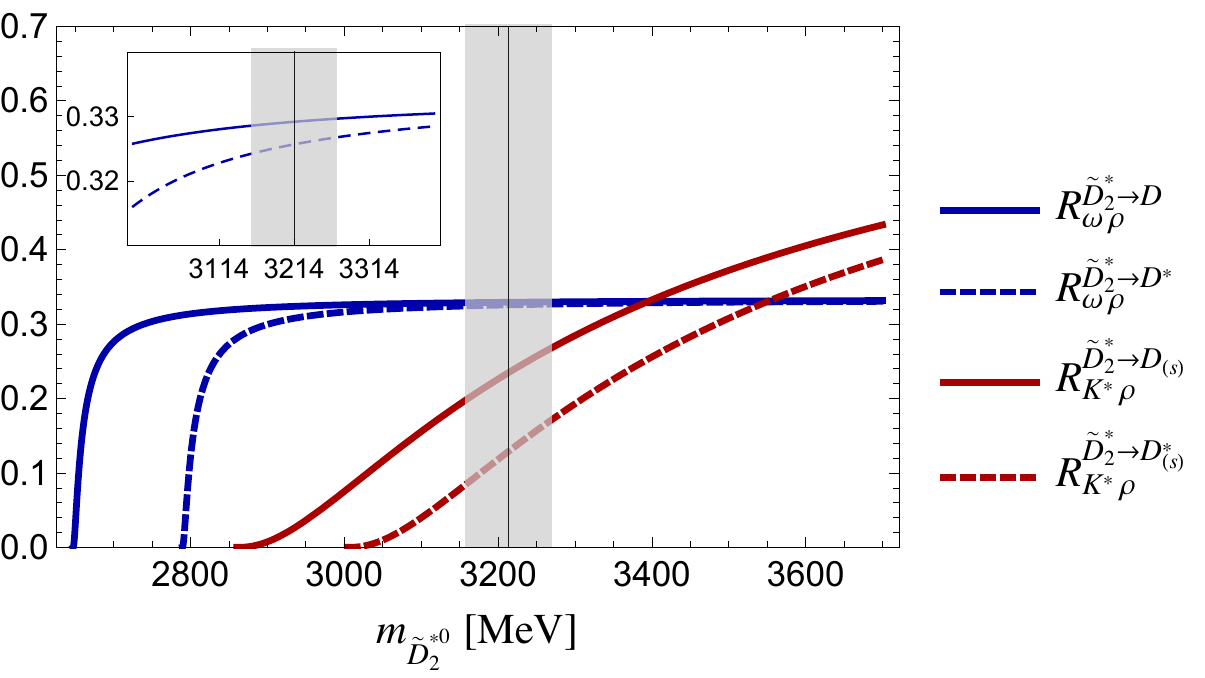}\hskip 0.3cm
\includegraphics[width=0.45\textwidth] {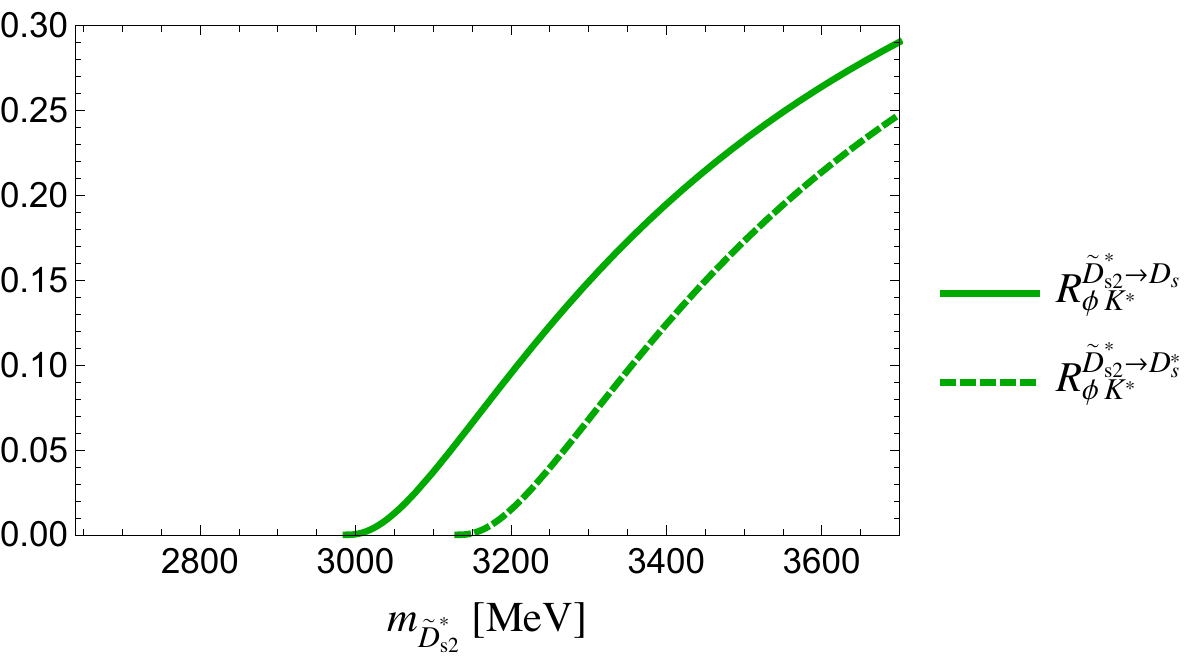}
\\
\includegraphics[width=0.45\textwidth] {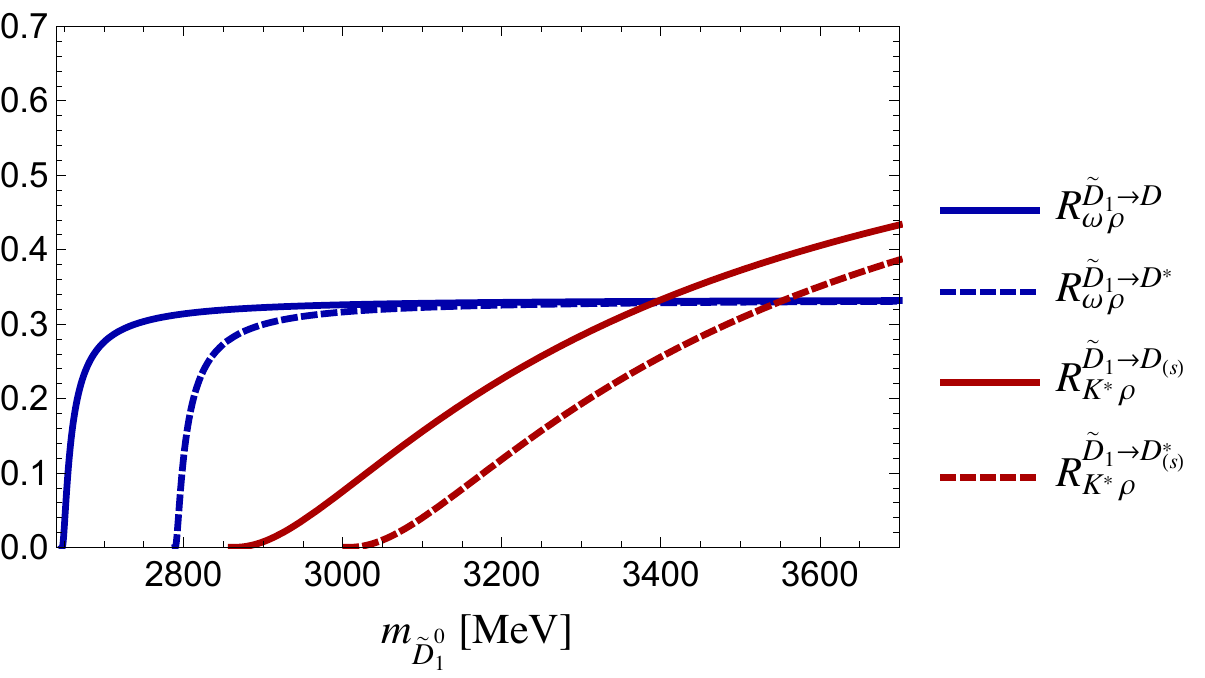}
\hskip 0.3cm
 \includegraphics[width=0.45\textwidth] {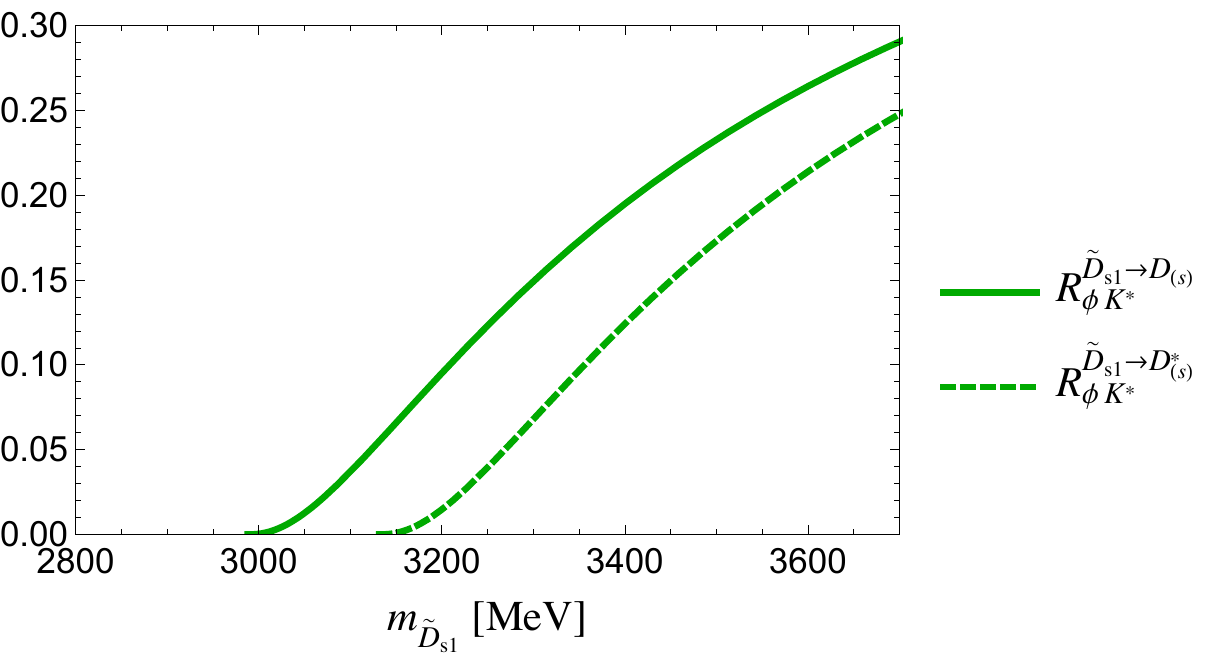}
 \caption{
Ratios in  Eqs.~(\ref{RDOmegaRho0})-(\ref{RDstarKstarRho0})   (left panels) and  ~(\ref{RsDOmegaKstar})-(\ref{RsDstarOmegaKstar}) (right panels),  evaluated when the decaying particle is  ${\tilde D}_{(s)2}^*$ (top row) and ${\tilde D}_{(s)1}$  (bottom row) belonging to ${\tilde T}$.
}
  \label{ratioTtildeD1}
 \end{center}
\end{figure}

\begin{figure}[b]
 \begin{center}
  \includegraphics[width=0.4\textwidth] {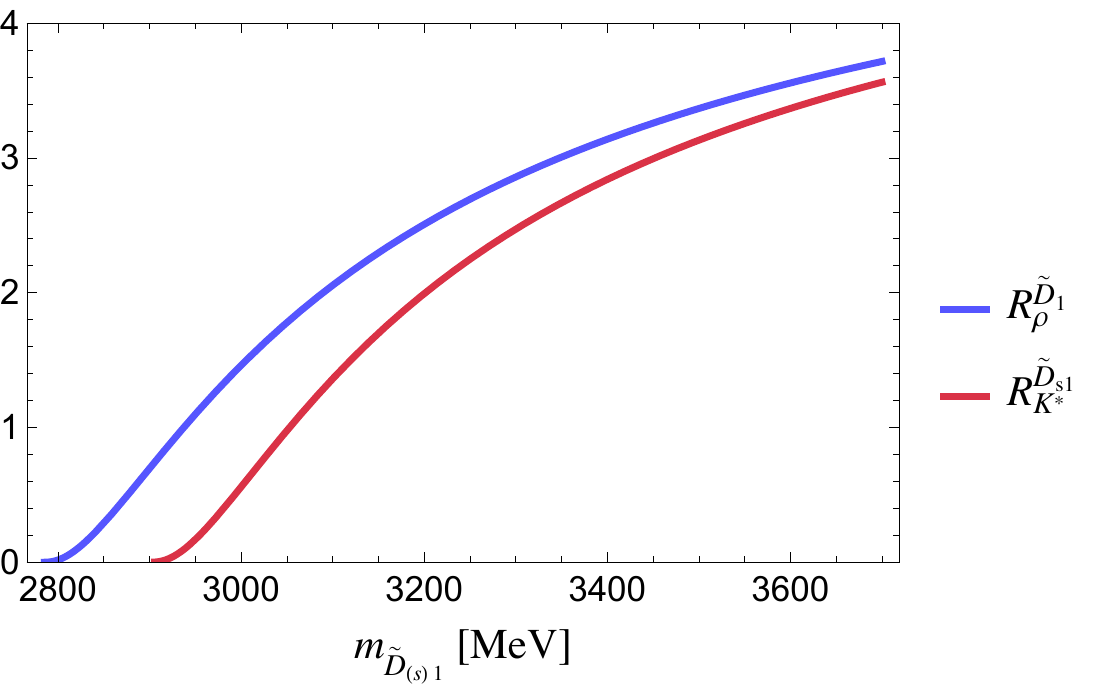}\hskip0.3cm
    \includegraphics[width=0.4\textwidth] {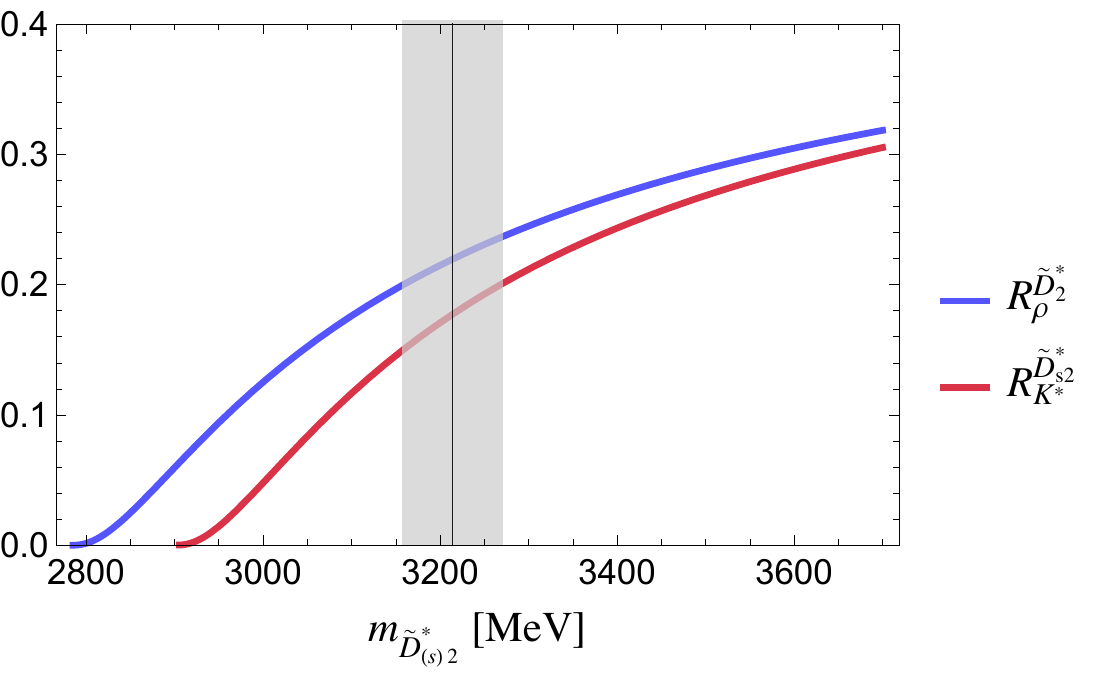}
 \vspace*{0mm}
 \caption{ Ratios (\ref{RsameRhopiu})-(\ref{RsameKstar}) when the decaying particle belongs to the ${\tilde T}$ doublet. }
  \label{TtildesameV}
 \end{center}
\end{figure}

\begin{figure}[t]
 \begin{center}
  \includegraphics[width=0.4\textwidth] {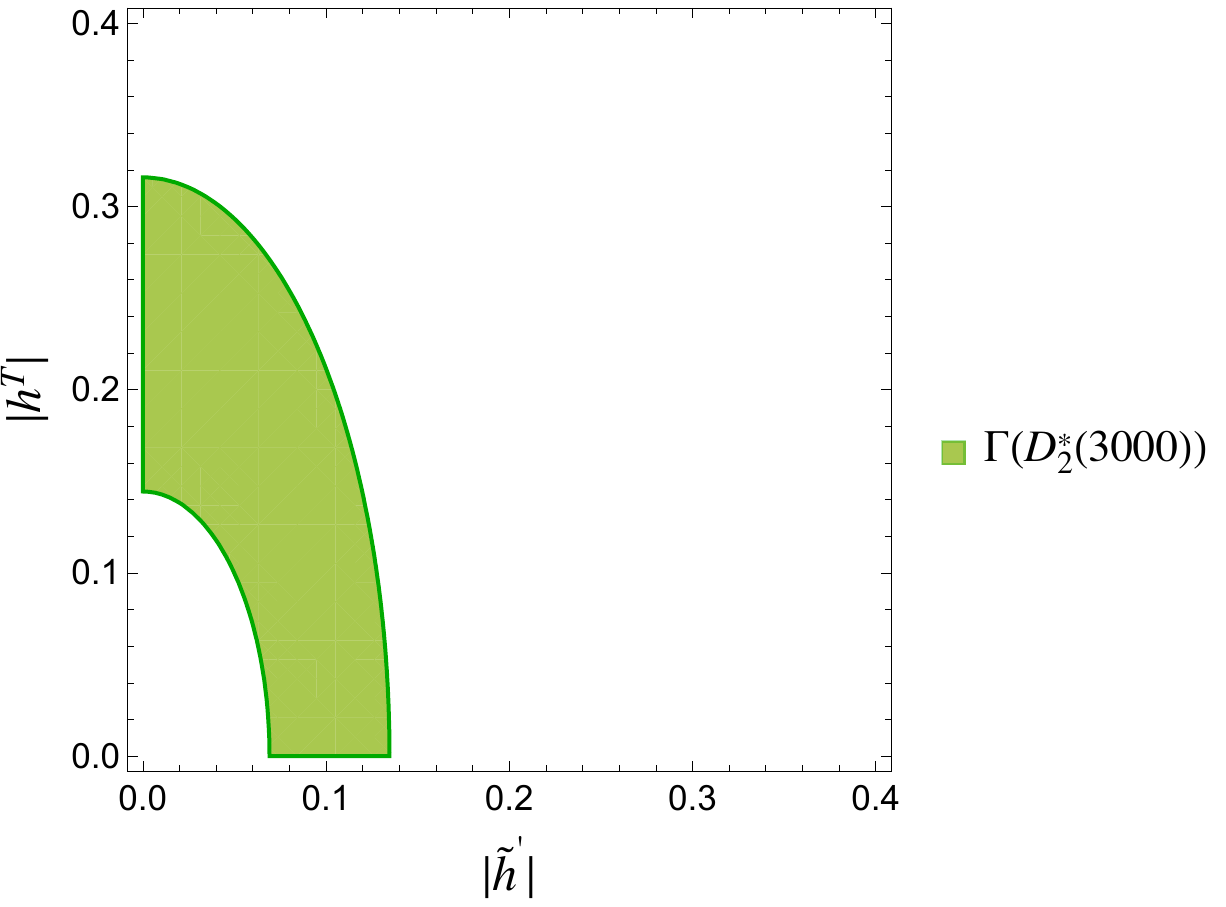}
\vspace*{0mm}
 \caption{Bounds on the couplings $\tilde h^\prime $ in (\ref{lagT}) and $h^T$ in (\ref{LHTV2}) from the measured width of  $D^{*}_2(3000)$,  assuming that the state belongs to  $\tilde T$.  }
  \label{regplotTtilde}
 \end{center}
\end{figure}

We analyze the  ${\tilde T}$ doublet before   $F$  since there is a state that can fit in both of them,  and this sequence in the discussion is convenient.
For each one of the two states in the ${\tilde T}$ spin doublet
we construct   ratios of decay rates independent of strong couplings.
A $J^P=2^+$ meson   has been observed  \cite{Aaij:2016fma},  $D_2^*(3000)$,  that could fit in the  ${\tilde T}$  or  in the $F$  doublet. Hence, we  compute  the various ratios varying the mass of the decaying particle,  then we specialize to the mass of the candidate, as shown in Fig.~\ref{ratioTtildeD1}.
For  $D_2^*(3000)$ belonging to this doublet we predict
\bea
R_{\omega \rho}^{D_2^{*0}(3000) \to D}&=&(33.0 \pm 0.1)\times 10^{-2} \,\, , \hskip1.2cm
R_{\omega \rho}^{D_2^{*0}(3000) \to D^*}=(32.6 \pm 0.2) \times 10^{-2} \,\, ,\nn \\
R_{K^* \rho}^{D_2^{*0}(3000) \to D_{(s)}}&=&(23.5 \pm 3.6) \times 10^{-2} \,\, , \hskip1cm
R_{K^* \rho}^{D_2^{*0}(3000) \to D_{(s)}^*}=(13.0 \pm 4.5) \times 10^{-2}\nn \,\,.
\eea
Ratios of decay rates involving the same final vector meson  are plotted  in Fig.~\ref{TtildesameV}.
For $D_2^*(3000)$ belonging to ${\tilde T}$  we predict
$R_\rho^{{\tilde D}_2^*}=0.22 \pm 0.02$.
%
If the $D_2^*(3000)$ full width   is saturated by the modes
$D^{(*)0}\pi^0,\,D^{(*)+}\pi^-$, $D^{(*)0} \eta$, $D^{(*)}_s K^-$, $D^{(*)+} \rho^-$, $D^{(*)0}\rho^0$, $D^{(*)0} \omega$, $D^{(*)}_s K^{*-}$,
 the two couplings $h^T$ in (\ref{LHTV2}) and $\tilde h^\prime$ in (\ref{lagT}) can be constrained to the region in Fig.~\ref{regplotTtilde}, with the bounds: 
$|\tilde h^\prime|<0.135$ and $|h^T|<0.29$.

 \subsection{$F$ doublet} 
\begin{figure}[b]
 \begin{center}
  \includegraphics[width=0.4\textwidth] {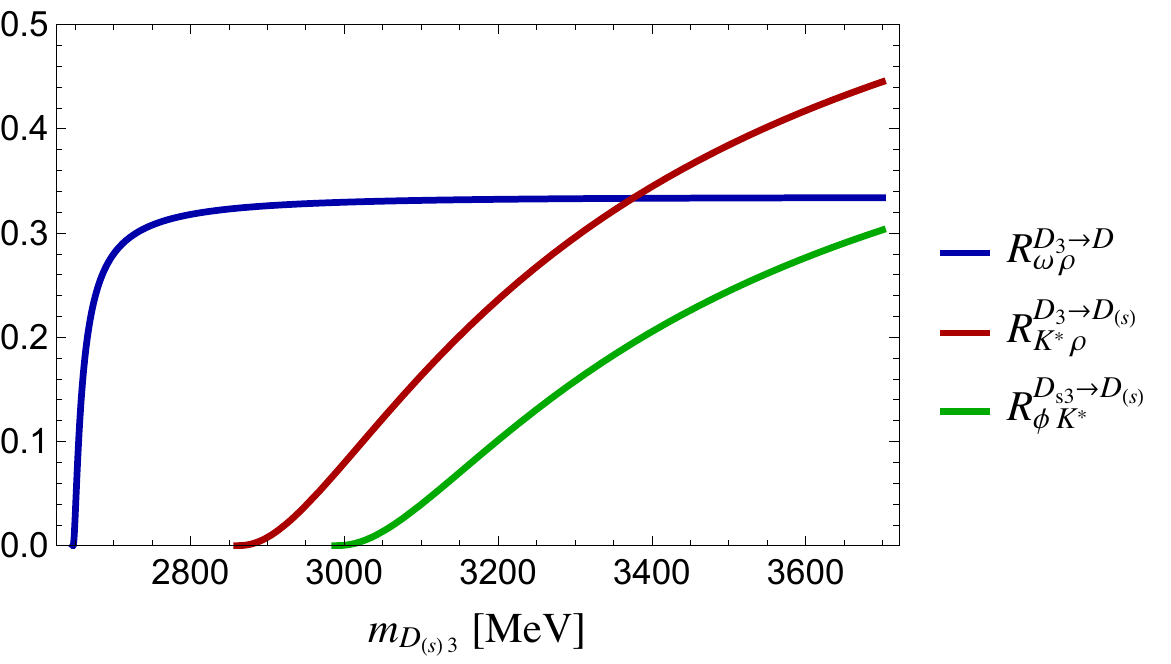}
 \vspace*{0mm}
 \caption{ Ratio (\ref{RP3FV1V2}) for different  final states when the decaying particle is $D_{(s)3}$ in the $F$ doublet. }
  \label{ratioP3FV1V2}
 \end{center}
\end{figure}

The single ratio  independent of the coupling constant in the effective Lagrangian Eq.~(\ref{RP3FV1V2})  for  $D_3$ and for its stange partner $D_{s3}$ is shown in Fig.~\ref{ratioP3FV1V2}, obtaining
 $R_{\omega \rho}^{D_3 \to D}<R_{K^* \rho}^{D_3 \to D_{(s)}}$ for a decaying particle mass  below 3.38 GeV.

 \subsection{More about $D_2^*(3000)$}
The LHCb assignment for this particle is  $J^P=2^+$ \cite{Aaij:2016fma}, and mass and width  are compatible with $D_J^{*0}(3000)$  \cite{Aaij:2013sza}   (Tables \ref{tabLHCb13} and  \ref{tabLHCb16}). It could be
identified with the lowest lying $J^P=2^+$    $n=2$ state $ {\tilde D}_2^*$ in $\tilde T$ doublet,  or   with $D_2^{\prime *}$ belonging to the $n=1$ $F$ doublet.
Predictions for the masses of the two states have been worked out in quark models. For example, using the chiral quark model developed in \cite{Manohar:1983md,Goity:1998jr}, the values $m_{{\tilde D}_2^*}=3035$ GeV and $m_{D_2^{\prime *}}=3101$ GeV have been predicted  \cite{DiPierro:2001dwf}.
As for the identification of $D_2^*(3000)$, no consensus is reached adopting  variants of the quark model.  Using a model with instantaneous Bethe-Salpeter potential, identification with  $ {\tilde D}_2^*$ is supported
 \cite{Li:2017zng},   while $D_2^*(3000)$ is preferably  interpreted as  $D_2^{\prime *}$ on the basis of the  $^3P_0$ model for strong decays \cite{Yu:2016mez}.

The two possibilities lead to different predictions for the  $P^{(*)}M$ and to $P^{(*)}V$ widths.
Possible   transitions to  $ P^{(*)}M$ are
$D_2^{*0}(3000) \to D^{(*)0} \pi^0$, $D^{(*)+} \pi^-$, $D^{(*)0} \eta,\, D_s^{(*)} K^- $, leading to the strong coupling  independent ratios:
\bea
R_\pi^0 &=& \frac{\Gamma(D_2^{*0}(3000) \to D^{*0} \pi^0)+\Gamma(D_2^{*0}(3000) \to D^{*+} \pi^-)}{\Gamma(D_2^{*0}(3000) \to D^{0} \pi^0)+\Gamma(D_2^{*0}(3000) \to D^{+} \pi^-)} \,\,\, ,
\label{RpiD2} \\
R_\eta^0 &=& \frac{\Gamma(D_2^{*0}(3000) \to D^{0} \eta)}{\Gamma(D_2^{*0}(3000) \to D^{0} \pi^0)+\Gamma(D_2^{*0}(3000) \to D^{+} \pi^-)} \,\,\, ,
\label{RetaD2} \\
R_\eta^{*0} &=& \frac{\Gamma(D_2^{*0}(3000) \to D^{*0} \eta)}{\Gamma(D_2^{*0}(3000) \to D^{0} \pi^0)+\Gamma(D_2^{*0}(3000) \to D^{+} \pi^-)} \,\,\, ,
\label{RstaretaD2} \\
R_K^0 &=& \frac{\Gamma(D_2^{*0}(3000) \to D_s K^-)}{\Gamma(D_2^{*0}(3000) \to D^{0} \pi^0)+\Gamma(D_2^{*0}(3000) \to D^{+} \pi^-)} \,\,\, ,
\label{RKD2}
\eea
\bea
R_K^{*0} &=& \frac{\Gamma(D_2^{*0}(3000) \to D_s^* K^-)}{\Gamma(D_2^{*0}(3000) \to D^{0} \pi^0)+\Gamma(D_2^{*0}(3000) \to D^{+} \pi^-)} \,\,\, .
\label{RstarKD2} 
\eea
Decay modes of the the strange partner of $D_2^*(3000)$ are
$D_{s2}^* \to D_s^{(*)}\eta$, $D^{(*)+}K_S$, $D^{(*)0}K^+$,
with  ratios :
\bea
R_{s,K}^* &=& \frac{\Gamma(D_{s2}^{*+} \to D^{*+} K_S)+\Gamma(D_{s2}^{*+} \to D^{*0} K^+)}{\Gamma(D_{s2}^{*+} \to D^{+} K_S)+\Gamma(D_{s2}^{*+} \to D^{0} K^+)} \,\,\, ,
\label{RstarKDs2} \\
R_{s,\eta} &=&\frac{\Gamma(D_{s2}^{*+} \to D_s \eta)}{\Gamma(D_{s2}^{*+} \to D^{+} K_S)+\Gamma(D_{s2}^{*+} \to D^{0} K^+)} \,\,\, ,
\label{RetaDs2} \\
R_{s,\eta}^* &=&\frac{\Gamma(D_{s2}^{*+} \to D_s^* \eta)}{\Gamma(D_{s2}^{*+} \to D^{+} K_S)+\Gamma(D_{s2}^{*+} \to D^{0} K^+)} \,\,\, .
\label{RstaretaDs2} 
\eea
The  results   are different if one identifies $D_2^*(3000)$ with ${\tilde D}_2^*$ in the $\tilde T$  or with  $D^{\prime *}_2$ in the $F$ doublet.
We fix the  $D_2^{*0}$ mass to the value  in Table \ref{tabLHCb16} with  the errors combined in quadrature: $m_{D^{*0}_{2}}=3214 \pm 57$ MeV, and
for the strange partner  we assume $m_{D^*_{s2}}=m_{D^{*0}_{2}}+100$ MeV  enlarging the uncertainty: $m_{D^*_{s2}}=3314 \pm 70$ MeV.
The ratios (\ref{RpiD2})-(\ref{RstarKD2}) and (\ref{RstarKDs2})-(\ref{RstaretaDs2}) for the two  classifications  are  in Tables \ref{ratiosD20} and  \ref{ratiosD2s}.
 \begin{table*}[h!]
\centering  \caption{Ratios in Eqs.~(\ref{RpiD2})-(\ref{RstarKD2}) for two different classifications of $D_2^*(3000)$.}
    \label{ratiosD20}
   \begin{tabular}
{ccccccc} \hline \hline
doublet &state  &  \,\,$R_{\pi}^0$\,\,   &\,\, $R_\eta^0$\,\, &\,\, $R_\eta^{*0}$\,\,& \,\,$R_K^0 $\,\,& \,\,$R_K^{*0} $\,\,
\\
      \hline
  $\tilde T$ (n=2) &   \,\, ${\tilde D}^{*0}_2$\,\, &  \,\,$1.06 \pm 0.03$\,\, & \,\, $0.29 \pm 0.01$\,\,&  \,\, $0.27 \pm 0.02 $   \,\,& \,\, $0.35 \pm 0.020 $  \,\,& \,\, $0.30 \pm 0.03 $  \,\,
  \\
\hline
 $F$ (n=1) &    \,\, $D^{\prime*0}_2$\,\, &  \,\,$0.40 \pm 0.015$\,\, & \,\, $0.31 \pm 0.01$\,\,&  \,\, $0.11 \pm 0.01 $   \,\,& \,\, $ 0.33 \pm 0.02$  \,\,& \,\, $0.11 \pm 0.01 $  \,\,
 \\
\hline \hline
    \end{tabular}
\end{table*}
 \begin{table*}[h!]
\centering  \caption{ Ratios defined in Eqs.~(\ref{RstarKDs2})-(\ref{RstaretaDs2}).}
   \label{ratiosD2s}
   \begin{tabular}
{ccccc} \hline \hline
doublet &state   & \,\,$R_{s,K}^* $\,\,& \,\, $R_{s,\eta}$\,\, &\,\, $R_{s,\eta}^*$\,\,
\\ \hline
  $\tilde T$ (n=2)&  \,\, ${\tilde D}^*_{s2}$\,\, & \,\, $ 1.02 \pm 0.04$  \,\,  \,\, & \,\, $0.31 \pm 0.01$\,\,&  \,\, $0.29 \pm 0.03 $   \,\,
  \\ \hline
 $F$ (n=1)  & \,\, $D^{\prime*}_{s2}$\,\,&  \,\, $0.40 \pm 0.02 $   \,\, & \,\, $0.28 \pm 0.01 $  \,\, & \,\, $0.10 \pm 0.01$\,\,
\\
\hline \hline
    \end{tabular}
\end{table*}
 $R_{\pi}^0$ and $R_{s,K}^* $ have the highest sensitivity to  the two classifications.

The two assignments lead to predictions for the spin partner of $D_2^*(3000)$. For  $D_2^*(3000)$  identified with ${\tilde D}_2^*$, the spin partner is the $J^P=1^+$ state ${\tilde D}_1$,  while the spin partner of $D^{\prime *}_2$ is $D_3$ with $J^P=3^+$. In the two cases we construct the   ratios of decay widths
\bea
R_{SP}^{\tilde T}&=&\frac{\Gamma({\tilde D}_1^0 \to D^{*+} \pi^-)+\Gamma({\tilde D}_1^0 \to D^{*0} \pi^0)}{\Gamma({\tilde D}_2^{*0} \to D^{*+} \pi^-)+\Gamma({\tilde D}_2^{*0} \to D^{*0} \pi^0) } \,\,\, , \label{R12}\\
R_{SP}^{F}&=&\frac{\Gamma( D_3^0 \to D^{*+} \pi^-)+\Gamma(D_3^0 \to D^{*0} \pi^0)}{\Gamma(D^{\prime *0}_2 \to D^{*+} \pi^-)+\Gamma(D^{\prime *0}_2 \to D^{*0} \pi^0)} \,\,\, .  \label{R32}
\eea
Varying conservatively the mass of ${\tilde D}_1$ in the  range $[m_{D_2^*(3000)}-100 \, {\rm MeV},  \,m_{D_2^*(3000)}]$ and the mass of   $D_3$ in $[m_{D_2^*(3000)},  \,m_{D_2^*(3000)}+100 \, {\rm MeV}]$
we obtain:
\be
1.2 \leq R_{SP}^{\tilde T} \leq 1.7 \,\,\, ,  \,\,\,\, 1.7 \leq R_{SP}^{\tilde T}  \leq 2.6 \,\,\,.
\ee

\section{Numerical analysis: Beauty }\label{sec:numericsB}
\begin{figure}[b]
 \begin{center}
 \includegraphics[width=0.45\textwidth] {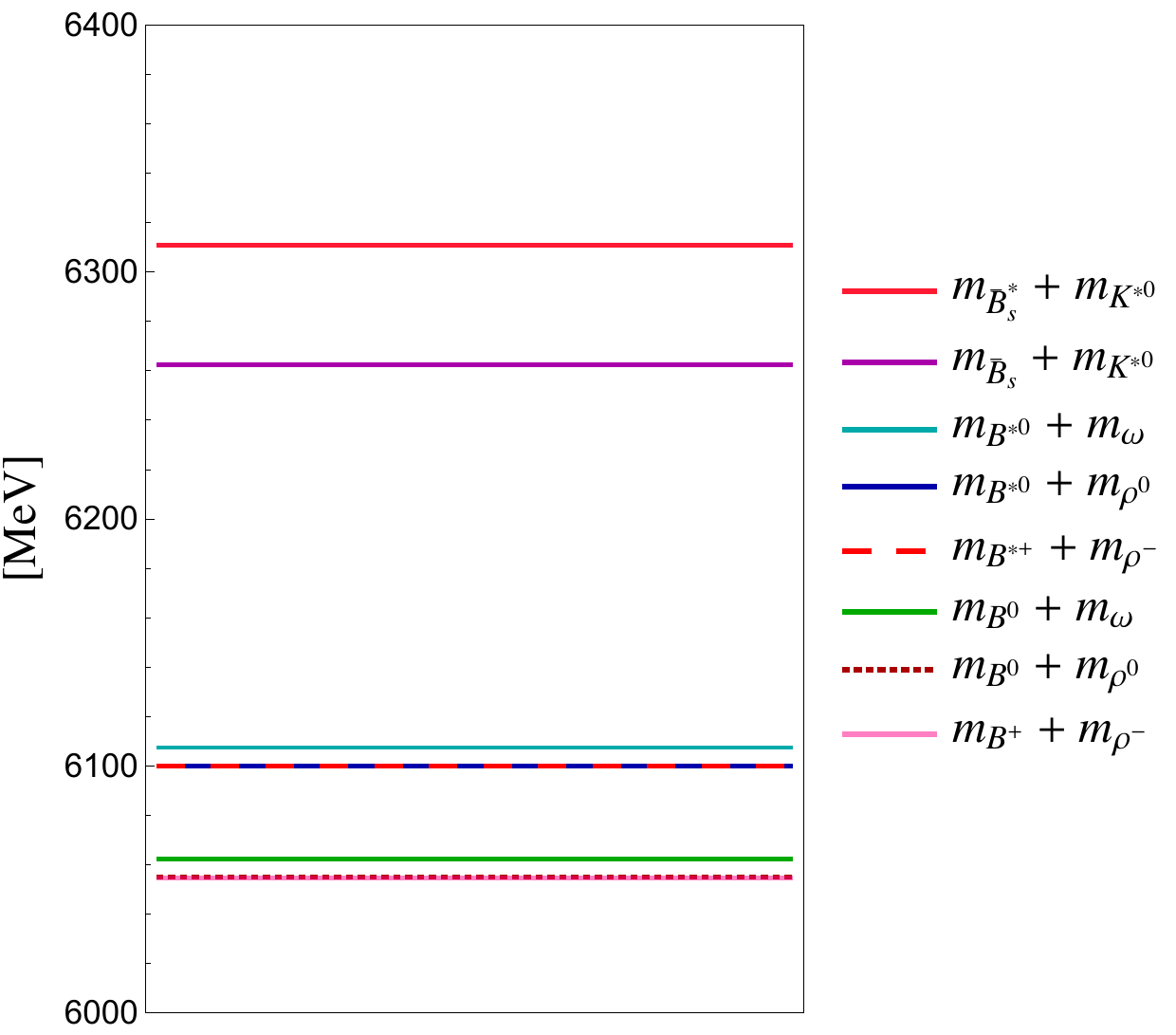}\hspace*{0.3cm}
  \includegraphics[width=0.45\textwidth] {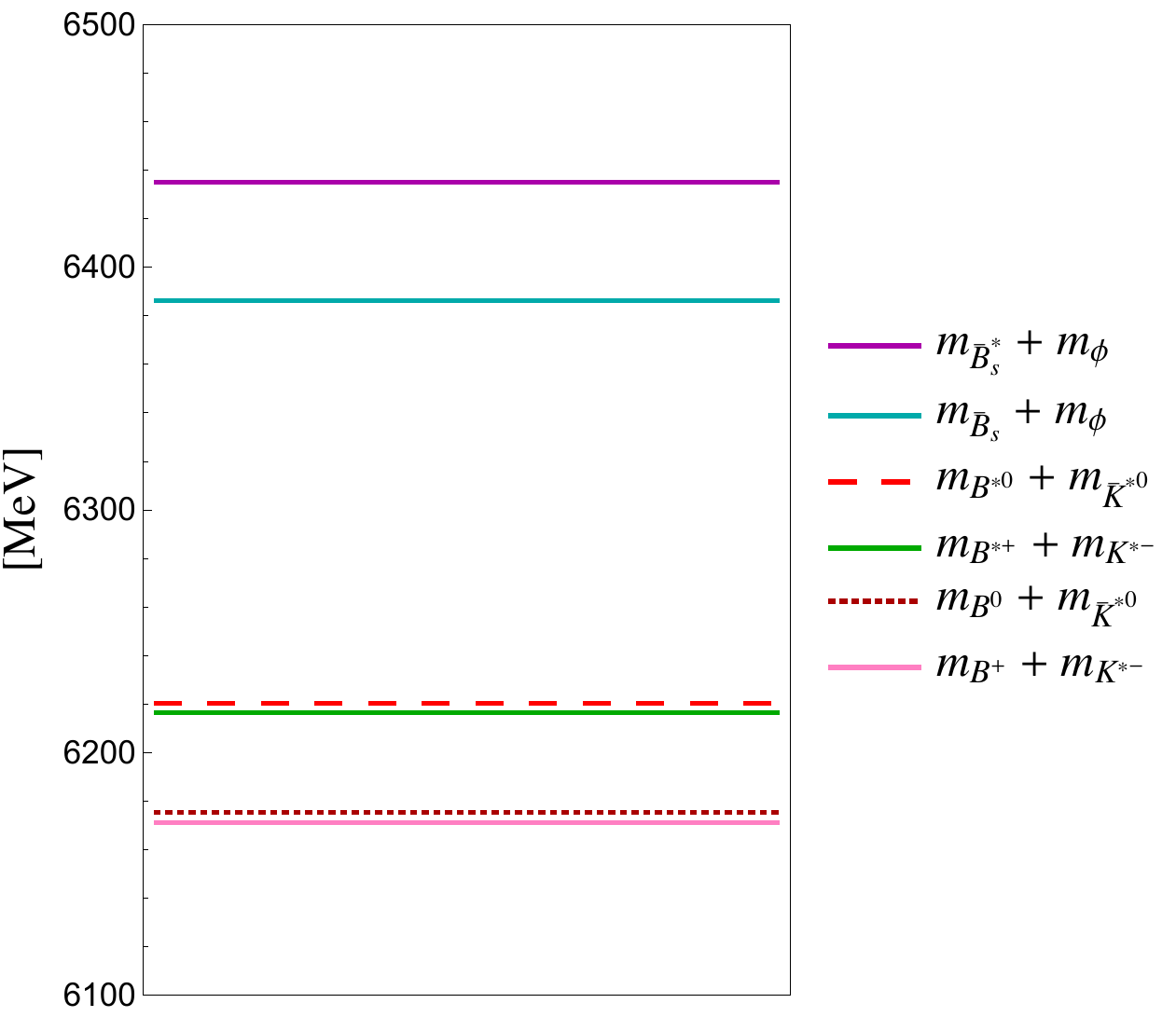}\hspace*{0.3cm}
%
 \caption{ $PV$ and $P^*V$ thresholds for neutral non-strange (left)  and strange  beauty mesons (right panel). }
  \label{thB}
 \end{center}
\end{figure}

The flavour symmetry allows to extend the analysis to the beauty sector. The ${\cal H}_i \to P^{(*)}V$ thresholds,   for ${\cal H}_i$  a  neutral beauty  or a beauty-strange meson, are displayed in 
 Fig.~\ref{thB}. 
No one of the observed excited beauty mesons are above the $P^{(*)}V$ thresholds, therefore our predictions hold for higher excitations.
We define ratios of decay widths for charged and  for  neutral non-strange decaying beauty meson:
\bea
R^{{\cal H}_i^+ \to B}_{\omega \rho}&=&\frac{\Gamma ({\cal H}_i^+ \to B^+ \omega)}{\Gamma ({\cal H}_i^+ \to B^+ \rho^0)+\Gamma ({\cal H}_i^+ \to B^0 \rho^+)}   \,\,\, ,
\label{RBOmegaRhopiu}\\
R^{{\cal H}_i^+ \to B_{(s)}}_{K^* \rho}&=&\frac{\Gamma ({\cal H}_i^+ \to {\bar B}_s K^{*+})}{\Gamma ({\cal H}_i^+ \to B^+ \rho^0)+\Gamma ({\cal H}_i^+ \to B^0 \rho^+)}   \,\,\, ,
\label{RBKstarRhopiu}\\
R^{{\cal H}_i^+ \to B^*}_{\omega \rho}&=&\frac{\Gamma ({\cal H}_i^+ \to B^{*+} \omega)}{\Gamma ({\cal H}_i^+ \to B^{*+} \rho^0)+\Gamma ({\cal H}_i^+ \to B^{*0} \rho^+)}   \,\,\, ,
\label{RBstarOmegaRhopiu}\\
R^{{\cal H}_i^+ \to B_{(s)}^*}_{K^* \rho}&=&\frac{\Gamma ({\cal H}_i^+ \to {\bar B}_s^*K^{*+})}{\Gamma ({\cal H}_i^+ \to B^{*+} \rho^0)+\Gamma ({\cal H}_i^+ \to B^{*0} \rho^+)}  \,\,\, ,
\label{RBstarKstarRhopiu}
\eea
and
\bea
R^{{\cal H}_i^0 \to B}_{\omega \rho}&=&\frac{\Gamma ({\cal H}_i^0 \to B^0 \omega)}{\Gamma({\cal H}_i^0 \to B^0 \rho^0)+\Gamma ({\cal H}_i^0 \to B^+ \rho^-)}  \,\,\, ,
\label{RBOmegaRho0}\\
R^{{\cal H}_i^0 \to B_{(s)}}_{K^* \rho}&=&\frac{\Gamma ({\cal H}_i^0 \to {\bar B}_s K^{*0})}{\Gamma({\cal H}_i^0 \to B^0 \rho^0)+\Gamma ({\cal H}_i^0 \to B^+ \rho^-)}  \,\,\, ,
\label{RBKstarRho0}\\
R^{{\cal H}_i^0 \to B^*}_{\omega \rho}&=&\frac{\Gamma ({\cal H}_i^0 \to B^{*0} \omega)}{\Gamma ({\cal H}_i^0 \to B^{*0} \rho^0)+\Gamma ({\cal H}_i^0 \to B^{*+} \rho^-)}  \,\,\, ,
\label{RBstarOmegaRho0}\\
R^{{\cal H}_i^0 \to B_{(s)}^*}_{K^* \rho}&=&\frac{\Gamma ({\cal H}_i^0 \to {\bar B}_s^* K^{*0})}{\Gamma ({\cal H}_i^0 \to B^{*0} \rho^0)+\Gamma ({\cal H}_i^0 \to B^{*+} \rho^-)}  \,\,\, .
\label{RBstarKstarRho0}
\eea
Ratios of decay widths can also be constructed for beauty mesons with strangeness:
\bea
R^{{\cal H}_{is}\to B_{(s)}}_{ \phi K^*}&=& \frac{\Gamma ({\cal H}_{is} \to {\bar B}_s \phi)}{\Gamma ({\cal H}_{is} \to B^0 {\bar K}^{*0})+\Gamma ({\cal H}_{is} \to B^+ K^{*-})}  \,\,\, , \label{RsBphiKstar}\\
R^{{\cal H}_{is}\to B_{(s)}^*}_{ \phi K^*}&=& \frac{\Gamma ({\cal H}_{is} \to {\bar B}_s^* \phi)}{\Gamma ({\cal H}_{is} \to B^{*+} {\bar K}^{*0})+\Gamma ({\cal H}_{is} \to B^{*+} K^{*-})}  \,\,\, . \label{RsBstarphiKstar}
\eea
Ratios of decay widths  with the same final $V$ meson are also independent of strong couplings:
\bea
R^{{\cal H}_i^+}_{ \rho}&=&\frac{\Gamma ({\cal H}_i^+ \to B^{*+} \rho^0)+\Gamma ({\cal H}_i^+ \to B^{*0} \rho^+)} {\Gamma({\cal H}_i^+ \to B^+ \rho^0)+\Gamma ({\cal H}_i^+ \to B^0 \rho^+)}  \,\,\, ,
\label{RBsameRhopiu}\\
R^{{\cal H}_i^0}_{ \rho}&=&\frac{\Gamma ({\cal H}_i^0 \to B^{*0} \rho^0)+\Gamma ({\cal H}_i^0 \to B^{*+} \rho^-)} {\Gamma({\cal H}_i^0 \to B^0 \rho^0)+\Gamma ({\cal H}_i^0 \to B^+ \rho^-)}  \,\,\, ,
\label{RBsameRho0}\\
R^{{\cal H}_{is}}_{K^*}&=&\frac{\Gamma ({\cal H}_{is} \to B^{*+} K^{*-})+\Gamma ({\cal H}_{is} \to B^{*0} {\bar K}^{*0})}{\Gamma ({\cal H}_{is} \to B^+ K^{*-})+\Gamma ({\cal H}_{is} \to B^0 {\bar K}^{*0})}  \,\,\, . \label{RBsameKstar}
\eea

\vskip 1.cm

\noindent{\bf $H$ doublet}

\begin{figure}[t]
 \begin{center}
 \includegraphics[width=0.45\textwidth] {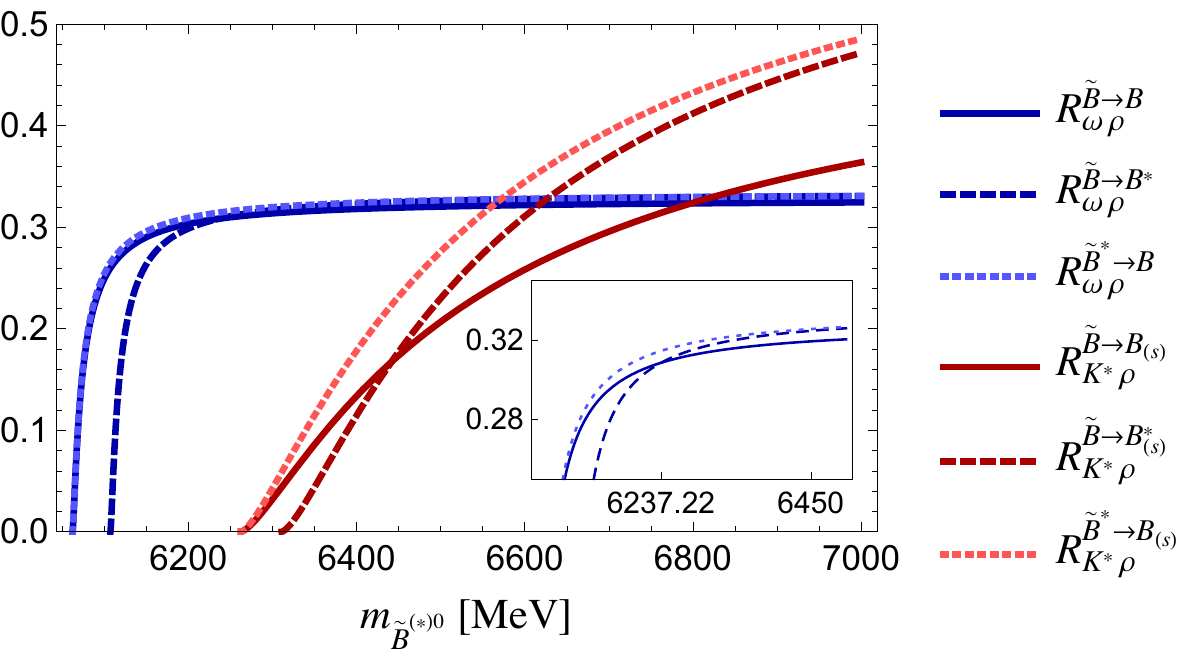}\hskip 0.3cm
\includegraphics[width=0.45\textwidth] {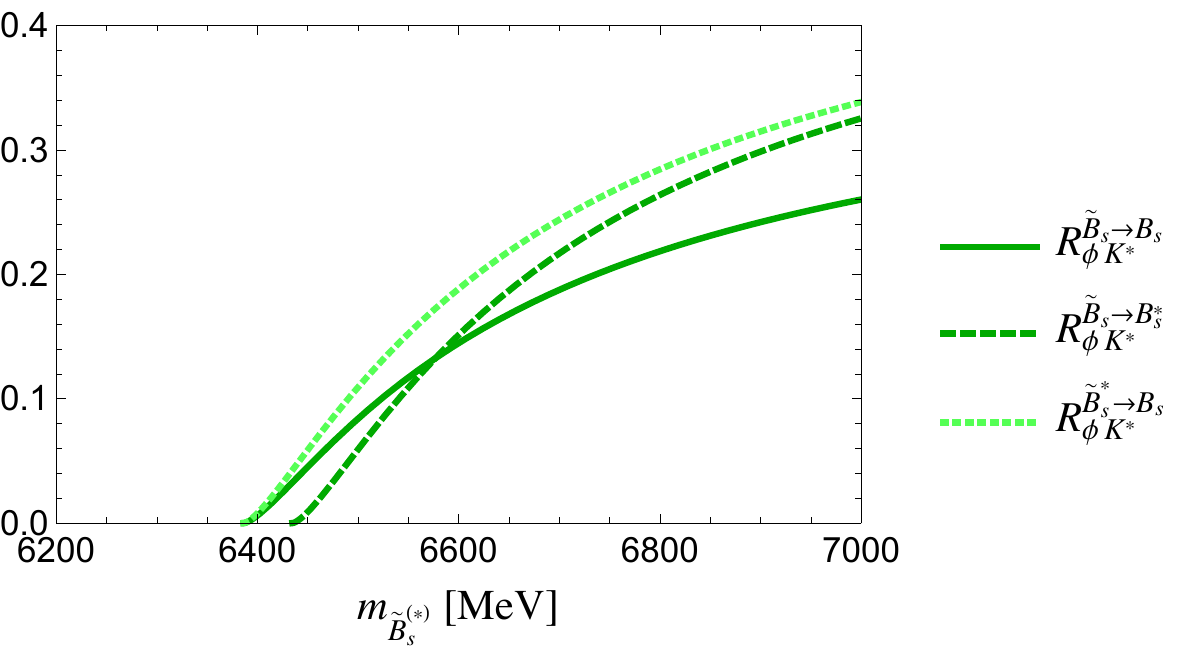}
\\
\includegraphics[width=0.45\textwidth] {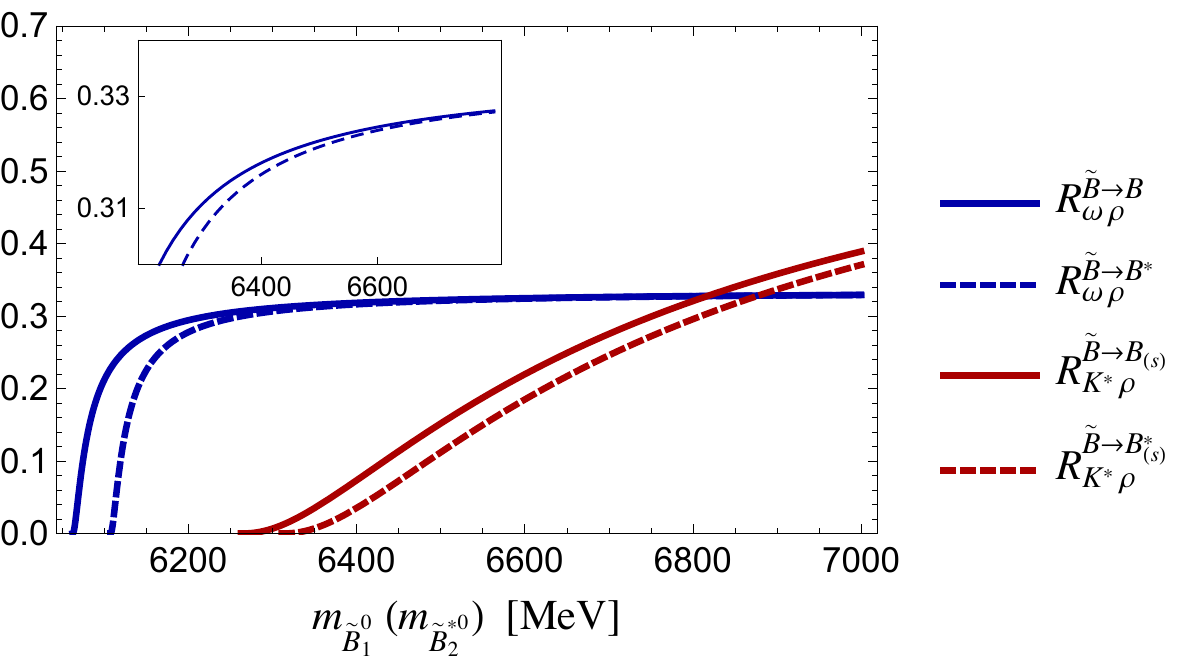}
\hskip 0.3cm
 \includegraphics[width=0.45\textwidth] {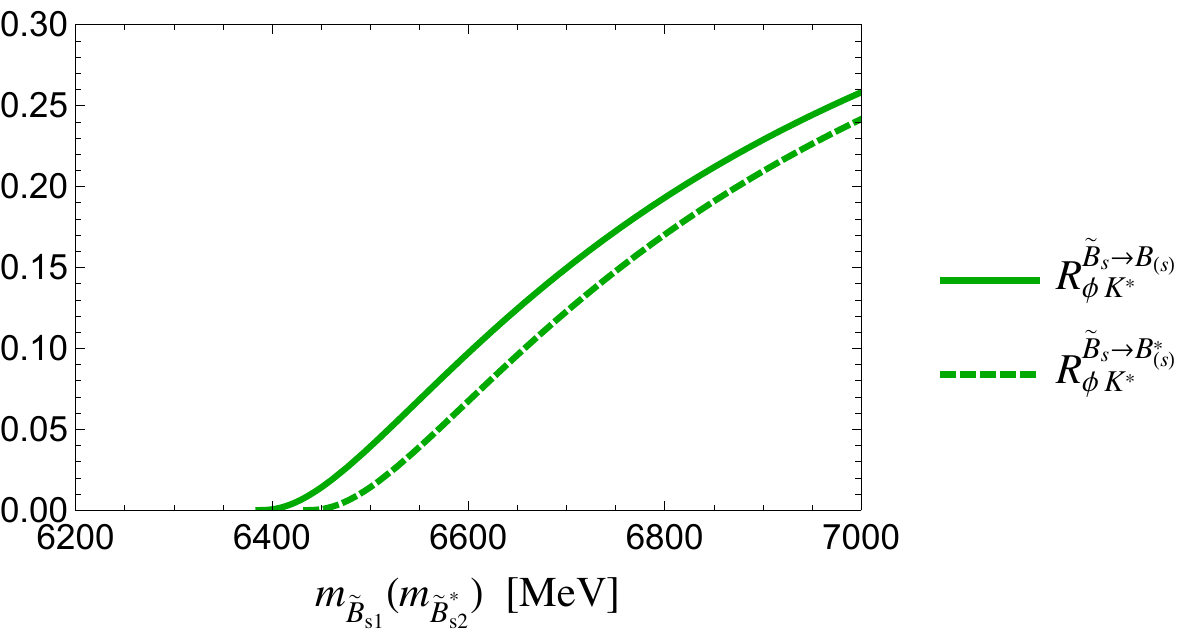}
 \\
\includegraphics[width=0.45\textwidth] {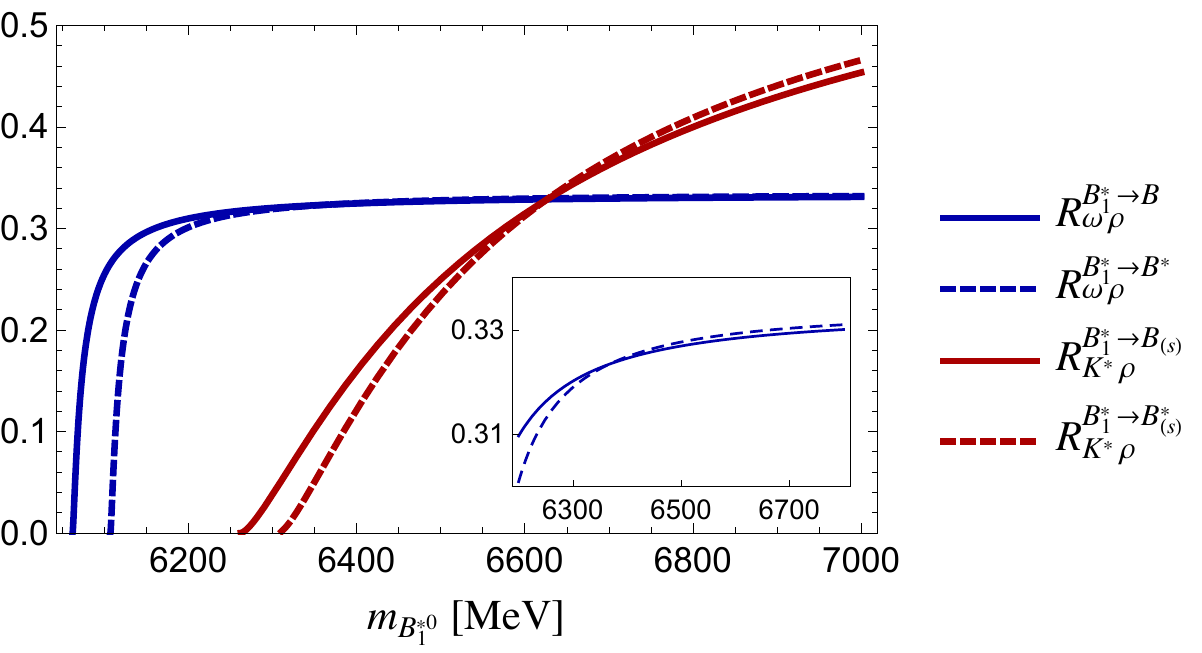}
\hskip 0.3cm
 \includegraphics[width=0.45\textwidth] {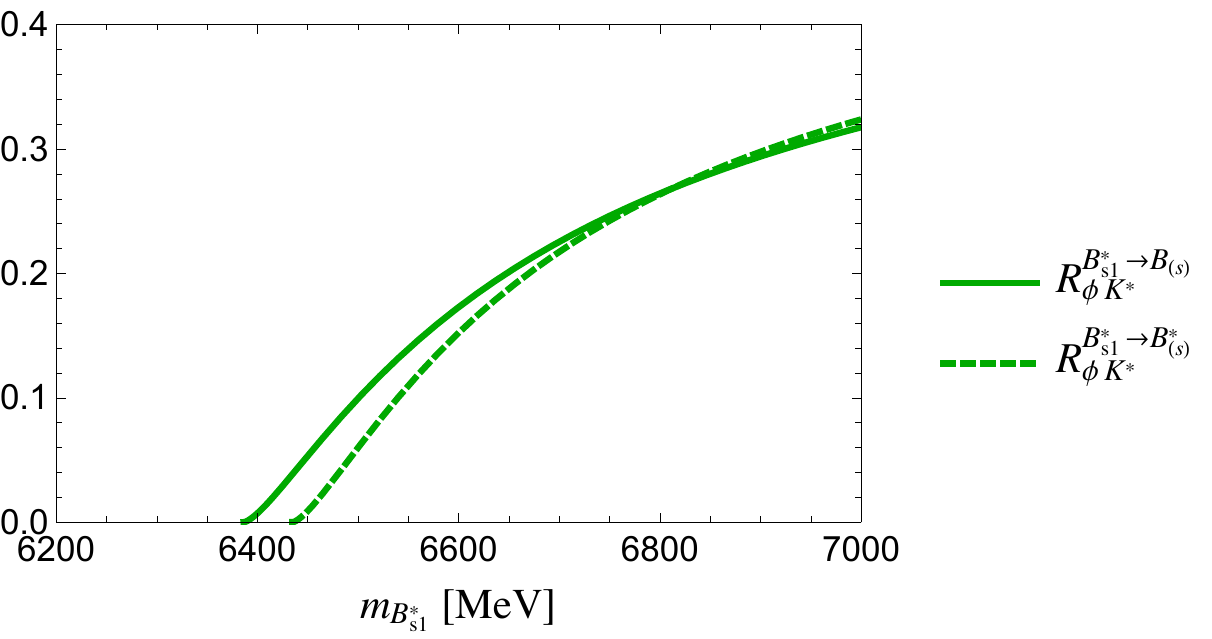}
\\
\includegraphics[width=0.45\textwidth] {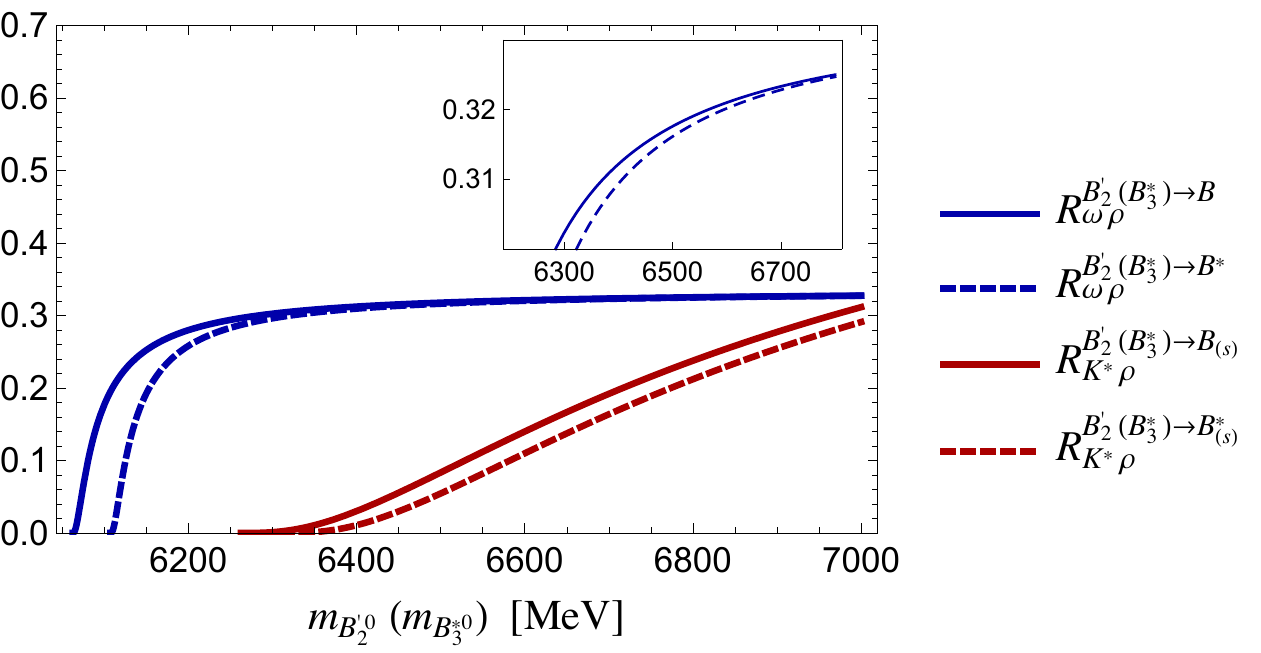}
\hskip 0.3cm
 \includegraphics[width=0.45\textwidth] {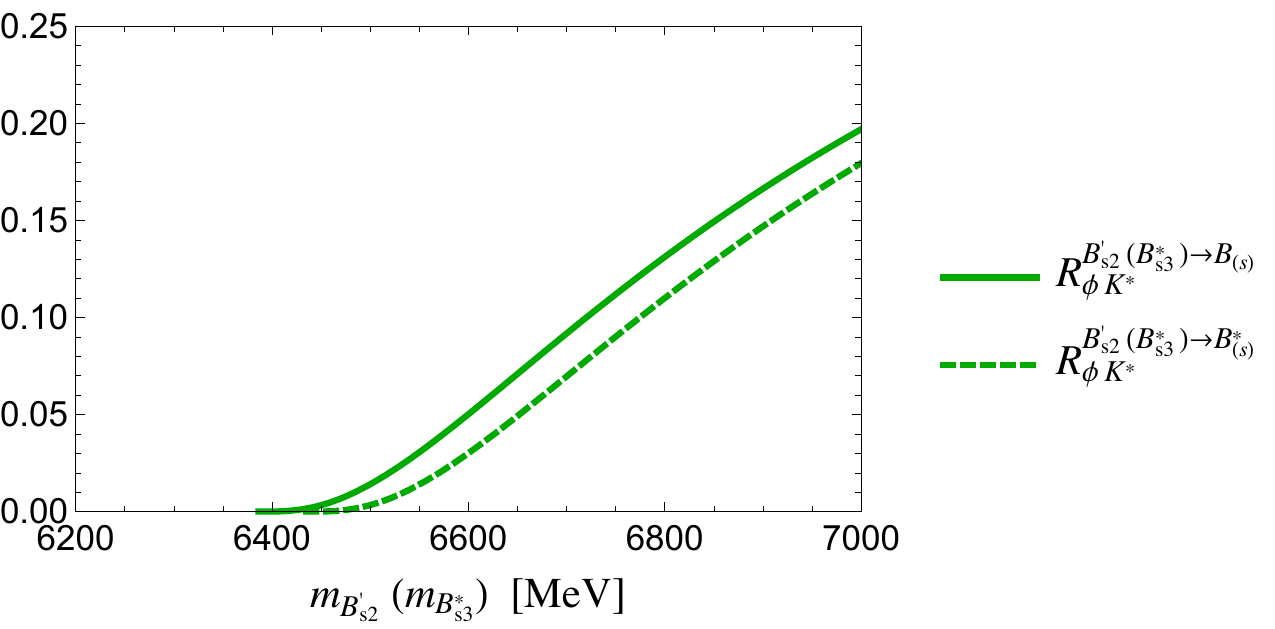}
 \caption{Ratios (\ref{RBOmegaRho0})-(\ref{RBstarKstarRho0}) (left) and  (\ref{RsBphiKstar}), (\ref{RsBstarphiKstar}) (right panels) for  decaying particles belonging to the  $ b \bar q$ ($b \bar s$) doublet  ${\tilde H}$ (top row),  ${\tilde T}$ (second row),  ${\tilde X}$ (third row) and  ${\tilde X^\prime}$ (bottom row). }
 \label{ratioHtildeB}
 \end{center}
\end{figure}

\vskip 0.2cm \noindent
The ratios  (\ref{RBOmegaRho0})-(\ref{RBstarKstarRho0})    and (\ref{RsBphiKstar}), (\ref{RsBstarphiKstar}) evaluated when the decaying particle in  ${\tilde H}$, are   in Fig.~\ref{ratioHtildeB}. 
For $m_{{\tilde B}^0}<6237.22$ MeV one has $R^{{\tilde B} \to B}_{\omega \rho}>R^{{\tilde B} \to B^*}_{\omega \rho}$.  
In  ${\tilde B}_s$ decays,   for $m_{{\tilde B}_s}>6576.8$ MeV one predicts $R^{{\tilde B}_s \to B_s}_{\phi K^*}<R^{{\tilde B}_s \to B^*_s}_{\phi K^*}$. 
Other ratios  show similar features, namely
 $R^{{\tilde B} \to B_{(s)}}_{K^* \rho}<R^{{\tilde B} \to B_{(s)}^*}_{K^* \rho}$ for $m_{{\tilde B}^0}>6441.3$ MeV.

\vskip 0.4cm
\noindent {\bf ${\tilde T}$ doublet}

\begin{figure}[t]
 \begin{center}
 \includegraphics[width=0.38\textwidth] {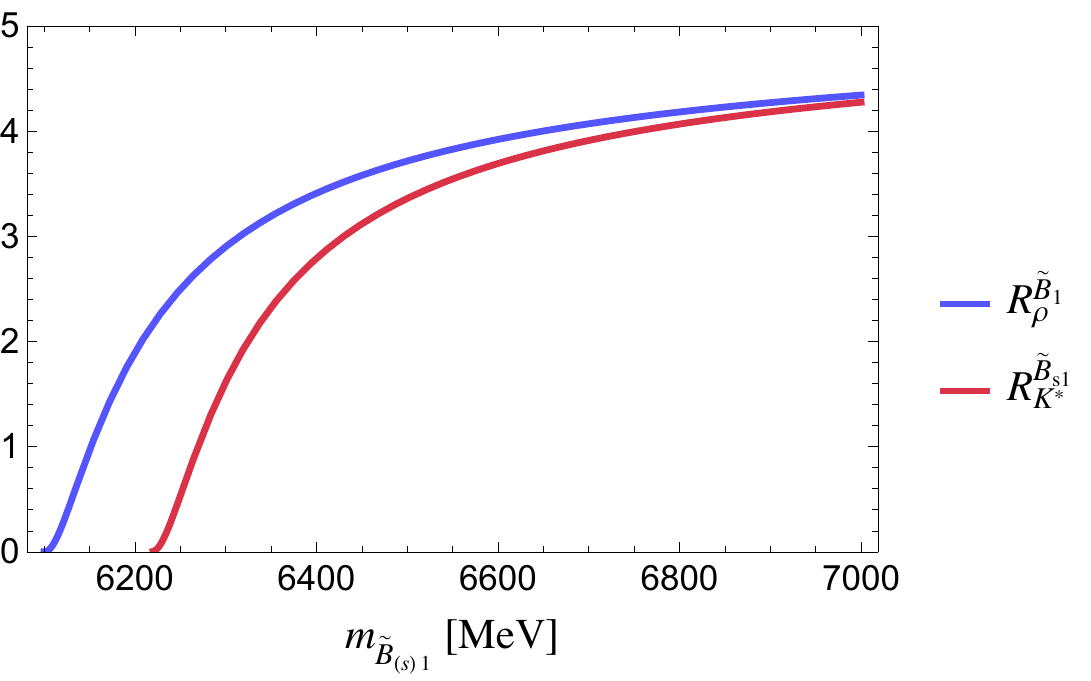}
\includegraphics[width=0.38\textwidth] {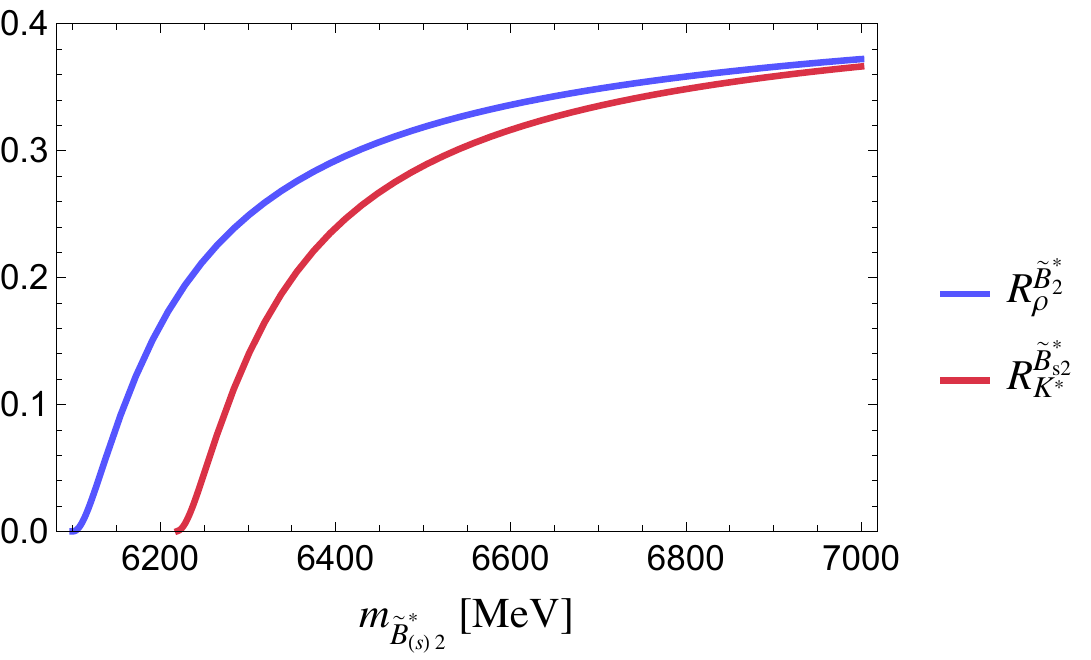}
\\
 \includegraphics[width=0.38\textwidth] {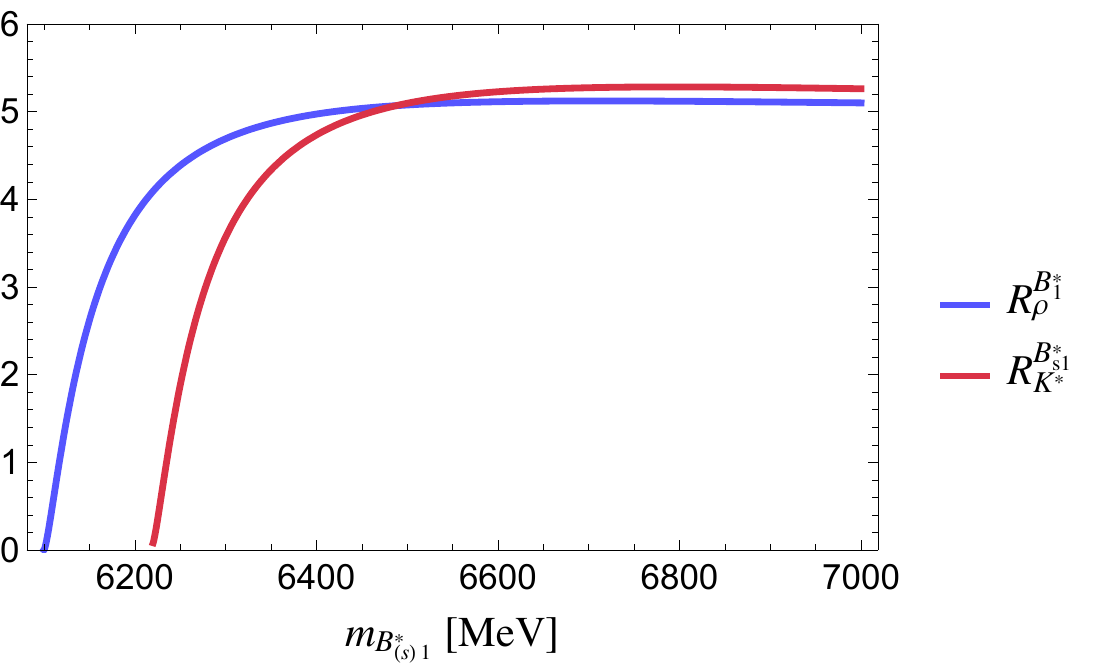}
\includegraphics[width=0.38\textwidth] {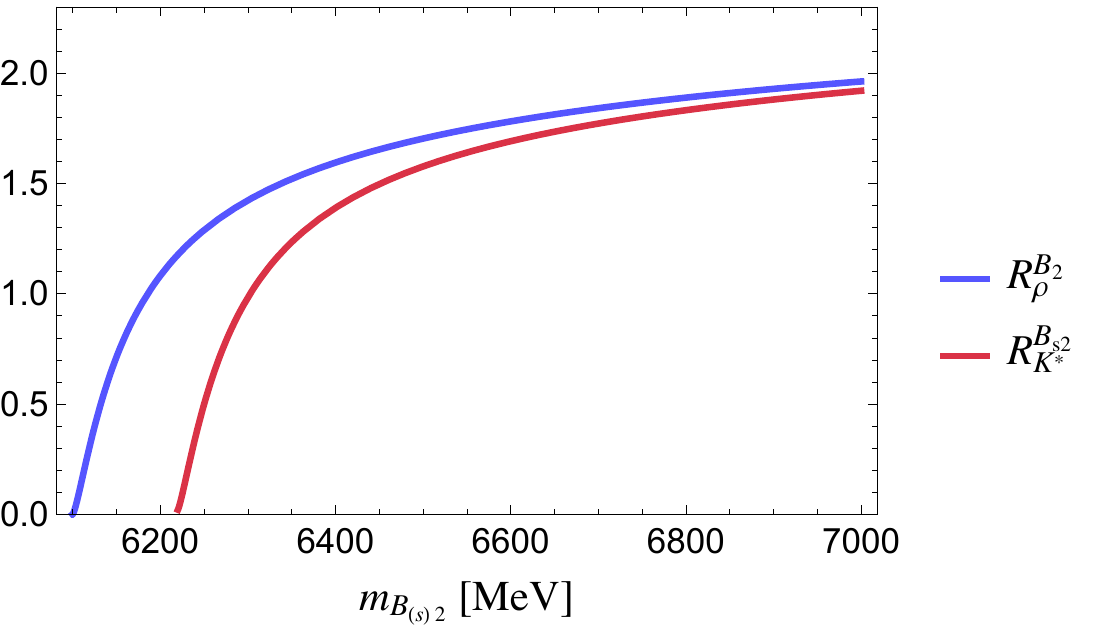}
\\
 \includegraphics[width=0.38\textwidth] {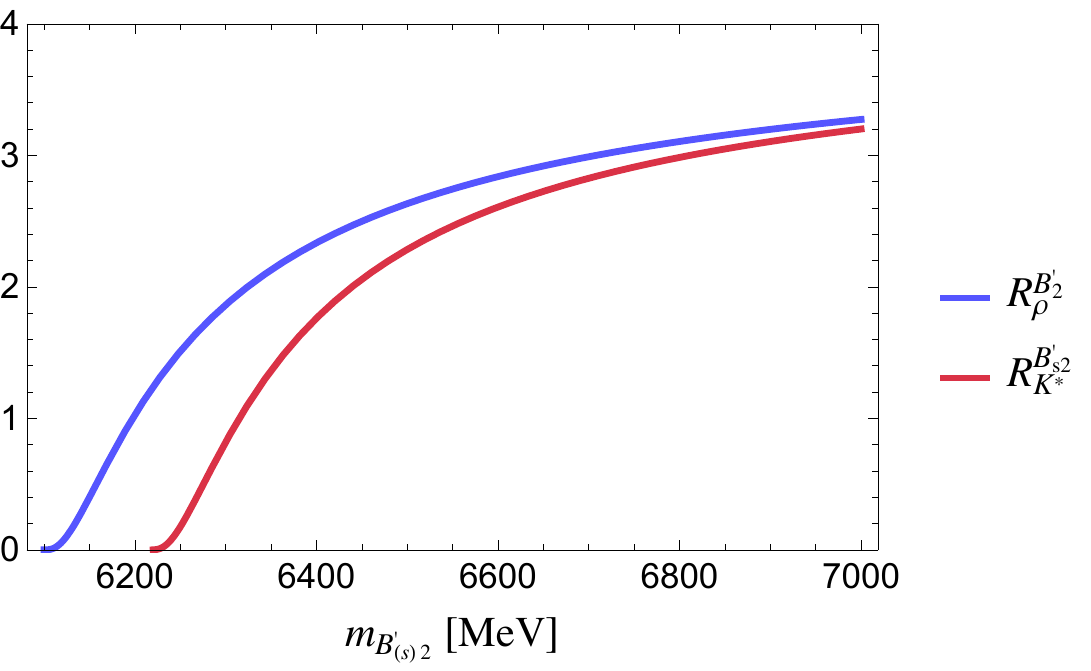}
\includegraphics[width=0.38\textwidth] {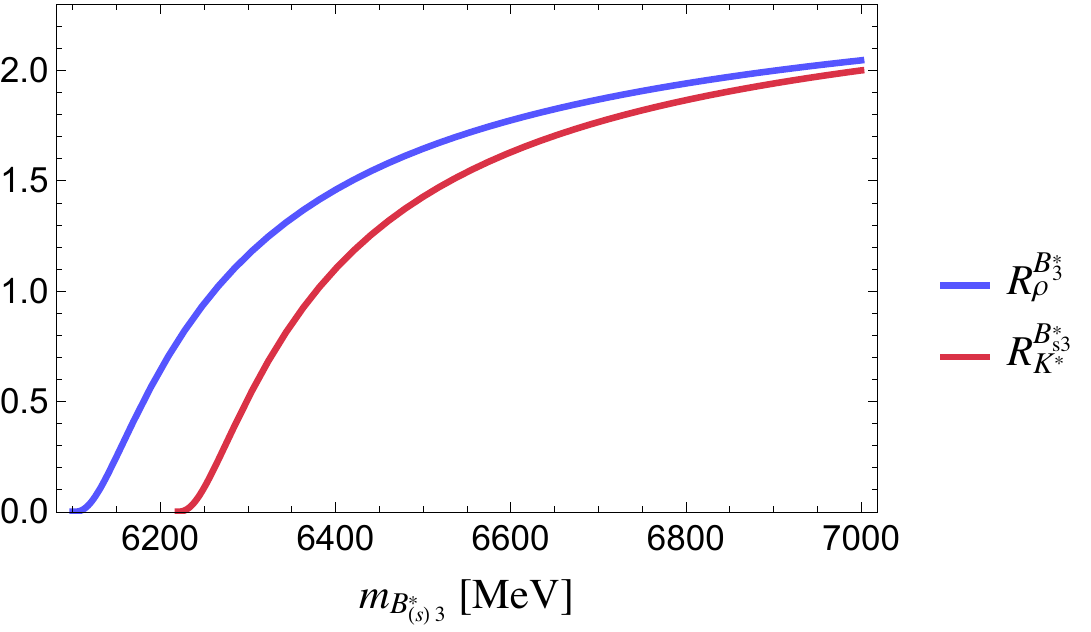}
 \caption{ Ratios (\ref{RBsameRhopiu}), (\ref{RBsameKstar}) for decaying particles in the $ b \bar q$ ($b \bar s$) ${\tilde T}$ doublet (top row), ${\tilde X}$  (middle row) and  ${\tilde X^\prime}$  (bottom row).  }
  \label{TtildeBsameV}
 \end{center}
\end{figure}

\vskip 0.2cm \noindent
Presenting the results in Fig.~\ref{ratioHtildeB}  we do not distinguish the decaying ${\tilde B}_1$ or ${\tilde B}_2^*$, which have the same expressions for the ratios.
 The observables in (\ref{RBsameRhopiu}), (\ref{RBsameKstar})  are displayed in Fig.~\ref{TtildeBsameV}.

\vskip 0.4cm
\noindent{\bf $X$ doublet}

\vskip 0.2cm \noindent
Ratios of decay rates for beauty mesons in the $X$ doublet are  in Fig.~\ref{ratioHtildeB}.
When the decaying particle is $B_1^*$,  the two ratios $R^{B_1^* \to B}_{\omega \rho}$ and $R^{B_1^* \to B^*}_{\omega \rho}$  become almost  coincident for  $m_{B_1^{*0}}\simeq 6367$ MeV.
Ratios involving the same final light vector meson are in Fig.~\ref{TtildeBsameV}.

\vskip 0.4cm
\noindent {\bf $X^\prime$ doublet}

\vskip 0.2cm \noindent
The considered ratios  have the same expressions for the two members in the $X^\prime$ doublet.  Those defined in Eqs.~(\ref{RBOmegaRho0})-(\ref{RBstarKstarRho0})   and (\ref{RsBphiKstar}), (\ref{RsBstarphiKstar})  are displayed in Fig.~\ref{ratioHtildeB}, those with the same final vector meson in  Fig.~\ref{TtildeBsameV}.

\vskip 0.4cm
\noindent {\bf $F$ doublet}

\begin{figure}[h!]
 \begin{center}
  \includegraphics[width=0.45\textwidth] {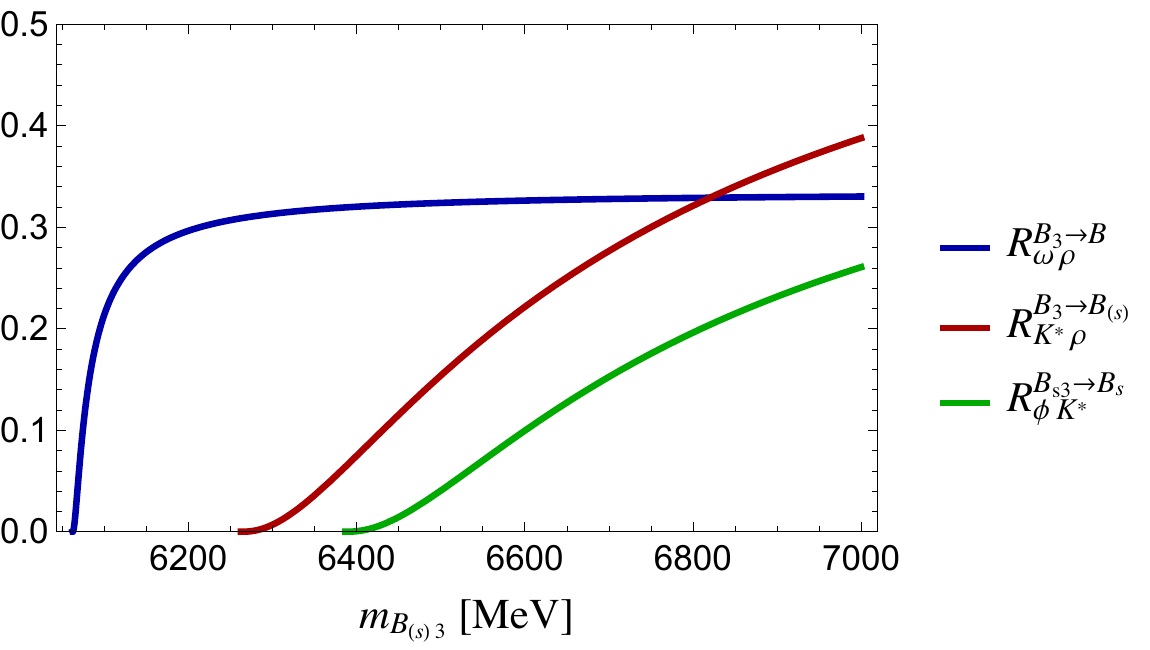}
 \caption{ Ratio (\ref{RP3FV1V2}) for several final states, for  a decaying  $B_{(s)3}$ in the $F$ doublet. }
  \label{ratioP3FV1VB2}
 \end{center}
\end{figure}

\vskip 0.2cm \noindent
For this doublet there is only one ratio  independent of the coupling constants, the one in Eq.~(\ref{RP3FV1V2}) for spin 3 meson.
The results displayed in Fig.~\ref{ratioP3FV1VB2} show  the hierachy $R^{B_3 \to B}_{\omega \rho}>R^{B_3 \to B_{(s)}}_{K^* \rho}$    for $m_{B_3}<3375$ MeV.
 
\section{Conclusions}
The construction of  a QCD-based framework to classify the excited  resonances with open charm and beauty and to describe their decays is needed in view of the ongoing and  forthcoming esperimental 
investigations. Since orbital and radial excitations  can be  above the  thresholds for  decays to light vector mesons, we have worked out
 effective Lagrangian terms  governing the strong transition  of a heavy meson  to a light vector meson and  a member of the lowest-lying heavy-light spin doublet, in the HQ limit.  We have defined observables 
independent of the  couplings in the Lagrangian, and made predictions  varying the mass of the decaying particle. The HQ limit is considered as the guideline for the description in the actual cases.
Our methods can be  exploited for a few  observed  states  with uncertain identification,   namely  $D^*_2(3000)$ for which we have compared  predictions corresponding to two different classifications.
Among the various tasks left to new analyses there are the computation of the various strong couplings and the classification of the subleading Lagrangian terms, which is particularly interesting  in  case of charm.

\vspace*{0.5cm}
\noindent {\bf Acknowledgments.} 
We thank A. Palano for  discussions. Work  carried out within the INFN project QFT-HEP.
\bibliography{ref}

\begin{thebibliography}{66}%
\makeatletter
\providecommand \@ifxundefined [1]{%
 \@ifx{#1\undefined}
}%
\providecommand \@ifnum [1]{%
 \ifnum #1\expandafter \@firstoftwo
 \else \expandafter \@secondoftwo
 \fi
}%
\providecommand \@ifx [1]{%
 \ifx #1\expandafter \@firstoftwo
 \else \expandafter \@secondoftwo
 \fi
}%
\providecommand \natexlab [1]{#1}%
\providecommand \enquote  [1]{``#1''}%
\providecommand \bibnamefont  [1]{#1}%
\providecommand \bibfnamefont [1]{#1}%
\providecommand \citenamefont [1]{#1}%
\providecommand \href@noop [0]{\@secondoftwo}%
\providecommand \href [0]{\begingroup \@sanitize@url \@href}%
\providecommand \@href[1]{\@@startlink{#1}\@@href}%
\providecommand \@@href[1]{\endgroup#1\@@endlink}%
\providecommand \@sanitize@url [0]{\catcode `\\12\catcode `\$12\catcode
  `\&12\catcode `\#12\catcode `\^12\catcode `\_12\catcode `\%12\relax}%
\providecommand \@@startlink[1]{}%
\providecommand \@@endlink[0]{}%
\providecommand \url  [0]{\begingroup\@sanitize@url \@url }%
\providecommand \@url [1]{\endgroup\@href {#1}{\urlprefix }}%
\providecommand \urlprefix  [0]{URL }%
\providecommand \Eprint [0]{\href }%
\providecommand \doibase [0]{http://dx.doi.org/}%
\providecommand \selectlanguage [0]{\@gobble}%
\providecommand \bibinfo  [0]{\@secondoftwo}%
\providecommand \bibfield  [0]{\@secondoftwo}%
\providecommand \translation [1]{[#1]}%
\providecommand \BibitemOpen [0]{}%
\providecommand \bibitemStop [0]{}%
\providecommand \bibitemNoStop [0]{.\EOS\space}%
\providecommand \EOS [0]{\spacefactor3000\relax}%
\providecommand \BibitemShut  [1]{\csname bibitem#1\endcsname}%
\let\auto@bib@innerbib\@empty
\bibitem [{\citenamefont {Chen}\ \emph {et~al.}(2017)\citenamefont {Chen},
  \citenamefont {Chen}, \citenamefont {Liu}, \citenamefont {Liu},\ and\
  \citenamefont {Zhu}}]{Chen:2016spr}%
  \BibitemOpen
  \bibfield  {author} {\bibinfo {author} {\bibfnamefont {H.-X.}\ \bibnamefont
  {Chen}}, \bibinfo {author} {\bibfnamefont {W.}~\bibnamefont {Chen}}, \bibinfo
  {author} {\bibfnamefont {X.}~\bibnamefont {Liu}}, \bibinfo {author}
  {\bibfnamefont {Y.-R.}\ \bibnamefont {Liu}}, \ and\ \bibinfo {author}
  {\bibfnamefont {S.-L.}\ \bibnamefont {Zhu}},\ }\href {\doibase
  10.1088/1361-6633/aa6420} {\bibfield  {journal} {\bibinfo  {journal} {Rept.
  Prog. Phys.}\ }\textbf {\bibinfo {volume} {80}},\ \bibinfo {pages} {076201}
  (\bibinfo {year} {2017})}\BibitemShut {NoStop}%
\bibitem [{\citenamefont {Patrignani}\ \emph {et~al.}(2016)\citenamefont
  {Patrignani} \emph {et~al.}}]{Patrignani:2016xqp}%
  \BibitemOpen
  \bibfield  {author} {\bibinfo {author} {\bibfnamefont {C.}~\bibnamefont
  {Patrignani}} \emph {et~al.} (\bibinfo {collaboration} {Particle Data
  Group}),\ }\href {\doibase 10.1088/1674-1137/40/10/100001} {\bibfield
  {journal} {\bibinfo  {journal} {Chin. Phys.}\ }\textbf {\bibinfo {volume}
  {C40}},\ \bibinfo {pages} {100001} (\bibinfo {year} {2016})}\BibitemShut
  {NoStop}%
\bibitem [{\citenamefont {Aaij}\ \emph
  {et~al.}(2017{\natexlab{a}})\citenamefont {Aaij} \emph
  {et~al.}}]{Aaij:2017nav}%
  \BibitemOpen
  \bibfield  {author} {\bibinfo {author} {\bibfnamefont {R.}~\bibnamefont
  {Aaij}} \emph {et~al.} (\bibinfo {collaboration} {LHCb}),\ }\href {\doibase
  10.1103/PhysRevLett.118.182001} {\bibfield  {journal} {\bibinfo  {journal}
  {Phys. Rev. Lett.}\ }\textbf {\bibinfo {volume} {118}},\ \bibinfo {pages}
  {182001} (\bibinfo {year} {2017}{\natexlab{a}})}\BibitemShut {NoStop}%
\bibitem [{\citenamefont {Aaij}\ \emph
  {et~al.}(2017{\natexlab{b}})\citenamefont {Aaij} \emph
  {et~al.}}]{Aaij:2017ueg}%
  \BibitemOpen
  \bibfield  {author} {\bibinfo {author} {\bibfnamefont {R.}~\bibnamefont
  {Aaij}} \emph {et~al.} (\bibinfo {collaboration} {LHCb}),\ }\href {\doibase
  10.1103/PhysRevLett.119.112001} {\bibfield  {journal} {\bibinfo  {journal}
  {Phys. Rev. Lett.}\ }\textbf {\bibinfo {volume} {119}},\ \bibinfo {pages}
  {112001} (\bibinfo {year} {2017}{\natexlab{b}})}\BibitemShut {NoStop}%
\bibitem [{\citenamefont {del Amo~Sanchez}\ \emph {et~al.}(2010)\citenamefont
  {del Amo~Sanchez} \emph {et~al.}}]{delAmoSanchez:2010vq}%
  \BibitemOpen
  \bibfield  {author} {\bibinfo {author} {\bibfnamefont {P.}~\bibnamefont {del
  Amo~Sanchez}} \emph {et~al.} (\bibinfo {collaboration} {BaBar}),\ }\href
  {\doibase 10.1103/PhysRevD.82.111101} {\bibfield  {journal} {\bibinfo
  {journal} {Phys. Rev.}\ }\textbf {\bibinfo {volume} {D82}},\ \bibinfo {pages}
  {111101} (\bibinfo {year} {2010})}\BibitemShut {NoStop}%
\bibitem [{\citenamefont {Aaij}\ \emph {et~al.}(2013)\citenamefont {Aaij} \emph
  {et~al.}}]{Aaij:2013sza}%
  \BibitemOpen
  \bibfield  {author} {\bibinfo {author} {\bibfnamefont {R.}~\bibnamefont
  {Aaij}} \emph {et~al.} (\bibinfo {collaboration} {LHCb}),\ }\href {\doibase
  10.1007/JHEP09(2013)145} {\bibfield  {journal} {\bibinfo  {journal} {JHEP}\
  }\textbf {\bibinfo {volume} {09}},\ \bibinfo {pages} {145} (\bibinfo {year}
  {2013})}\BibitemShut {NoStop}%
\bibitem [{\citenamefont {Aaij}\ \emph {et~al.}(2016)\citenamefont {Aaij} \emph
  {et~al.}}]{Aaij:2016fma}%
  \BibitemOpen
  \bibfield  {author} {\bibinfo {author} {\bibfnamefont {R.}~\bibnamefont
  {Aaij}} \emph {et~al.} (\bibinfo {collaboration} {LHCb}),\ }\href {\doibase
  10.1103/PhysRevD.94.072001} {\bibfield  {journal} {\bibinfo  {journal} {Phys.
  Rev.}\ }\textbf {\bibinfo {volume} {D94}},\ \bibinfo {pages} {072001}
  (\bibinfo {year} {2016})}\BibitemShut {NoStop}%
\bibitem [{\citenamefont {Aaij}\ \emph {et~al.}(2012)\citenamefont {Aaij} \emph
  {et~al.}}]{Aaij:2012pc}%
  \BibitemOpen
  \bibfield  {author} {\bibinfo {author} {\bibfnamefont {R.}~\bibnamefont
  {Aaij}} \emph {et~al.} (\bibinfo {collaboration} {LHCb}),\ }\href {\doibase
  10.1007/JHEP10(2012)151} {\bibfield  {journal} {\bibinfo  {journal} {JHEP}\
  }\textbf {\bibinfo {volume} {10}},\ \bibinfo {pages} {151} (\bibinfo {year}
  {2012})}\BibitemShut {NoStop}%
\bibitem [{\citenamefont {Lees}\ \emph {et~al.}(2015)\citenamefont {Lees} \emph
  {et~al.}}]{Lees:2014abp}%
  \BibitemOpen
  \bibfield  {author} {\bibinfo {author} {\bibfnamefont {J.~P.}\ \bibnamefont
  {Lees}} \emph {et~al.} (\bibinfo {collaboration} {BaBar}),\ }\href {\doibase
  10.1103/PhysRevD.91.052002} {\bibfield  {journal} {\bibinfo  {journal} {Phys.
  Rev.}\ }\textbf {\bibinfo {volume} {D91}},\ \bibinfo {pages} {052002}
  (\bibinfo {year} {2015})}\BibitemShut {NoStop}%
\bibitem [{\citenamefont {Aaij}\ \emph {et~al.}(2014)\citenamefont {Aaij} \emph
  {et~al.}}]{Aaij:2014xza}%
  \BibitemOpen
  \bibfield  {author} {\bibinfo {author} {\bibfnamefont {R.}~\bibnamefont
  {Aaij}} \emph {et~al.} (\bibinfo {collaboration} {LHCb}),\ }\href {\doibase
  10.1103/PhysRevLett.113.162001} {\bibfield  {journal} {\bibinfo  {journal}
  {Phys. Rev. Lett.}\ }\textbf {\bibinfo {volume} {113}},\ \bibinfo {pages}
  {162001} (\bibinfo {year} {2014})}\BibitemShut {NoStop}%
\bibitem [{\citenamefont {Aubert}\ \emph {et~al.}(2009)\citenamefont {Aubert}
  \emph {et~al.}}]{Aubert:2009ah}%
  \BibitemOpen
  \bibfield  {author} {\bibinfo {author} {\bibfnamefont {B.}~\bibnamefont
  {Aubert}} \emph {et~al.} (\bibinfo {collaboration} {BaBar}),\ }\href
  {\doibase 10.1103/PhysRevD.80.092003} {\bibfield  {journal} {\bibinfo
  {journal} {Phys. Rev.}\ }\textbf {\bibinfo {volume} {D80}},\ \bibinfo {pages}
  {092003} (\bibinfo {year} {2009})}\BibitemShut {NoStop}%
\bibitem [{\citenamefont {Aaij}\ \emph
  {et~al.}(2015{\natexlab{a}})\citenamefont {Aaij} \emph
  {et~al.}}]{Aaij:2015sqa}%
  \BibitemOpen
  \bibfield  {author} {\bibinfo {author} {\bibfnamefont {R.}~\bibnamefont
  {Aaij}} \emph {et~al.} (\bibinfo {collaboration} {LHCb}),\ }\href {\doibase
  10.1103/PhysRevD.92.032002} {\bibfield  {journal} {\bibinfo  {journal} {Phys.
  Rev.}\ }\textbf {\bibinfo {volume} {D92}},\ \bibinfo {pages} {032002}
  (\bibinfo {year} {2015}{\natexlab{a}})}\BibitemShut {NoStop}%
\bibitem [{\citenamefont {Aaltonen}\ \emph {et~al.}(2014)\citenamefont
  {Aaltonen} \emph {et~al.}}]{Aaltonen:2013atp}%
  \BibitemOpen
  \bibfield  {author} {\bibinfo {author} {\bibfnamefont {T.~A.}\ \bibnamefont
  {Aaltonen}} \emph {et~al.} (\bibinfo {collaboration} {CDF}),\ }\href
  {\doibase 10.1103/PhysRevD.90.012013} {\bibfield  {journal} {\bibinfo
  {journal} {Phys. Rev.}\ }\textbf {\bibinfo {volume} {D90}},\ \bibinfo {pages}
  {012013} (\bibinfo {year} {2014})}\BibitemShut {NoStop}%
\bibitem [{\citenamefont {Aaij}\ \emph
  {et~al.}(2015{\natexlab{b}})\citenamefont {Aaij} \emph
  {et~al.}}]{Aaij:2015qla}%
  \BibitemOpen
  \bibfield  {author} {\bibinfo {author} {\bibfnamefont {R.}~\bibnamefont
  {Aaij}} \emph {et~al.} (\bibinfo {collaboration} {LHCb}),\ }\href {\doibase
  10.1007/JHEP04(2015)024} {\bibfield  {journal} {\bibinfo  {journal} {JHEP}\
  }\textbf {\bibinfo {volume} {04}},\ \bibinfo {pages} {024} (\bibinfo {year}
  {2015}{\natexlab{b}})}\BibitemShut {NoStop}%
\bibitem [{\citenamefont {Sirunyan}\ \emph {et~al.}(2018)\citenamefont
  {Sirunyan} \emph {et~al.}}]{Sirunyan:2018grk}%
  \BibitemOpen
  \bibfield  {author} {\bibinfo {author} {\bibfnamefont {A.~M.}\ \bibnamefont
  {Sirunyan}} \emph {et~al.} (\bibinfo {collaboration} {CMS}),\ }\href@noop {}
  {\  (\bibinfo {year} {2018})},\ \Eprint {http://arxiv.org/abs/1809.03578}
  {arXiv:1809.03578 [hep-ex]} \BibitemShut {NoStop}%
\bibitem [{\citenamefont {Colangelo}\ \emph {et~al.}(2012)\citenamefont
  {Colangelo}, \citenamefont {De~Fazio}, \citenamefont {Giannuzzi},\ and\
  \citenamefont {Nicotri}}]{Colangelo:2012xi}%
  \BibitemOpen
  \bibfield  {author} {\bibinfo {author} {\bibfnamefont {P.}~\bibnamefont
  {Colangelo}}, \bibinfo {author} {\bibfnamefont {F.}~\bibnamefont {De~Fazio}},
  \bibinfo {author} {\bibfnamefont {F.}~\bibnamefont {Giannuzzi}}, \ and\
  \bibinfo {author} {\bibfnamefont {S.}~\bibnamefont {Nicotri}},\ }\href
  {\doibase 10.1103/PhysRevD.86.054024} {\bibfield  {journal} {\bibinfo
  {journal} {Phys. Rev.}\ }\textbf {\bibinfo {volume} {D86}},\ \bibinfo {pages}
  {054024} (\bibinfo {year} {2012})}\BibitemShut {NoStop}%
\bibitem [{\citenamefont {Wang}(2016)}]{Wang:2016ewb}%
  \BibitemOpen
  \bibfield  {author} {\bibinfo {author} {\bibfnamefont {Z.-G.}\ \bibnamefont
  {Wang}},\ }\href {\doibase 10.1088/0253-6102/66/6/671} {\bibfield  {journal}
  {\bibinfo  {journal} {Commun. Theor. Phys.}\ }\textbf {\bibinfo {volume}
  {66}},\ \bibinfo {pages} {671} (\bibinfo {year} {2016})}\BibitemShut
  {NoStop}%
\bibitem [{\citenamefont {Gupta}\ and\ \citenamefont
  {Upadhyay}(2018)}]{Gupta:2018zlg}%
  \BibitemOpen
  \bibfield  {author} {\bibinfo {author} {\bibfnamefont {P.}~\bibnamefont
  {Gupta}}\ and\ \bibinfo {author} {\bibfnamefont {A.}~\bibnamefont
  {Upadhyay}},\ }\href {\doibase 10.1103/PhysRevD.97.014015} {\bibfield
  {journal} {\bibinfo  {journal} {Phys. Rev.}\ }\textbf {\bibinfo {volume}
  {D97}},\ \bibinfo {pages} {014015} (\bibinfo {year} {2018})}\BibitemShut
  {NoStop}%
\bibitem [{\citenamefont {Casalbuoni}\ \emph {et~al.}(1992)\citenamefont
  {Casalbuoni}, \citenamefont {Deandrea}, \citenamefont {Di~Bartolomeo},
  \citenamefont {Gatto}, \citenamefont {Feruglio},\ and\ \citenamefont
  {Nardulli}}]{Casalbuoni:1992gi}%
  \BibitemOpen
  \bibfield  {author} {\bibinfo {author} {\bibfnamefont {R.}~\bibnamefont
  {Casalbuoni}}, \bibinfo {author} {\bibfnamefont {A.}~\bibnamefont
  {Deandrea}}, \bibinfo {author} {\bibfnamefont {N.}~\bibnamefont
  {Di~Bartolomeo}}, \bibinfo {author} {\bibfnamefont {R.}~\bibnamefont
  {Gatto}}, \bibinfo {author} {\bibfnamefont {F.}~\bibnamefont {Feruglio}}, \
  and\ \bibinfo {author} {\bibfnamefont {G.}~\bibnamefont {Nardulli}},\ }\href
  {\doibase 10.1016/0370-2693(92)91189-G} {\bibfield  {journal} {\bibinfo
  {journal} {Phys. Lett.}\ }\textbf {\bibinfo {volume} {B292}},\ \bibinfo
  {pages} {371} (\bibinfo {year} {1992})}\BibitemShut {NoStop}%
\bibitem [{\citenamefont {Casalbuoni}\ \emph {et~al.}(1997)\citenamefont
  {Casalbuoni}, \citenamefont {Deandrea}, \citenamefont {Di~Bartolomeo},
  \citenamefont {Gatto}, \citenamefont {Feruglio},\ and\ \citenamefont
  {Nardulli}}]{Casalbuoni:1996pg}%
  \BibitemOpen
  \bibfield  {author} {\bibinfo {author} {\bibfnamefont {R.}~\bibnamefont
  {Casalbuoni}}, \bibinfo {author} {\bibfnamefont {A.}~\bibnamefont
  {Deandrea}}, \bibinfo {author} {\bibfnamefont {N.}~\bibnamefont
  {Di~Bartolomeo}}, \bibinfo {author} {\bibfnamefont {R.}~\bibnamefont
  {Gatto}}, \bibinfo {author} {\bibfnamefont {F.}~\bibnamefont {Feruglio}}, \
  and\ \bibinfo {author} {\bibfnamefont {G.}~\bibnamefont {Nardulli}},\ }\href
  {\doibase 10.1016/S0370-1573(96)00027-0} {\bibfield  {journal} {\bibinfo
  {journal} {Phys. Rept.}\ }\textbf {\bibinfo {volume} {281}},\ \bibinfo
  {pages} {145} (\bibinfo {year} {1997})}\BibitemShut {NoStop}%
\bibitem [{\citenamefont {Neubert}(1994)}]{Neubert:1993mb}%
  \BibitemOpen
  \bibfield  {author} {\bibinfo {author} {\bibfnamefont {M.}~\bibnamefont
  {Neubert}},\ }\href {\doibase 10.1016/0370-1573(94)90091-4} {\bibfield
  {journal} {\bibinfo  {journal} {Phys. Rept.}\ }\textbf {\bibinfo {volume}
  {245}},\ \bibinfo {pages} {259} (\bibinfo {year} {1994})}\BibitemShut
  {NoStop}%
\bibitem [{\citenamefont {Falk}(1992)}]{Falk:1991nq}%
  \BibitemOpen
  \bibfield  {author} {\bibinfo {author} {\bibfnamefont {A.~F.}\ \bibnamefont
  {Falk}},\ }\href {\doibase 10.1016/0550-3213(92)90004-U} {\bibfield
  {journal} {\bibinfo  {journal} {Nucl. Phys.}\ }\textbf {\bibinfo {volume}
  {B378}},\ \bibinfo {pages} {79} (\bibinfo {year} {1992})}\BibitemShut
  {NoStop}%
\bibitem [{\citenamefont {Wise}(1992)}]{Wise:1992hn}%
  \BibitemOpen
  \bibfield  {author} {\bibinfo {author} {\bibfnamefont {M.~B.}\ \bibnamefont
  {Wise}},\ }\href {\doibase 10.1103/PhysRevD.45.R2188} {\bibfield  {journal}
  {\bibinfo  {journal} {Phys. Rev.}\ }\textbf {\bibinfo {volume} {D45}},\
  \bibinfo {pages} {R2188} (\bibinfo {year} {1992})}\BibitemShut {NoStop}%
\bibitem [{\citenamefont {Burdman}\ and\ \citenamefont
  {Donoghue}(1992)}]{Burdman:1992gh}%
  \BibitemOpen
  \bibfield  {author} {\bibinfo {author} {\bibfnamefont {G.}~\bibnamefont
  {Burdman}}\ and\ \bibinfo {author} {\bibfnamefont {J.~F.}\ \bibnamefont
  {Donoghue}},\ }\href {\doibase 10.1016/0370-2693(92)90068-F} {\bibfield
  {journal} {\bibinfo  {journal} {Phys. Lett.}\ }\textbf {\bibinfo {volume}
  {B280}},\ \bibinfo {pages} {287} (\bibinfo {year} {1992})}\BibitemShut
  {NoStop}%
\bibitem [{\citenamefont {Cho}(1992)}]{Cho:1992gg}%
  \BibitemOpen
  \bibfield  {author} {\bibinfo {author} {\bibfnamefont {P.~L.}\ \bibnamefont
  {Cho}},\ }\href {\doibase 10.1016/0370-2693(92)91314-Y} {\bibfield  {journal}
  {\bibinfo  {journal} {Phys. Lett.}\ }\textbf {\bibinfo {volume} {B285}},\
  \bibinfo {pages} {145} (\bibinfo {year} {1992})}\BibitemShut {NoStop}%
\bibitem [{\citenamefont {Yan}\ \emph {et~al.}(1992)\citenamefont {Yan},
  \citenamefont {Cheng}, \citenamefont {Cheung}, \citenamefont {Lin},
  \citenamefont {Lin},\ and\ \citenamefont {Yu}}]{Yan:1992gz}%
  \BibitemOpen
  \bibfield  {author} {\bibinfo {author} {\bibfnamefont {T.-M.}\ \bibnamefont
  {Yan}}, \bibinfo {author} {\bibfnamefont {H.-Y.}\ \bibnamefont {Cheng}},
  \bibinfo {author} {\bibfnamefont {C.-Y.}\ \bibnamefont {Cheung}}, \bibinfo
  {author} {\bibfnamefont {G.-L.}\ \bibnamefont {Lin}}, \bibinfo {author}
  {\bibfnamefont {Y.~C.}\ \bibnamefont {Lin}}, \ and\ \bibinfo {author}
  {\bibfnamefont {H.-L.}\ \bibnamefont {Yu}},\ }\href {\doibase
  10.1103/PhysRevD.46.1148, 10.1103/PhysRevD.55.5851} {\bibfield  {journal}
  {\bibinfo  {journal} {Phys. Rev.}\ }\textbf {\bibinfo {volume} {D46}},\
  \bibinfo {pages} {1148} (\bibinfo {year} {1992})},\ \bibinfo {note}
  {[Erratum: Phys. Rev.D55,5851(1997)]}\BibitemShut {NoStop}%
\bibitem [{\citenamefont {Casalbuoni}\ \emph {et~al.}(1993)\citenamefont
  {Casalbuoni}, \citenamefont {Deandrea}, \citenamefont {Di~Bartolomeo},
  \citenamefont {Gatto}, \citenamefont {Feruglio},\ and\ \citenamefont
  {Nardulli}}]{Casalbuoni:1992dx}%
  \BibitemOpen
  \bibfield  {author} {\bibinfo {author} {\bibfnamefont {R.}~\bibnamefont
  {Casalbuoni}}, \bibinfo {author} {\bibfnamefont {A.}~\bibnamefont
  {Deandrea}}, \bibinfo {author} {\bibfnamefont {N.}~\bibnamefont
  {Di~Bartolomeo}}, \bibinfo {author} {\bibfnamefont {R.}~\bibnamefont
  {Gatto}}, \bibinfo {author} {\bibfnamefont {F.}~\bibnamefont {Feruglio}}, \
  and\ \bibinfo {author} {\bibfnamefont {G.}~\bibnamefont {Nardulli}},\ }\href
  {\doibase 10.1016/0370-2693(93)90895-O} {\bibfield  {journal} {\bibinfo
  {journal} {Phys. Lett.}\ }\textbf {\bibinfo {volume} {B299}},\ \bibinfo
  {pages} {139} (\bibinfo {year} {1993})}\BibitemShut {NoStop}%
\bibitem [{\citenamefont {Colangelo}\ \emph
  {et~al.}(1994{\natexlab{a}})\citenamefont {Colangelo}, \citenamefont
  {De~Fazio},\ and\ \citenamefont {Nardulli}}]{Colangelo:1994jc}%
  \BibitemOpen
  \bibfield  {author} {\bibinfo {author} {\bibfnamefont {P.}~\bibnamefont
  {Colangelo}}, \bibinfo {author} {\bibfnamefont {F.}~\bibnamefont {De~Fazio}},
  \ and\ \bibinfo {author} {\bibfnamefont {G.}~\bibnamefont {Nardulli}},\
  }\href {\doibase 10.1016/0370-2693(94)90607-6} {\bibfield  {journal}
  {\bibinfo  {journal} {Phys. Lett.}\ }\textbf {\bibinfo {volume} {B334}},\
  \bibinfo {pages} {175} (\bibinfo {year} {1994}{\natexlab{a}})}\BibitemShut
  {NoStop}%
\bibitem [{\citenamefont {Colangelo}\ \emph
  {et~al.}(1994{\natexlab{b}})\citenamefont {Colangelo}, \citenamefont
  {Nardulli}, \citenamefont {Deandrea}, \citenamefont {Di~Bartolomeo},
  \citenamefont {Gatto},\ and\ \citenamefont {Feruglio}}]{Colangelo:1994es}%
  \BibitemOpen
  \bibfield  {author} {\bibinfo {author} {\bibfnamefont {P.}~\bibnamefont
  {Colangelo}}, \bibinfo {author} {\bibfnamefont {G.}~\bibnamefont {Nardulli}},
  \bibinfo {author} {\bibfnamefont {A.}~\bibnamefont {Deandrea}}, \bibinfo
  {author} {\bibfnamefont {N.}~\bibnamefont {Di~Bartolomeo}}, \bibinfo {author}
  {\bibfnamefont {R.}~\bibnamefont {Gatto}}, \ and\ \bibinfo {author}
  {\bibfnamefont {F.}~\bibnamefont {Feruglio}},\ }\href {\doibase
  10.1016/0370-2693(94)91148-7} {\bibfield  {journal} {\bibinfo  {journal}
  {Phys. Lett.}\ }\textbf {\bibinfo {volume} {B339}},\ \bibinfo {pages} {151}
  (\bibinfo {year} {1994}{\natexlab{b}})}\BibitemShut {NoStop}%
\bibitem [{\citenamefont {Belyaev}\ \emph {et~al.}(1995)\citenamefont
  {Belyaev}, \citenamefont {Braun}, \citenamefont {Khodjamirian},\ and\
  \citenamefont {Ruckl}}]{Belyaev:1994zk}%
  \BibitemOpen
  \bibfield  {author} {\bibinfo {author} {\bibfnamefont {V.~M.}\ \bibnamefont
  {Belyaev}}, \bibinfo {author} {\bibfnamefont {V.~M.}\ \bibnamefont {Braun}},
  \bibinfo {author} {\bibfnamefont {A.}~\bibnamefont {Khodjamirian}}, \ and\
  \bibinfo {author} {\bibfnamefont {R.}~\bibnamefont {Ruckl}},\ }\href
  {\doibase 10.1103/PhysRevD.51.6177} {\bibfield  {journal} {\bibinfo
  {journal} {Phys. Rev.}\ }\textbf {\bibinfo {volume} {D51}},\ \bibinfo {pages}
  {6177} (\bibinfo {year} {1995})}\BibitemShut {NoStop}%
\bibitem [{\citenamefont {Colangelo}\ \emph {et~al.}(1995)\citenamefont
  {Colangelo}, \citenamefont {De~Fazio}, \citenamefont {Nardulli},
  \citenamefont {Di~Bartolomeo},\ and\ \citenamefont
  {Gatto}}]{Colangelo:1995ph}%
  \BibitemOpen
  \bibfield  {author} {\bibinfo {author} {\bibfnamefont {P.}~\bibnamefont
  {Colangelo}}, \bibinfo {author} {\bibfnamefont {F.}~\bibnamefont {De~Fazio}},
  \bibinfo {author} {\bibfnamefont {G.}~\bibnamefont {Nardulli}}, \bibinfo
  {author} {\bibfnamefont {N.}~\bibnamefont {Di~Bartolomeo}}, \ and\ \bibinfo
  {author} {\bibfnamefont {R.}~\bibnamefont {Gatto}},\ }\href {\doibase
  10.1103/PhysRevD.52.6422} {\bibfield  {journal} {\bibinfo  {journal} {Phys.
  Rev.}\ }\textbf {\bibinfo {volume} {D52}},\ \bibinfo {pages} {6422} (\bibinfo
  {year} {1995})}\BibitemShut {NoStop}%
\bibitem [{\citenamefont {Colangelo}\ and\ \citenamefont
  {De~Fazio}(1998)}]{Colangelo:1997rp}%
  \BibitemOpen
  \bibfield  {author} {\bibinfo {author} {\bibfnamefont {P.}~\bibnamefont
  {Colangelo}}\ and\ \bibinfo {author} {\bibfnamefont {F.}~\bibnamefont
  {De~Fazio}},\ }\href {\doibase 10.1007/s100529800787, 10.1007/s100520050222}
  {\bibfield  {journal} {\bibinfo  {journal} {Eur. Phys. J.}\ }\textbf
  {\bibinfo {volume} {C4}},\ \bibinfo {pages} {503} (\bibinfo {year}
  {1998})}\BibitemShut {NoStop}%
\bibitem [{\citenamefont {Becirevic}\ \emph {et~al.}(2009)\citenamefont
  {Becirevic}, \citenamefont {Blossier}, \citenamefont {Chang},\ and\
  \citenamefont {Haas}}]{Becirevic:2009yb}%
  \BibitemOpen
  \bibfield  {author} {\bibinfo {author} {\bibfnamefont {D.}~\bibnamefont
  {Becirevic}}, \bibinfo {author} {\bibfnamefont {B.}~\bibnamefont {Blossier}},
  \bibinfo {author} {\bibfnamefont {E.}~\bibnamefont {Chang}}, \ and\ \bibinfo
  {author} {\bibfnamefont {B.}~\bibnamefont {Haas}},\ }\href {\doibase
  10.1016/j.physletb.2009.07.031} {\bibfield  {journal} {\bibinfo  {journal}
  {Phys. Lett.}\ }\textbf {\bibinfo {volume} {B679}},\ \bibinfo {pages} {231}
  (\bibinfo {year} {2009})}\BibitemShut {NoStop}%
\bibitem [{\citenamefont {Becirevic}\ \emph {et~al.}(2012)\citenamefont
  {Becirevic}, \citenamefont {Chang},\ and\ \citenamefont
  {Le~Yaouanc}}]{Becirevic:2012zza}%
  \BibitemOpen
  \bibfield  {author} {\bibinfo {author} {\bibfnamefont {D.}~\bibnamefont
  {Becirevic}}, \bibinfo {author} {\bibfnamefont {E.}~\bibnamefont {Chang}}, \
  and\ \bibinfo {author} {\bibfnamefont {A.}~\bibnamefont {Le~Yaouanc}},\
  }\href@noop {} {\  (\bibinfo {year} {2012})},\ \Eprint
  {http://arxiv.org/abs/1203.0167} {arXiv:1203.0167 [hep-lat]} \BibitemShut
  {NoStop}%
\bibitem [{\citenamefont {Bando}\ \emph {et~al.}(1985)\citenamefont {Bando},
  \citenamefont {Kugo},\ and\ \citenamefont {Yamawaki}}]{Bando:1985rf}%
  \BibitemOpen
  \bibfield  {author} {\bibinfo {author} {\bibfnamefont {M.}~\bibnamefont
  {Bando}}, \bibinfo {author} {\bibfnamefont {T.}~\bibnamefont {Kugo}}, \ and\
  \bibinfo {author} {\bibfnamefont {K.}~\bibnamefont {Yamawaki}},\ }\href
  {\doibase 10.1016/0550-3213(85)90647-9} {\bibfield  {journal} {\bibinfo
  {journal} {Nucl. Phys.}\ }\textbf {\bibinfo {volume} {B259}},\ \bibinfo
  {pages} {493} (\bibinfo {year} {1985})}\BibitemShut {NoStop}%
\bibitem [{\citenamefont {Bando}\ \emph {et~al.}(1988)\citenamefont {Bando},
  \citenamefont {Kugo},\ and\ \citenamefont {Yamawaki}}]{Bando:1987br}%
  \BibitemOpen
  \bibfield  {author} {\bibinfo {author} {\bibfnamefont {M.}~\bibnamefont
  {Bando}}, \bibinfo {author} {\bibfnamefont {T.}~\bibnamefont {Kugo}}, \ and\
  \bibinfo {author} {\bibfnamefont {K.}~\bibnamefont {Yamawaki}},\ }\href
  {\doibase 10.1016/0370-1573(88)90019-1} {\bibfield  {journal} {\bibinfo
  {journal} {Phys. Rept.}\ }\textbf {\bibinfo {volume} {164}},\ \bibinfo
  {pages} {217} (\bibinfo {year} {1988})}\BibitemShut {NoStop}%
\bibitem [{\citenamefont {Georgi}(1989)}]{Georgi:1989gp}%
  \BibitemOpen
  \bibfield  {author} {\bibinfo {author} {\bibfnamefont {H.}~\bibnamefont
  {Georgi}},\ }\href {\doibase 10.1103/PhysRevLett.63.1917} {\bibfield
  {journal} {\bibinfo  {journal} {Phys. Rev. Lett.}\ }\textbf {\bibinfo
  {volume} {63}},\ \bibinfo {pages} {1917} (\bibinfo {year}
  {1989})}\BibitemShut {NoStop}%
\bibitem [{\citenamefont {Georgi}(1990)}]{Georgi:1989xy}%
  \BibitemOpen
  \bibfield  {author} {\bibinfo {author} {\bibfnamefont {H.}~\bibnamefont
  {Georgi}},\ }\href {\doibase 10.1016/0550-3213(90)90210-5} {\bibfield
  {journal} {\bibinfo  {journal} {Nucl. Phys.}\ }\textbf {\bibinfo {volume}
  {B331}},\ \bibinfo {pages} {311} (\bibinfo {year} {1990})}\BibitemShut
  {NoStop}%
\bibitem [{\citenamefont {Ko}(1993)}]{Ko:1993fn}%
  \BibitemOpen
  \bibfield  {author} {\bibinfo {author} {\bibfnamefont {P.}~\bibnamefont
  {Ko}},\ }\href {\doibase 10.1103/PhysRevD.47.1964} {\bibfield  {journal}
  {\bibinfo  {journal} {Phys. Rev.}\ }\textbf {\bibinfo {volume} {D47}},\
  \bibinfo {pages} {1964} (\bibinfo {year} {1993})}\BibitemShut {NoStop}%
\bibitem [{\citenamefont {Schechter}\ and\ \citenamefont
  {Subbaraman}(1993)}]{Schechter:1992ue}%
  \BibitemOpen
  \bibfield  {author} {\bibinfo {author} {\bibfnamefont {J.}~\bibnamefont
  {Schechter}}\ and\ \bibinfo {author} {\bibfnamefont {A.}~\bibnamefont
  {Subbaraman}},\ }\href {\doibase 10.1103/PhysRevD.48.332} {\bibfield
  {journal} {\bibinfo  {journal} {Phys. Rev.}\ }\textbf {\bibinfo {volume}
  {D48}},\ \bibinfo {pages} {332} (\bibinfo {year} {1993})}\BibitemShut
  {NoStop}%
\bibitem [{\citenamefont {Kitazawa}\ and\ \citenamefont
  {Kurimoto}(1994)}]{Kitazawa:1993bk}%
  \BibitemOpen
  \bibfield  {author} {\bibinfo {author} {\bibfnamefont {N.}~\bibnamefont
  {Kitazawa}}\ and\ \bibinfo {author} {\bibfnamefont {T.}~\bibnamefont
  {Kurimoto}},\ }\href {\doibase 10.1016/0370-2693(94)00047-6} {\bibfield
  {journal} {\bibinfo  {journal} {Phys. Lett.}\ }\textbf {\bibinfo {volume}
  {B323}},\ \bibinfo {pages} {65} (\bibinfo {year} {1994})}\BibitemShut
  {NoStop}%
\bibitem [{\citenamefont {Cremmer}\ and\ \citenamefont
  {Julia}(1978)}]{Cremmer:1978ds}%
  \BibitemOpen
  \bibfield  {author} {\bibinfo {author} {\bibfnamefont {E.}~\bibnamefont
  {Cremmer}}\ and\ \bibinfo {author} {\bibfnamefont {B.}~\bibnamefont
  {Julia}},\ }\href {\doibase 10.1016/0370-2693(78)90303-9} {\bibfield
  {journal} {\bibinfo  {journal} {Phys. Lett.}\ }\textbf {\bibinfo {volume}
  {B80}},\ \bibinfo {pages} {48} (\bibinfo {year} {1978})}\BibitemShut
  {NoStop}%
\bibitem [{\citenamefont {Cremmer}\ and\ \citenamefont
  {Julia}(1979)}]{Cremmer:1979up}%
  \BibitemOpen
  \bibfield  {author} {\bibinfo {author} {\bibfnamefont {E.}~\bibnamefont
  {Cremmer}}\ and\ \bibinfo {author} {\bibfnamefont {B.}~\bibnamefont
  {Julia}},\ }\href {\doibase 10.1016/0550-3213(79)90331-6} {\bibfield
  {journal} {\bibinfo  {journal} {Nucl. Phys.}\ }\textbf {\bibinfo {volume}
  {B159}},\ \bibinfo {pages} {141} (\bibinfo {year} {1979})}\BibitemShut
  {NoStop}%
\bibitem [{\citenamefont {Weinberg}(1968)}]{Weinberg:1968de}%
  \BibitemOpen
  \bibfield  {author} {\bibinfo {author} {\bibfnamefont {S.}~\bibnamefont
  {Weinberg}},\ }\href {\doibase 10.1103/PhysRev.166.1568} {\bibfield
  {journal} {\bibinfo  {journal} {Phys. Rev.}\ }\textbf {\bibinfo {volume}
  {166}},\ \bibinfo {pages} {1568} (\bibinfo {year} {1968})}\BibitemShut
  {NoStop}%
\bibitem [{\citenamefont {Callan}\ \emph {et~al.}(1969)\citenamefont {Callan},
  \citenamefont {Coleman}, \citenamefont {Wess},\ and\ \citenamefont
  {Zumino}}]{Callan:1969sn}%
  \BibitemOpen
  \bibfield  {author} {\bibinfo {author} {\bibfnamefont {C.~G.}\ \bibnamefont
  {Callan}, \bibfnamefont {Jr.}}, \bibinfo {author} {\bibfnamefont {S.~R.}\
  \bibnamefont {Coleman}}, \bibinfo {author} {\bibfnamefont {J.}~\bibnamefont
  {Wess}}, \ and\ \bibinfo {author} {\bibfnamefont {B.}~\bibnamefont
  {Zumino}},\ }\href {\doibase 10.1103/PhysRev.177.2247} {\bibfield  {journal}
  {\bibinfo  {journal} {Phys. Rev.}\ }\textbf {\bibinfo {volume} {177}},\
  \bibinfo {pages} {2247} (\bibinfo {year} {1969})}\BibitemShut {NoStop}%
\bibitem [{\citenamefont {Yamawaki}(1987)}]{Yamawaki:1986zz}%
  \BibitemOpen
  \bibfield  {author} {\bibinfo {author} {\bibfnamefont {K.}~\bibnamefont
  {Yamawaki}},\ }\href {\doibase 10.1103/PhysRevD.35.412} {\bibfield  {journal}
  {\bibinfo  {journal} {Phys. Rev.}\ }\textbf {\bibinfo {volume} {D35}},\
  \bibinfo {pages} {412} (\bibinfo {year} {1987})}\BibitemShut {NoStop}%
\bibitem [{\citenamefont {Ecker}\ \emph {et~al.}(1989)\citenamefont {Ecker},
  \citenamefont {Gasser}, \citenamefont {Leutwyler}, \citenamefont {Pich},\
  and\ \citenamefont {de~Rafael}}]{Ecker:1989yg}%
  \BibitemOpen
  \bibfield  {author} {\bibinfo {author} {\bibfnamefont {G.}~\bibnamefont
  {Ecker}}, \bibinfo {author} {\bibfnamefont {J.}~\bibnamefont {Gasser}},
  \bibinfo {author} {\bibfnamefont {H.}~\bibnamefont {Leutwyler}}, \bibinfo
  {author} {\bibfnamefont {A.}~\bibnamefont {Pich}}, \ and\ \bibinfo {author}
  {\bibfnamefont {E.}~\bibnamefont {de~Rafael}},\ }\href {\doibase
  10.1016/0370-2693(89)91627-4} {\bibfield  {journal} {\bibinfo  {journal}
  {Phys. Lett.}\ }\textbf {\bibinfo {volume} {B223}},\ \bibinfo {pages} {425}
  (\bibinfo {year} {1989})}\BibitemShut {NoStop}%
\bibitem [{\citenamefont {Tanabashi}(1996)}]{Tanabashi:1995nz}%
  \BibitemOpen
  \bibfield  {author} {\bibinfo {author} {\bibfnamefont {M.}~\bibnamefont
  {Tanabashi}},\ }\href {\doibase 10.1016/0370-2693(96)00827-1} {\bibfield
  {journal} {\bibinfo  {journal} {Phys. Lett.}\ }\textbf {\bibinfo {volume}
  {B384}},\ \bibinfo {pages} {218} (\bibinfo {year} {1996})}\BibitemShut
  {NoStop}%
\bibitem [{\citenamefont {Birse}(1996)}]{Birse:1996hd}%
  \BibitemOpen
  \bibfield  {author} {\bibinfo {author} {\bibfnamefont {M.~C.}\ \bibnamefont
  {Birse}},\ }\href {\doibase 10.1007/s002180050105} {\bibfield  {journal}
  {\bibinfo  {journal} {Z. Phys.}\ }\textbf {\bibinfo {volume} {A355}},\
  \bibinfo {pages} {231} (\bibinfo {year} {1996})}\BibitemShut {NoStop}%
\bibitem [{\citenamefont {Harada}\ and\ \citenamefont
  {Yamawaki}(2003)}]{Harada:2003jx}%
  \BibitemOpen
  \bibfield  {author} {\bibinfo {author} {\bibfnamefont {M.}~\bibnamefont
  {Harada}}\ and\ \bibinfo {author} {\bibfnamefont {K.}~\bibnamefont
  {Yamawaki}},\ }\href {\doibase 10.1016/S0370-1573(03)00139-X} {\bibfield
  {journal} {\bibinfo  {journal} {Phys. Rept.}\ }\textbf {\bibinfo {volume}
  {381}},\ \bibinfo {pages} {1} (\bibinfo {year} {2003})}\BibitemShut {NoStop}%
\bibitem [{\citenamefont {Kawarabayashi}\ and\ \citenamefont
  {Suzuki}(1966)}]{Kawarabayashi:1966kd}%
  \BibitemOpen
  \bibfield  {author} {\bibinfo {author} {\bibfnamefont {K.}~\bibnamefont
  {Kawarabayashi}}\ and\ \bibinfo {author} {\bibfnamefont {M.}~\bibnamefont
  {Suzuki}},\ }\href {\doibase 10.1103/PhysRevLett.16.255} {\bibfield
  {journal} {\bibinfo  {journal} {Phys. Rev. Lett.}\ }\textbf {\bibinfo
  {volume} {16}},\ \bibinfo {pages} {255} (\bibinfo {year} {1966})}\BibitemShut
  {NoStop}%
\bibitem [{\citenamefont {Riazuddin}\ and\ \citenamefont
  {Fayyazuddin}(1966)}]{Riazuddin:1966sw}%
  \BibitemOpen
  \bibfield  {author} {\bibinfo {author} {\bibnamefont {Riazuddin}}\ and\
  \bibinfo {author} {\bibnamefont {Fayyazuddin}},\ }\href {\doibase
  10.1103/PhysRev.147.1071} {\bibfield  {journal} {\bibinfo  {journal} {Phys.
  Rev.}\ }\textbf {\bibinfo {volume} {147}},\ \bibinfo {pages} {1071} (\bibinfo
  {year} {1966})}\BibitemShut {NoStop}%
\bibitem [{\citenamefont {Jenkins}\ \emph {et~al.}(1995)\citenamefont
  {Jenkins}, \citenamefont {Manohar},\ and\ \citenamefont
  {Wise}}]{Jenkins:1995vb}%
  \BibitemOpen
  \bibfield  {author} {\bibinfo {author} {\bibfnamefont {E.~E.}\ \bibnamefont
  {Jenkins}}, \bibinfo {author} {\bibfnamefont {A.~V.}\ \bibnamefont
  {Manohar}}, \ and\ \bibinfo {author} {\bibfnamefont {M.~B.}\ \bibnamefont
  {Wise}},\ }\href {\doibase 10.1103/PhysRevLett.75.2272} {\bibfield  {journal}
  {\bibinfo  {journal} {Phys. Rev. Lett.}\ }\textbf {\bibinfo {volume} {75}},\
  \bibinfo {pages} {2272} (\bibinfo {year} {1995})}\BibitemShut {NoStop}%
\bibitem [{\citenamefont {Boyd}\ and\ \citenamefont
  {Grinstein}()}]{Boyd:1994pa}%
  \BibitemOpen
  \bibfield  {author} {\bibinfo {author} {\bibfnamefont {C.~G.}\ \bibnamefont
  {Boyd}}\ and\ \bibinfo {author} {\bibfnamefont {B.}~\bibnamefont
  {Grinstein}},\ }\href@noop {} {\bibinfo  {journal} {Nucl. Phys.}\
  }\BibitemShut {NoStop}%
\bibitem [{\citenamefont {Luke}\ and\ \citenamefont
  {Manohar}(1992)}]{Luke:1992cs}%
  \BibitemOpen
\bibfield  {journal} {  }\bibfield  {author} {\bibinfo {author} {\bibfnamefont
  {M.~E.}\ \bibnamefont {Luke}}\ and\ \bibinfo {author} {\bibfnamefont {A.~V.}\
  \bibnamefont {Manohar}},\ }\href {\doibase 10.1016/0370-2693(92)91786-9}
  {\bibfield  {journal} {\bibinfo  {journal} {Phys. Lett.}\ }\textbf {\bibinfo
  {volume} {B286}},\ \bibinfo {pages} {348} (\bibinfo {year}
  {1992})}\BibitemShut {NoStop}%
\bibitem [{\citenamefont {Falk}\ and\ \citenamefont
  {Mehen}(1996)}]{Falk:1995th}%
  \BibitemOpen
  \bibfield  {author} {\bibinfo {author} {\bibfnamefont {A.~F.}\ \bibnamefont
  {Falk}}\ and\ \bibinfo {author} {\bibfnamefont {T.}~\bibnamefont {Mehen}},\
  }\href {\doibase 10.1103/PhysRevD.53.231} {\bibfield  {journal} {\bibinfo
  {journal} {Phys. Rev.}\ }\textbf {\bibinfo {volume} {D53}},\ \bibinfo {pages}
  {231} (\bibinfo {year} {1996})}\BibitemShut {NoStop}%
\bibitem [{\citenamefont {Colangelo}\ \emph
  {et~al.}(2006{\natexlab{a}})\citenamefont {Colangelo}, \citenamefont
  {De~Fazio},\ and\ \citenamefont {Ferrandes}}]{Colangelo:2005gb}%
  \BibitemOpen
  \bibfield  {author} {\bibinfo {author} {\bibfnamefont {P.}~\bibnamefont
  {Colangelo}}, \bibinfo {author} {\bibfnamefont {F.}~\bibnamefont {De~Fazio}},
  \ and\ \bibinfo {author} {\bibfnamefont {R.}~\bibnamefont {Ferrandes}},\
  }\href {\doibase 10.1016/j.physletb.2006.01.021} {\bibfield  {journal}
  {\bibinfo  {journal} {Phys. Lett.}\ }\textbf {\bibinfo {volume} {B634}},\
  \bibinfo {pages} {235} (\bibinfo {year} {2006}{\natexlab{a}})}\BibitemShut
  {NoStop}%
\bibitem [{\citenamefont {Brodzicka}\ \emph {et~al.}(2008)\citenamefont
  {Brodzicka} \emph {et~al.}}]{Brodzicka:2007aa}%
  \BibitemOpen
  \bibfield  {author} {\bibinfo {author} {\bibfnamefont {J.}~\bibnamefont
  {Brodzicka}} \emph {et~al.} (\bibinfo {collaboration} {Belle}),\ }\href
  {\doibase 10.1103/PhysRevLett.100.092001} {\bibfield  {journal} {\bibinfo
  {journal} {Phys. Rev. Lett.}\ }\textbf {\bibinfo {volume} {100}},\ \bibinfo
  {pages} {092001} (\bibinfo {year} {2008})}\BibitemShut {NoStop}%
\bibitem [{\citenamefont {Aubert}\ \emph {et~al.}(2006)\citenamefont {Aubert}
  \emph {et~al.}}]{Aubert:2006mh}%
  \BibitemOpen
  \bibfield  {author} {\bibinfo {author} {\bibfnamefont {B.}~\bibnamefont
  {Aubert}} \emph {et~al.} (\bibinfo {collaboration} {BaBar}),\ }\href
  {\doibase 10.1103/PhysRevLett.97.222001} {\bibfield  {journal} {\bibinfo
  {journal} {Phys. Rev. Lett.}\ }\textbf {\bibinfo {volume} {97}},\ \bibinfo
  {pages} {222001} (\bibinfo {year} {2006})}\BibitemShut {NoStop}%
\bibitem [{\citenamefont {Colangelo}\ \emph {et~al.}(2008)\citenamefont
  {Colangelo}, \citenamefont {De~Fazio}, \citenamefont {Nicotri},\ and\
  \citenamefont {Rizzi}}]{Colangelo:2007ds}%
  \BibitemOpen
  \bibfield  {author} {\bibinfo {author} {\bibfnamefont {P.}~\bibnamefont
  {Colangelo}}, \bibinfo {author} {\bibfnamefont {F.}~\bibnamefont {De~Fazio}},
  \bibinfo {author} {\bibfnamefont {S.}~\bibnamefont {Nicotri}}, \ and\
  \bibinfo {author} {\bibfnamefont {M.}~\bibnamefont {Rizzi}},\ }\href
  {\doibase 10.1103/PhysRevD.77.014012} {\bibfield  {journal} {\bibinfo
  {journal} {Phys. Rev.}\ }\textbf {\bibinfo {volume} {D77}},\ \bibinfo {pages}
  {014012} (\bibinfo {year} {2008})}\BibitemShut {NoStop}%
\bibitem [{\citenamefont {Colangelo}\ \emph
  {et~al.}(2006{\natexlab{b}})\citenamefont {Colangelo}, \citenamefont
  {De~Fazio},\ and\ \citenamefont {Nicotri}}]{Colangelo:2006rq}%
  \BibitemOpen
  \bibfield  {author} {\bibinfo {author} {\bibfnamefont {P.}~\bibnamefont
  {Colangelo}}, \bibinfo {author} {\bibfnamefont {F.}~\bibnamefont {De~Fazio}},
  \ and\ \bibinfo {author} {\bibfnamefont {S.}~\bibnamefont {Nicotri}},\ }\href
  {\doibase 10.1016/j.physletb.2006.09.018} {\bibfield  {journal} {\bibinfo
  {journal} {Phys. Lett.}\ }\textbf {\bibinfo {volume} {B642}},\ \bibinfo
  {pages} {48} (\bibinfo {year} {2006}{\natexlab{b}})}\BibitemShut {NoStop}%
\bibitem [{\citenamefont {Manohar}\ and\ \citenamefont
  {Georgi}(1984)}]{Manohar:1983md}%
  \BibitemOpen
  \bibfield  {author} {\bibinfo {author} {\bibfnamefont {A.}~\bibnamefont
  {Manohar}}\ and\ \bibinfo {author} {\bibfnamefont {H.}~\bibnamefont
  {Georgi}},\ }\href {\doibase 10.1016/0550-3213(84)90231-1} {\bibfield
  {journal} {\bibinfo  {journal} {Nucl. Phys.}\ }\textbf {\bibinfo {volume}
  {B234}},\ \bibinfo {pages} {189} (\bibinfo {year} {1984})}\BibitemShut
  {NoStop}%
\bibitem [{\citenamefont {Goity}\ and\ \citenamefont
  {Roberts}(1999)}]{Goity:1998jr}%
  \BibitemOpen
  \bibfield  {author} {\bibinfo {author} {\bibfnamefont {J.~L.}\ \bibnamefont
  {Goity}}\ and\ \bibinfo {author} {\bibfnamefont {W.}~\bibnamefont
  {Roberts}},\ }\href {\doibase 10.1103/PhysRevD.60.034001} {\bibfield
  {journal} {\bibinfo  {journal} {Phys. Rev.}\ }\textbf {\bibinfo {volume}
  {D60}},\ \bibinfo {pages} {034001} (\bibinfo {year} {1999})}\BibitemShut
  {NoStop}%
\bibitem [{\citenamefont {Di~Pierro}\ and\ \citenamefont
  {Eichten}(2001)}]{DiPierro:2001dwf}%
  \BibitemOpen
  \bibfield  {author} {\bibinfo {author} {\bibfnamefont {M.}~\bibnamefont
  {Di~Pierro}}\ and\ \bibinfo {author} {\bibfnamefont {E.}~\bibnamefont
  {Eichten}},\ }\href {\doibase 10.1103/PhysRevD.64.114004} {\bibfield
  {journal} {\bibinfo  {journal} {Phys. Rev.}\ }\textbf {\bibinfo {volume}
  {D64}},\ \bibinfo {pages} {114004} (\bibinfo {year} {2001})}\BibitemShut
  {NoStop}%
\bibitem [{\citenamefont {Li}\ \emph {et~al.}(2018)\citenamefont {Li},
  \citenamefont {Wang}, \citenamefont {Jiang}, \citenamefont {Tan},
  \citenamefont {Li}, \citenamefont {Wang},\ and\ \citenamefont
  {Chang}}]{Li:2017zng}%
  \BibitemOpen
  \bibfield  {author} {\bibinfo {author} {\bibfnamefont {S.-C.}\ \bibnamefont
  {Li}}, \bibinfo {author} {\bibfnamefont {T.}~\bibnamefont {Wang}}, \bibinfo
  {author} {\bibfnamefont {Y.}~\bibnamefont {Jiang}}, \bibinfo {author}
  {\bibfnamefont {X.}~\bibnamefont {Tan}}, \bibinfo {author} {\bibfnamefont
  {Q.}~\bibnamefont {Li}}, \bibinfo {author} {\bibfnamefont {G.-L.}\
  \bibnamefont {Wang}}, \ and\ \bibinfo {author} {\bibfnamefont {C.-H.}\
  \bibnamefont {Chang}},\ }\href {\doibase 10.1103/PhysRevD.97.054002}
  {\bibfield  {journal} {\bibinfo  {journal} {Phys. Rev.}\ }\textbf {\bibinfo
  {volume} {D97}},\ \bibinfo {pages} {054002} (\bibinfo {year}
  {2018})}\BibitemShut {NoStop}%
\bibitem [{\citenamefont {Yu}\ \emph {et~al.}(2016)\citenamefont {Yu},
  \citenamefont {Wang},\ and\ \citenamefont {Li}}]{Yu:2016mez}%
  \BibitemOpen
  \bibfield  {author} {\bibinfo {author} {\bibfnamefont {G.~L.}\ \bibnamefont
  {Yu}}, \bibinfo {author} {\bibfnamefont {Z.~G.}\ \bibnamefont {Wang}}, \ and\
  \bibinfo {author} {\bibfnamefont {Z.~Y.}\ \bibnamefont {Li}},\ }\href
  {\doibase 10.1103/PhysRevD.94.074024} {\bibfield  {journal} {\bibinfo
  {journal} {Phys. Rev.}\ }\textbf {\bibinfo {volume} {D94}},\ \bibinfo {pages}
  {074024} (\bibinfo {year} {2016})}\BibitemShut {NoStop}%
\end{thebibliography}%

\end{document}